\documentclass[sn-basic]{sn-jnl}

\usepackage{graphicx}%
\usepackage{multirow}%
\usepackage{amsmath,amssymb,amsfonts}%
\usepackage{amsthm}%
\usepackage{mathrsfs}%
\usepackage[title]{appendix}%
\usepackage{xcolor}%
\usepackage{textcomp}%
\usepackage{manyfoot}%
\usepackage{booktabs}%
\usepackage{algorithm}%
\usepackage{algorithmicx}%
\usepackage{algpseudocode}%
\usepackage{listings}%
\usepackage{upgreek}%
\usepackage[center]{subfigure}%

\newcommand{\vecx}{\ensuremath{\boldsymbol{x}}}

\newcommand{\veck}{\ensuremath{\boldsymbol{k}}}
\newcommand{\vecf}{\ensuremath{\boldsymbol{f}}}
\newcommand{\vecg}{\ensuremath{\boldsymbol{g}}}
\newcommand{\vecu}{\ensuremath{\boldsymbol{u}}}
\newcommand{\vecU}{\ensuremath{\boldsymbol{U}}}
\newcommand{\vecE}{\ensuremath{\boldsymbol{E}}}
\newcommand{\vecB}{\ensuremath{\boldsymbol{B}}}
\newcommand{\vecJ}{\ensuremath{\boldsymbol{J}}}
\newcommand{\vecnab}{\ensuremath{\boldsymbol{\nabla}}}

\newcommand{\dd}{\ensuremath{\mathrm{d}}}

\newcommand{\FT}[1]{\ensuremath{\widehat{#1}}}

\newcommand{\vect}[1]{\ensuremath{\boldsymbol{#1}}}
\newcommand{\tens}[1]{\ensuremath{\boldsymbol{\mathsf{#1}}}}

\newcommand{\frho}{\ensuremath{\langle\rho\rangle}}
\newcommand{\fX}{\ensuremath{\tilde{X}}}
\newcommand{\fU}{\ensuremath{\tilde{U}}}
\newcommand{\fE}{\ensuremath{\tilde{E}}}
\newcommand{\fEu}{\ensuremath{\tilde{E}^\mathrm{u}}}
\newcommand{\fEb}{\ensuremath{\tilde{E}^\mathrm{b}}}
\newcommand{\fB}{\ensuremath{\overline{B}}}
\newcommand{\fJ}{\ensuremath{\overline{J}}}
\newcommand{\fe}{\ensuremath{\tilde{e}}}
\newcommand{\fP}{\ensuremath{\langle P\rangle}}

\newcommand{\fvecu}{\ensuremath{\tilde{\vecu}}}
\newcommand{\fvecU}{\ensuremath{\tilde{\vecU}}}
\newcommand{\fvecE}{\ensuremath{\overline{\vecE}}}
\newcommand{\fvecB}{\ensuremath{\overline{\vecB}}}
\newcommand{\fvecJ}{\ensuremath{\overline{\vecJ}}}

\newcommand{\fh}{\ensuremath{\tilde{h}}}
\newcommand{\Fconv}{\ensuremath{\boldsymbol{\mathfrak{F}}^{\mathrm{(conv)}}}}
\newcommand{\Fdiff}{\ensuremath{\boldsymbol{\mathfrak{F}}^{\mathrm{(kin)}}}}
\newcommand{\Fpress}{\ensuremath{\boldsymbol{\mathfrak{F}}^{\mathrm{(press)}}}}
\newcommand{\Fchem}{\ensuremath{\boldsymbol{\mathfrak{F}}^{\mathrm{(chem)}}}}

\newcommand{\Reyn}{\ensuremath{\mathrm{Re}}}

\newcommand{\adot}{\ensuremath{\dot{a}}}

\begin{document}

\title{Large eddy simulations in astrophysics}


\author[1]{\fnm{Wolfram} \sur{Schmidt-Br\"uckner}}\email{wolfram.schmidt-brueckner@uni-hamburg.de}

\affil[1]{\orgdiv{Hamburger Sternwarte}, \orgname{Universit\"at Hamburg}, \orgaddress{\street{Gojenbergsweg 112}, \city{Hamburg}, \postcode{21029}, \country{Germany}}}


\abstract{In this review, the methodology of large eddy simulations (LES) is introduced and
applications in astrophysics are discussed. As theoretical framework,
the scale decomposition of the dynamical equations for compressible neutral fluids 
by means of spatial filtering is explained. For cosmological applications,
the filtered equations in co-moving coordinates are formulated. Moreover,
the decomposition is extended to magnetohydrodynamics (MHD).
While energy is dissipated through numerical diffusivities in 
implicit large eddy simulations (ILES), explicit subgrid-scale (SGS) models
are applied in LES to compute energy dissipation, mixing, and dynamo action
due to numerically unresolved turbulent eddies. 
The most commonly used models in astrophysics are the Smagorinsky model, 
the hydrodynamical SGS turbulence energy equation model,
and the non-linear structural model for both non-relativistic and relativistic MHD.
Model validation is carried out \emph{a priori} by testing correlations between
model and data for specific terms or \emph{a posteriori} by comparing turbulence statistics
in LES and ILES. Since most solvers in astrophysical simulation codes
have significant numerical diffusion, the additional effect of SGS models
is generally small. However, convergence with resolution increases in some cases. 
A recent example is magnetic field amplification in binary neutron star mergers.
For mesh-free codes, it has been shown that explicit modelling of turbulent diffusion 
of metals has a significant impact.
Moreover, SGS models can help to compute the turbulent velocity dispersion consistently
and to parameterize sub-resolution processes that are influenced 
by turbulence, such as the star formation efficiency in galaxy simulations.
}

\keywords{Large eddy simulations (LES), Numerical simulations, Turbulence, Hydrodynamics, Magnetohydrodynamics
\\
\\This article is a revised version of \url{https://doi.org/10.1007/lrca-2015-2}. \\
\textbf{Change summary} Major revision, updated and expanded.
}

\maketitle

\clearpage
\setcounter{tocdepth}{3} 
\tableofcontents



\section{Introduction}
\label{sec:intro}

Turbulent flows with high Reynolds numbers are often encountered in computational astrophysics.
Examples are the solar wind, stellar convection zones, supernovae, star-forming clouds, and 
probably the gas in galaxy clusters. For a broad overview of astrophysical turbulence,
see \cite{hille_lnp_2009}. This review concentrates on computational methods that
discretize and solve the fluid dynamics equations for a limited range of length scales, 
while small-scale effects are treated by a subgrid-scale model. Since turbulence is not
fully resolved, this approach is called large eddy simulation (LES). 

If no external forces act on a fluid, the velocity field $\vecu(t,\vecx)$ is governed by the partial differential equation
\begin{equation}
 \label{eq:navier_forcefree}
  \frac{\partial}{\partial t}\rho\vecu + 
  \vecnab\cdot\left(\rho\vecu\otimes\vecu\right)\ =
  -\vecnab P + \vecnab\cdot\tens{\sigma}\,,
\end{equation}
where $\otimes$ signifies the tensor product, $\rho$ and $P$ are the mass density and pressure, respectively, and $\tens{\sigma}$ is the viscous stress tensor. The Navier-Stokes equation and accompanying equations for density and energy will be covered in detail in Sect.~\ref{sc:decomp_navier}. In the incompressible limit ($\vecnab\cdot\vecu=0$), it simplifies to
\begin{equation}
 \label{eq:navier_incompress}
  \frac{\partial}{\partial t}\vecu + 
  \vecu\cdot\vecnab\vecu =
  -\frac{1}{\rho}\vecnab P + \nu\nabla^2\vecu\,. \\
\end{equation}
The coefficient $\nu$ represents the kinematic viscosity of the fluid. It can be understood as momentum diffusivity and is a microscopic property of the fluid. The viscous term on the right-hand side of the equation damps fluid motions, while the advection term on the left-hand side causes non-linear interactions between different scales. This effect is at the core of subgrid-scale modeling. From simple dimensional analysis, it follows that the advection term is of the order $V^2/L$, where $V$ is the characteristic velocity of the flow and $L$ its integral length scale. The importance of non-linear interactions relative to viscous damping, which is of the order $\nu V/L^2$, is specified by the Reynolds number:
\begin{equation}
	\label{eq:Re}
	\Reyn = \frac{VL}{\nu} \,.
\end{equation}
The flow becomes turbulent if the non-linear interactions are much stronger than viscous 
damping. Generally, this happens if $\Reyn$ reaches values greater than a few $10^3$
for flows with boundaries. But $\Reyn$ in astrophysical flows are typically much greater than that.
For instance, an estimate for the turbulent convection zone of the Sun is $\Reyn\sim 10^{14}$
\citep{Canuto94}.

In principle, we can also define a scale-dependent Reynolds number $\Reyn(\ell) = v'(\ell)\ell/\nu$, 
where $v'(\ell)$ is the typical magnitude of velocity fluctuations on the
length scale $\ell$. The length sale of strong viscous damping
is then given by $\Reyn(\ell_{\mathrm{K}})\sim 1$. For incompressible turbulence,
substitution of the Kolmogorov--Obukhov scaling law $v'(\ell)\sim(\epsilon\ell)^{1/3}$
yields \citep{Frisch}
\[
	\frac{\epsilon^{1/3}\ell_{\mathrm{K}}^{4/3}}{\nu}\sim 1\,.
\]
Since the mean dissipation rate is $\epsilon\sim V^3/L$, it follows that
\begin{equation}
	\frac{L}{\ell_{\mathrm{K}}}\sim\Reyn^{3/4}.
\end{equation}
The problem of high \Reyn\ is thus a problem of largely different length scales
or, equivalently, a high number of degrees of freedom.

In a numerical simulation of turbulence, the range of length scales is
limited by the grid scale $\Delta$, which is simply the linear size of the
grid cells in the case of finite-volume methods (the closest equivalent for
particle-based methods would be the smoothing length). Only if 
$\Delta\lesssim\ell_{\mathrm{K}}$, turbulence can be fully resolved in a 
so-called \emph{direct numerical simulation} (DNS).%
\footnote{
	In astrophysics, the term DNS is sometimes used in an improper sense, which actually
	corresponds to ILES (see below). We strictly distinguish between DNS and ILES here,
	which means that a DNS has to explicitly treat the \emph{physical} viscosity of
	the fluid, i.e., it solves the Navier--Stokes equations.}
However, DNS become infeasible even for moderately high \Reyn\ 
because the total amount of floating point operations (FLOPs) increases with
$(L/\Delta)^4\gtrsim (L/\ell_{\mathrm{K}})^4\sim \Reyn^{\,3}$. 
The scaling may differ for highly compressible turbulence, but the basic problem 
remains the same. 
For a DNS of solar convection over one dynamical time scale, it would be necessary to
perform very roughly $10^{42}$ FLOPs, which is far beyond the reach of even
exascale computing. 

In practice, however, it is neither feasible nor strictly necessary to account for 
all degrees of freedom in a simulation of high-\Reyn\ turbulence.
To reproduce certain statistical properties, a much coarser sampling of the
degrees of freedom can be quite sufficient for many applications.
This is why LES encompass
only the energy-containing scales and structures dominated by 
non-linear interactions and the turbulent cascade down to
a cut-off scale that can be much greater than the microscopic dissipation scale. 
The cut-off scale is roughly given by grid scale $\Delta$.\footnote{More precisely,
	$\Delta$ is the smallest represented scale. Since small multiples of the grid
	can be significantly affected by numerical truncation erros, the range of resolved scales 
	is constrained by the numerical method \citep{Garnier}.}
The defining criterion for LES is thus $L\gg\Delta\gg \ell_{\mathrm{K}}$ or, equivalently,
\[
	\Reyn\gg\Reyn(\Delta)\gg 1\,.
\]
Here, $\Reyn(\Delta)\sim v'(\Delta)\Delta/\nu$ is the Reynolds number of subgrid-scale turbulence.
The product $v'(\Delta)\Delta$ can be interpreted as turbulent viscosity of the numerically
unresolved eddies of size $\ell\lesssim \Delta$.
The effective Reynolds number of the numerically computed flow is therefore given by
\begin{equation}
	\label{eq:Re_eff}
	\Reyn_{\mathrm{eff}} = \frac{\Reyn}{\Reyn(\Delta)}\sim \frac{VL}{v'(\Delta)\Delta}\sim 
	\left(\frac{L}{\Delta}\right)^{4/3}\,.
\end{equation}
This means that LES reduces the number of degrees of freedom by replacing 
the microscopic viscosity $\nu$ by a turbulent viscosity
of the order $v'(\Delta)\Delta\gg \nu$. As a result, the purely non-linear turbulent dynamics of the ``large eddies''
is separated from microscopic dissipation.%
\footnote{For many applications, particularly in astrophysics,
  the definition used here is appropriate. In a broader sense, LES may
  include the case where microscopic dissipation is partially
  resolved. DNS can then be considered as limiting case of LES for
  $\Reyn(\Delta)\sim 1$.}
The biggest challenge when implementing this concept is
to find an appropriate model for the coupling between the small- and large-scale dynamics.

A mathematical framework for LES is based on the notion of a filter, which separates large-scale
($\ell\gtrsim\Delta$) from small-scale ($\ell\lesssim\Delta$) fluctuations. 
Filters can be used to decompose the equations of fluid dynamics into equations for smoothed
variables, which have a very similar mathematical structure as the unfiltered equations, and equations for
second-order moments of the fluctuations. 
The latter are interpreted as subgrid-scale variables. 
In Sect.~\ref{sec:separation}, we will carry out the decomposition of the compressible
Navier--Stokes equation by applying the filter formalism of \cite{Germano92}.
This formalism comprises the Reynolds-averaged Navier--Stokes (RANS)
equations as limiting case if the filter length is comparable to the integral length scale of the flow.
This method is equivalent to numerically solving a mean-field theory for turbulent flow. 
Simulations based on the RANS equations work with low $\Reyn_{\mathrm{eff}}$, while
LES have high $\Reyn_{\mathrm{eff}}$. In principle, second-order moments can be expressed in terms
of higher-order moments. Since this would entail an infinite hierarchy of moments, 
the set of variables is limited by introducing closures. Usually, one attempts to find
closures for the second-order moments by expressing them in terms of the filtered variables.
This is what is called a subgrid-scale (SGS) model.%
\footnote{In astrophysics, the term subgrid-scale model may
  comprise models that capture sub-resolution physics other than
  turbulence. A typical example are star-formation models in galaxy
  simulations.}
For example, a multi-equation second-order closure model for turbulent convection is formulated in 
\cite{Canuto94}.
Much simpler, yet often employed is the one-equation model for the
SGS turbulence energy $K$, i.e., the local kinetic energy of 
numerically unresolved turbulent eddies.
For this reason, it is sometimes called the $K$-equation model. It can be simplified further 
by assuming local equilibrium between turbulence production and dissipation (Smagorinsky model).
Closures for the transport and source terms in the SGS turbulence energy equation are presented 
in some detail in Sect.~\ref{sec:subgrid}, followed by a discussion of how the closure coefficients 
can be determined (Sect.~\ref{sec:closure}). Of particular importance is the prediction of the
local turbulent viscosity, which is is given by $\Delta\sqrt{K}$ times a dimension-less coefficient.
The turbulent viscosity is required to calculate the turbulent stresses, which enter the 
equations for the filtered variables analogous to the viscous stresses in the unfiltered
Navier--Stokes equations (see Sect.~\ref{sec:turb_stress}).
 
Filtering the dynamical equations is usually considered to be equivalent to numerical discretization.
The filter length can then be identified with the grid scale $\Delta$. Since the
numerical truncation errors of finite difference or finite volume schemes for the
computation of compressible astrophysical flows are more or less diffusion-like terms, 
they produce a numerical viscosity that effectively reduces the Reynolds number to a value comparable to Eq.~(\ref{eq:Re_eff}).
It is actually a common assumption that numerical viscosity approximates the
turbulent viscosity on the grid scale. This leads to the notion of an \emph{implicit} large
eddy simulation (ILES) \citep{Garnier}, 
which is widely used for simulating turbulent flows in astrophysics. 
For example, ILES has been applied to extensively study 
energy spectra and scaling laws of forced compressible turbulence 
\citep{KritNor07,SchmFeder08,KritWag13,federrath_sonic_2021}. 
Since inertial-range scaling is independent of the dissipation mechanism, be it microscopic, turbulent
or due to numerical viscosity, ILES should be able predict scaling laws provided that the 
dynamical range is large enough. However, even in idealized simulations the inertial subrange
is quite narrow for computationally feasible resolutions because the bottleneck
effect distorts the spectrum over a large range of high wave numbers
below the Nyquist wave number \citep{Falko94,SyPort00,DobHaug03,SchmHille06}. 
Moreover, there are indications that large-scale forcing has an impact on smaller scales 
in the strongly supersonic regime.
\citep{federrath_universality_2013,schmidt_kinetic_2019}.
The intrinsic numerical dissipation of finite-volume methods is also the reason why
explicit SGS model have only a marginal effect on turbulence statistics
\citep{SchmFeder11,grete_comparative_2017}. However, this can be different
with higher-order schemes \citep{HauBrand06} or mesh-free methods \citep{rennehan_mixing_2021-1}.

Isotropic hydrodynamical turbulence is just the simplest case. In virtually all astrophysical systems,
there is a complex interplay between fluid dynamics and other physical processes. 
Magnetic fields in particular play an important role in systems as diverse as neutron stars and galaxy clusters
(see \citealt{BrandSub05,Shukurov_Subramanian_2021} for comprehensive overviews). 
The interaction of a magnetic field with partially or fully ionized gas introduces several difficulties,
such as anisotropies and potentially strong back-reaction from smaller to larger scales.
Magnetohydrodynamical turbulence in the interstellar medium (ISM), for example,
extends to the supersonic and super-Alfv\'{e}nic regimes. The exchange of energy between
different scales in these regimes is far more complex than in the simple picture of a 
hydrodynamical turbulent cascade \citep{grete_transfer_2017,beattie_magnetized_2024}.
Moreover, plasmas become collisionless for high temperatures and low densities. A typical example is the
solar corona. Since the MHD description is not applicable in this case, kinetic methods have to
be employed to model dissipative processes such as turbulent reconnection \citep{Buech07,ZweiYa09}. 
However, MHD can provide a reasonable approximation on length scales that are sufficiently large 
compared to the characteristic scales of kinetic processes. Notwithstanding these difficulties, 
some progress has been made with MHD LES using the structural modelling approach (see Sections~\ref{sec:struc} and~\ref{sec:collapse_merge}).

Subgrid scale models offer unique possibilities for modelling physical processes 
that are influenced by turbulence. An example is turbulent deflagration, where the
turbulent diffusivity predicted by the SGS model dominates the effective flame propagation 
speed in underresolved numerical simulations (Sect.~\ref{sec:SN_Ia}). 
Turbulent deflagration arguably plays a role at least in the initial phase of
thermonuclear explosions of white dwarfs, which is one of the
scenarios that are thought to produce type Ia supernovae \citep{Roepke18}.
A particularly promising field of application is the impact of star formation
and feedback processes on the evolution of galaxies. Turbulence
is thought to be an important factor in the regulation of star formation
in the ISM \citep{HenneFalg12}. 
Owing to the limited numerical resolution of simulations of whole galaxies, the
local star formation rate has to be parameterized. The SGS turbulence energy
can be used to couple a locally variable star formation efficiency to
environmental changes in galaxies (Sect.~\ref{sec:galaxy}).

In simulations of cosmological structure formation, which are discussed in Sect.~\ref{sec:clusters}, 
spatially inhomogeneous turbulence is driven by gravity on varying length scales \citep{Brueggen2015}.
Although adaptive methods, such as adaptive mesh refinement (AMR), 
are commonly applied to track down collapsing structures, it is difficult to
to resolve a wide range of length scales between the smallest driving scale and
the grid scale at the highest refinement level. In this situation, SGS effects can become fairly large.
However, the variable grid scale complicates the scale separation in AMR simulations because 
energy has to be transferred between the resolved and SGS energy variables if a region is refined or de-refined.
Sect.~\ref{sec:consrv} describes how to combine hydrodynamical LES and AMR. 
Another difficulty is that turbulence is mostly produced in collapsing structures, while the
surrounding low-density environment has little or no turbulence.
This entails the problem that the SGS model should dynamically adapt to conditions
ranging from laminar flow to developed turbulence. Inhomogeneous and non-stationary
turbulence can be treated by dynamical procedures for the calculation of closure coefficients 
or shear-improved SGS models, which decompose the numerically resolved flow into mean and 
fluctuating components. These techniques are briefly outlined in Sections~\ref{sec:dyn_proc} and~\ref{sec:clusters}. In different variants, adaptively refined LES 
have been applied to galaxy clusters, the intergalactic medium, isolated galaxies, 
and primordial atomic cooling haloes (Sections~\ref{sec:clusters} and~\ref{sec:collapse_merge})
Results from these simulations indicate that the contribution of the 
numerically unresolved turbulent pressure to the support against
gravity can reach a non-negligible fraction of the thermal pressure. Moreover,
the SGS model provides indicators of turbulence production and dissipation and
allows for the computation of the turbulent velocity dispersion. Furthermore, 
SGS turbulence can enhance the diffusion of metals expelled from galaxies 
(Section~\ref{sec:mixing_metals}).



\section{Scale separation}
\label{sec:separation}

Large eddy simulations are based on the notion of scale separation. Although
turbulence is a multi-scale phenomenon, with interactions among different 
length scales, a separation into smoothed and fluctuating components can
be rigorously defined by means of filter operators. Of course, the filtering
of non-linear terms gives rise to interactions between these components.
Filter operators were originally applied in the context of mean-field theories, 
but can be generalized to LES. For incompressible hydrodynamical turbulence, 
\cite{Germano92} introduced a general framework that encompasses 
mean field theories as limiting case.

The smoothed component of a generic field variable $q(\vecx,t)$ is defined 
by means of a spatial low-pass filter,
which is a convolution of $q$ with an appropriate filter kernel $G$ (see also Section 2.3 in \citealt{Garnier}):
\begin{equation}
  \label{eq:q_flt}
  \langle q\rangle_G(\vecx) =
  \int G(\vecx-\vecx')q(\vecx',t)\,\dd^{3}x'.
\end{equation}
A homogeneous isotropic low-pass filter has the following properties:
\begin{itemize}
\item The filter kernel is independent of direction:
	\[
		G(\vecx-\vecx') = G(r),\quad\text{where}\ r=|\vecx-\vecx'|\,.
	\]
\item Filtering smooths out fluctuations on length scales smaller
  than the filter length $\Delta_G$. Length scales that are large in comparison to
  $\Delta_G$ are not affected. This implies
  \[
    G(\vecx-\vecx') \sim \left\{\begin{array}{ll}
				1/\Delta_G^3 &\text{if}\ |\vecx-\vecx'|\ll \Delta_G,\\
				0 &\text{if}\ |\vecx-\vecx'|\gg \Delta_G.
				\end{array}\right.
  \]
\item The filter operator is linear, conserves constants, and commutes with spatial derivatives: 
	\[
		\langle\vecnab q\rangle_G = \vecnab\langle q\rangle_G.
	\]
\end{itemize}
For grid-based codes with Cartesian coordinates $x_i$, it is usually assumed that $\Delta_i=\Delta$ for all spatial dimensions and the isotropy condition is relaxed by assuming that 
\begin{equation}
  \label{eq:cartesian_filter}
  G(\vecx-\vecx') = \prod_{i=1}^3 G_i(x_i-x_i')\,,
\end{equation}
where $G_i(x_i-x_i')$ are identical one-dimensional filter kernels along the coordinate axes. In the following, the index $i$
is omitted and only one-dimensional kernels in $x$-direction are considered.

The simplest low-pass filter is the box or top-hat filter:
\begin{equation}
  \label{eq:box_filter}
  G(x-x')_{\mathrm{box}} = 
  	\left\{\begin{array}{ll}
		1/\Delta &\text{if}\ |x-x'|\le\Delta/2,\\
		0 &\text{otherwise}.
	\end{array}\right.
\end{equation}
The mean value of $q$ in a rectangular domain with periodic boundary conditions 
follows in the limit that $\Delta$ is the linear size of the domain in each dimension. 

The dual definition of the filter operation $\langle\,\,\rangle_G$ in Fourier space is just a multiplication:
\begin{equation}
	\langle\hat{q}\rangle_G(\veck,t)=\hat{q}(\veck,t)\FT{G}(\veck)\,.
\end{equation}
The transfer function $\FT{G}(\veck)$, which is the Fourier transform of the filter kernel in physical space,
drops rapidly to zero for wave numbers $k\gtrsim k_{\mathrm{c}}=\pi/\Delta_{G}$ in the case of a low-pass filter.
As a result, only Fourier modes $\hat{q}(\veck,t)$ with $k\lesssim k_{\mathrm{c}}$ contribute significantly to 
the corresponding filtered field $\langle q\rangle_G(\vecx)$ in physical space.

The simplest case is the sharp cut-off filter, for which
\begin{equation}
  \label{eq:spect_filter}
  \FT{G}_{\mathrm{sharp}}(k) = 
  	\left\{\begin{array}{ll}
		1 &\text{if}\ k\le k_{\mathrm{c}},\\
		0 &\text{otherwise}.
	\end{array}\right.
\end{equation}
The sharp cut-off filter, however, is not equivalent to the box filter, which has
the Fourier representation
\begin{equation}
	\label{eq:sharp_filter}
	\FT{G}_{\mathrm{box}}(k) = \frac{\sin(k\Delta/2)}{k\Delta/2}.
\end{equation}
A filter that is intermediate between these two cases is the Gaussian filter:
\begin{equation}
	\label{eq:gauss_filter}
	\FT{G}_\mathrm{gauss}(k) = \exp\left(-\frac{\Delta^2 k^2}{4\gamma}\right)\,.
\end{equation}
Its representation in physical space is also a Gaussian. The standard choice for the width parameter is $\gamma = 6$
\citep[see][]{VlayGrete16}.

\subsection{Decomposition of the compressible Navier--Stokes equations}
\label{sc:decomp_navier}

The compressible Navier--Stokes equations for the mass density $\rho$, the momentum density $\rho\vecu$, and
the energy density $\rho E$ of an electrically neutral fluid subject to gravitational and mechanical accelerations $\vecg$ and $\vecf$,
respectively, are
\begin{align}
 \label{eq:navier_rho}
  \frac{\partial}{\partial t}\rho + \vecnab\cdot(\vecu\rho)\, &= 0\,, \\
 \label{eq:navier_momt}
  \frac{\partial}{\partial t}\rho\vecu + 
  \vecnab\cdot\left(\rho\vecu\otimes\vecu\right)\, &=
   \rho(\vect{g} + \vect{f}) -\vecnab P + \vecnab\cdot\tens{\sigma}\,, \\
 \label{eq:navier_energy}
   \frac{\partial}{\partial t} \rho E + \vecnab\cdot(\rho\vecu E)\, &=
   \rho\vect{u}\cdot(\vect{g} + \vect{f}) -\vecnab\cdot(\vecu P) + \vecnab\cdot(\vecu\cdot\tens{\sigma})\,.
\end{align}
Thermal conduction is neglected here. The energy per unit mass can be expressed as
\begin{equation}
  \label{eq:e_tot}
    E = e + \frac{1}{2}u^{2}\,,
\end{equation}
where $e$ is the internal gas energy. For a perfect gas, 
$e$ is related to the gas pressure $P$ and the temperature $T$ via the ideal gas law:
\begin{equation}
  \label{eq:e_int}
    e = \frac{P}{(\gamma-1)\rho} = \frac{k_{\mathrm{B}}T}{(\gamma-1)\mu m_{\mathrm{H}}}\,,
\end{equation}
where $\gamma$ is the adiabatic exponent, $k_{\mathrm{B}}$ the Boltzmann constant, 
$\mu$ the mean molecular weight, and $m_{\mathrm{H}}$ the mass of the hydrogen atom.
The viscous stress tensor $\tens{\sigma}$ is defined by
\begin{equation}
  \label{eq:visc_stress}
  \sigma_{ij} =
    2\eta S_{ij}^{\,\ast} + \zeta d\delta_{ij} = 
    2\eta\left(S_{ij} - \frac{1}{3}d\delta_{ij}\right) + \zeta d\delta_{ij}\,,
\end{equation}
where the two coefficients $\eta=\rho\nu$ and $\zeta$ are, respectively, the dynamic shear and bulk
viscosities of the fluid,
\begin{equation}
    \label{eq:strain}
	S_{ij}=\frac{1}{2}\left(\frac{\partial u_i}{\partial x_j}+\frac{\partial u_j}{\partial x_i}\right)
\end{equation}
is the rate-of-strain tensor, and the trace $S_{ii}$ is equal to the divergence $d=\vecnab\cdot\vecu$.
The gravitational acceleration is given by $\vecg=-\vecnab\phi$, where the gravitational potential
$\phi$ is determined by the Poisson equation
\begin{equation}
  \label{eq:poisson}
	\nabla^2\phi = 4\pi G\rho
\end{equation}
with Newton's constant $G$.\footnote{
	For periodic boundary conditions with mean density $\rho_0$, the source term
	on the right hand side is $4\pi G(\rho-\rho_0)$.} 

Mean-field equations for compressible turbulence are derived in \cite{Canuto97}.
Much in the same way, a general low-pass filter $\langle\;\rangle_G$
can be applied to the system of PDEs~(\ref{eq:navier_rho})--(\ref{eq:navier_energy}).\footnote{
	Alternative formulations can be found in \cite{Garnier}, Section 2.4.}
For brevity, we omit the subscript $G$ in the following. 
Since $\langle\;\rangle$ commutes with differential operators, 
the smoothed mass density $\langle\rho\rangle$ obeys an equation of exactly
the same form as the continuity equation,
\begin{equation}
  \label{eq:dens_flt}
  \frac{\partial}{\partial t}\frho + \vect{\nabla}\cdot\frho\fvecu = 0\,,
\end{equation}
if we set $\langle\rho\vecu\rangle=\frho\fvecu$. This relation defines the 
\emph{Favre-filtered} velocity \citep{speziale_subgrid_1988}
\begin{equation}
  \label{eq:favre_flt}
  \fvecu = \frac{\langle\rho\vecu\rangle}{\frho}\,.
\end{equation}
If several chemical species contribute to the mass density,
the fractional density of each species is governed by
\begin{equation}
	  \frac{\partial}{\partial t}\rho X_n + \vecnab\cdot(\vecu\rho X_n) = 0\,,
\end{equation}
where 
\begin{equation}
	\sum_n X_n = 1\,.
\end{equation}
For example, $X_1= X$, $X_2 = Y$, and $X_3 = Z$ are commonly used to specfiy the
mass fractions of hydrogen, helium, and metals, respectively.

It turns out that the filtered continuity equation for an individual species cannot be expressed
in terms of the filtered quantities $\langle\rho X_n\rangle$ and $\fvecu$ alone
(except for the trivial case where the mass fractions $X_n$ are constant):
\begin{equation}
  \frac{\partial}{\partial t}\langle\rho X_n\rangle + \vect{\nabla}\cdot\langle\vecu\rho X_n\rangle = 0\,.
\end{equation}
However, this equation can be cast in a form that is similar to Eq.~(\ref{eq:dens_flt}):
\begin{equation}
  \label{eq:dens_spec_flt}
  \frac{\partial}{\partial t}\frho\fX_n + \vect{\nabla}\cdot\frho\fvecu\fX_n = 
  \vect{\nabla}\cdot\Fchem_n\,,
\end{equation}
where $\fX_n := \langle\rho X_n\rangle/\frho$ is the Favre-filtered mass fraction and
the flux on the right-hand side is identified as
\begin{equation}
  \label{eq:spec_flux}
  \Fchem_n = -\langle\vecu\rho X_n\rangle + \frac{\langle\rho\vecu\rangle\langle\rho X_n\rangle}{\frho} = 
  -\langle\rho\vecu X_n\rangle + \frho\fvecu\fX_n\,.
\end{equation}
As we will see shortly, the additional flux term occurring in Eq.~(\ref{eq:dens_spec_flt}) can be interpreted as 
turbulent transport term.

Let us next consider the momentum equation. Filtering results in
\[
  \frac{\partial}{\partial t}\frho\fvecu + 
  \vecnab\cdot\left\langle\rho\vecu\otimes\vecu\right\rangle =
   \left\langle\rho(-\vecnab\phi + \vect{f})\right\rangle -\vecnab \fP + \vecnab\cdot\langle\tens{\sigma}\rangle
\]
Owing to the non-linearities, we are also facing difficulties here. 
To obtain a PDE with the same basic structure as the unfiltered momentum equation,
the advection term on the left-hand side should read $\vecnab\cdot\left[\langle\rho\rangle\fvecu\otimes\fvecu\right]$.
Similar to the advection of species above, the solution is to split the filtered non-linear terms:
\begin{equation}
	\label{eq:turb_stress}
	\left\langle\rho\vecu\otimes\vecu\right\rangle
	= \langle\rho\rangle\fvecu\otimes\fvecu - \tens{\tau}(\rho\vecu,\vecu)\quad
	\text{where}\quad
	\tens{\tau} := 
	-\langle\rho\vecu\otimes\vecu\rangle + 
	\frac{\langle\rho\vecu\rangle \otimes \langle\rho\vecu\rangle}{\langle\rho\rangle}\,.
\end{equation} 
Since the Poisson equation~(\ref{eq:poisson}) is linear, 
the smoothed potential $\langle\phi\rangle$ is solely determined by $\langle\rho\rangle$. The self-gravity
term $\langle\rho\vecnab\phi\rangle$, however, has to be split by defining
\begin{equation}
	\label{eq:grav_sgs}
	\vect{\psi} := 
	-\langle\rho\vecnab\phi\rangle + \langle\rho\rangle\vecnab\langle\phi\rangle.
\end{equation}
The specific external force $\vect{f}$, on the other hand, usually varies only over the
largest scales of the system. This can be any type of "stirring" force or large-scale
gravitational forces acting on the system \citep[see also][]{WagFalk12}. 
If the filter length is small compared to these scales, 
$\langle\rho\vect{f}\rangle\simeq \langle\rho\rangle\vect{f}$ is a good approximation.
Thus, the filtered momentum equation can be cast into the following form 
\citep{MoinSqui91,Yoshi91,Germano92,Canuto97,SchmNie06b}:
\begin{equation}
  \label{eq:momt_flt}
  \frac{\partial}{\partial t}\frho\fvecu + 
  \vecnab\cdot\left[\langle\rho\rangle\fvecu\otimes\fvecu\right] =
   \langle\rho\rangle(-\vecnab\langle\phi\rangle + \vect{f}) -\vecnab \fP + 
   \vecnab\cdot\left[\langle\tens{\sigma}\rangle+\tens{\tau}\right]
   + \vect{\psi}\,.
\end{equation}

Now, what is the physical interpretation of the terms $\tens{\tau}$?
Let us first consider the weakly compressible limit. By assuming that $\rho$ varies only little over the filter length,
density factors can be pulled out of brackets. In this case, $\fvecu\simeq\langle\vecu\rangle$. 
By defining the fluctuation of the velocity as $\vect{u}'=\vecu-\fvecu$, it follows that
\[
    \tens{\tau} \simeq
    \rho\left[\langle\vecu\rangle \otimes \langle\vecu\rangle
    -\langle\langle\vecu\rangle \otimes \langle\vecu\rangle\rangle 
    - 2\langle\langle\vecu\rangle \otimes \vecu'\rangle
    - \langle\vecu'\otimes\vecu'\rangle\right]\,.
\]
If we further assume that $\langle\;\rangle$ is a Reynolds operator (see Section 3.3 in \citealp{Sagaut}), 
which is not generally true for filters but applies, for example, to global averages, filtered quantities 
can be pulled out of brackets and the above expression simplifies to the second-order moment of the
velocity fluctuation:
\[
    \tens{\tau} \simeq -\rho\langle\vecu'\otimes\vecu'\rangle\,.
\]
Although this simple relation holds only for a Reynolds operator in the weakly compressible limit,
$\tens{\tau}$ is generally interpreted as the stress tensor associated with
the turbulent velocity fluctuations below the filter length. 
For this reason, $\tens{\tau}$ is called the subgrid-scale turbulence stress tensor
in the context of LES. The non-linear interactions
of the filtered flow (the ``large eddies'') with small-scale fluctuations below the grid scale $\Delta$
are given by $\vecnab\cdot\tens{\tau}$ in Eq.~(\ref{eq:momt_flt}). Likewise, 
the term $\vect{\psi}$ defined by Eq.~(\ref{eq:grav_sgs}) accounts for the momentum transfer due self-gravitating
fluctuations in the density and the flux $\Fchem_n$ in Eq.~(\ref{eq:spec_flux}) expresses the coupling
between fluctuations in the concentration of a species and the turbulent flow.

The trace of $\tens{\tau}$ defines the fraction of kinetic energy 
on length scales smaller than the filter length:
\begin{equation}
  \label{eq:sgs_energy}
  \langle\rho\rangle K: = -\frac{1}{2}\tau_{ii} = \frac{1}{2}\langle\rho u^2\rangle - \frac{1}{2}\langle\rho\rangle\tilde{u}^2\,.
\end{equation}
If the filter length is the grid scale, $\rho K$ is called the subgrid-scale turbulence energy. The first term
on the right-hand side of Eq.~(\ref{eq:sgs_energy}) is the total kinetic energy, the second term the kinetic energy  on length scales greater than the filter length (i.e., the numerically resolved kinetic energy in LES).

In the limit of high Reynolds numbers, the viscous dissipation scale (also known as Kolmogorov scale)
is typically much smaller than the filter length. In this case, scaling arguments for incompressible turbulence 
imply $\langle\tens{\sigma}\rangle\ll\tens{\tau}$, i.e., the filtered viscous stresses are negligible
compared to the stresses associated with the turbulent velocity fluctuations.\footnote{
	See \citet{RoepSchm09} for a more detailed discussion.}
Since the scaling of compressible turbulence tends to be stiffer than for incompressible turbulence 
\citep{KritNor07,SchmFeder08}, 
one can reasonably assume that this conclusion is generally applicable. 
The filtered momentum equation~(\ref{eq:momt_flt}) thus can be written as
\begin{equation}
  \label{eq:momt_flt_high_Re}
  \frac{\partial}{\partial t}\frho\fvecu + 
  \vecnab\cdot\left[\langle\rho\rangle\fvecu\otimes\fvecu\right] =\,
   \langle\rho\rangle(-\vecnab\langle\phi\rangle + \vect{f}) -\vecnab\!\left(\fP+\frac{2}{3}\rho K\right)
   + \vecnab\cdot\tens{\tau}^\ast + \vect{\psi}\,.
\end{equation}
where $\tens{\tau}^\ast$ is the trace-free part of $\tens{\tau}$:
\begin{equation}
  \tau_{ij}^\ast=\tau_{ij}-\frac{1}{3}\tau_{kk}\delta_{ij}=\tau_{ij}+\frac{2}{3}\rho K\delta_{ij}\,.
\end{equation}
As one can see from Eq.~(\ref{eq:momt_flt_high_Re}), the trace of $\tens{\tau}$ is associated
with the turbulent pressure $\frac{2}{3}\rho K$ on the filter length scale. 

In contrast to the filtered momentum density, which can be expressed as $\langle\rho\vecu\rangle=\langle\rho\rangle\fvecu$, 
the energy density on length scales greater than the filter length is given by
\begin{equation}
	\label{eq:energy_total}
	\langle\rho\rangle\fE:=
	\langle\rho\rangle\left(\tilde{e} + \frac{1}{2}\tilde{u}^2\right)
	= \langle\rho E\rangle - \langle\rho\rangle K\,,
\end{equation}
where the second equality follows from Eqs..~(\ref{eq:e_tot}) and (\ref{eq:sgs_energy}). 
Consequently, $\langle\rho\rangle\fE\ne \langle\rho E\rangle$.%
\footnote{In \cite{Garnier}, Section 2.4.5, filtered total energy equations for
  various definitions of the total energy are formulated. In particular, 
  $\langle\rho\rangle\fE$ is identified with $\langle\rho E\rangle$ and a different symbol is
  used for ``computable'' energy $\tilde{e} + \frac{1}{2}\tilde{u}^2$. We do not
  follow this nomenclature here. }
A PDE for $\langle\rho\rangle\fE$
follows from the contraction of Eq.~(\ref{eq:momt_flt_high_Re}) with $\fvecu$
plus the filtered internal energy equation. The subtraction of this PDE from the 
filtered equation for the total energy yields the PDE for $\rho K$
\citep{MoinSqui91,Yoshi91,Germano92,Canuto97,SchmNie06b}.
In the limit of high Reynolds numbers, the resulting equations are:
\begin{align}
	\label{eq:pde_energy_res}
	\begin{split}
	\frac{\partial}{\partial t}\langle\rho\rangle\fE + \vecnab\cdot\langle\rho\rangle\fvecu\fE =\,
    	&\langle\rho\rangle\fvecu\cdot(-\vecnab\langle\phi\rangle + \vecf) \\
    	&+ \vecnab\cdot\left[-\fvecu\left(\fP+\frac{2}{3}\rho K\right) + \fvecu\cdot\tens{\tau}^\ast + \Fconv\right]\\
	&- \Sigma + \langle\rho\rangle(\epsilon + \lambda) + \fvecu\cdot\vect{\psi}\,,
	\end{split}\\
	\label{eq:pde_energy_sgs}
	\frac{\partial}{\partial t}\langle\rho\rangle K + \vecnab\cdot\langle\rho\rangle\fvecu K =\,
		&\Psi + \Sigma - \langle\rho\rangle(\epsilon + \lambda) + \vecnab\cdot\left[\Fdiff+\Fpress\right].
\end{align}
The additional source and transport terms resulting form the scale separation of energy are defined as follows.
\begin{itemize}
\item Gravitational energy injection on subgrid scales:
	\begin{equation}
		\Psi = 
		-\langle\rho\vecu\cdot\vecnab\phi\rangle + \fvecu\cdot\langle\rho\vecnab\phi\rangle =\
		-\langle\rho\fvecu\cdot\vecnab\phi\rangle + \langle\rho\rangle\fvecu\cdot\vecnab\langle\phi\rangle
		-\fvecu\cdot\vect{\psi}\,. 
	\end{equation} 
\item Rate of subgrid-scale turbulence energy production:%
\footnote{Also called turbulence energy flux, although this is
  not a transport term. In the incompressible limit, $\Sigma$
  corresponds to the energy transfer in spectral space.}
	\begin{equation}
		\Sigma = \tau_{ij}\tilde{S}_{ij}\,,
		\label{eq:sgs_prod}
	\end{equation} 
	where $\tau_{ij}$ is defined by Eqs..~(\ref{eq:turb_stress})
	and $\tilde{S}_{ij}$ is the rate-of-strain tensor associated
        with the Favre-filtered velocity:%
\footnote{The definition of $\tilde{S}_{ij}$ is a consequence of
  integration by parts of $\tilde{u}_i\partial_j\tau_{ij}$. The symbol
  $\tilde{S}_{ij}$ is used for convenience. It is important to keep in
  mind that $\tilde{S}_{ij}\ne \langle\rho
  S_{ij}\rangle/\langle\rho\rangle$ because $\partial_j\tilde{u}_i
  =\partial_j[\langle\rho u_i\rangle/\langle\rho\rangle]\ne
  \langle\rho\partial_j u_i\rangle/\langle\rho\rangle$.}
	\begin{equation}
	  \label{eq:strain_flt}
	  \tilde{S}_{ij}:=
	  \frac{1}{2}\left(\frac{\partial\tilde{u}_i}{\partial x_j}+\frac{\partial\tilde{u}_j}{\partial x_i}\right)\,.
	\end{equation}
\item Rate of viscous energy dissipation in the limit of high Reynolds
  numbers:%
\footnote{Since $S_{ij}$ is a velocity derivative, it is of the
  order of the velocity fluctuation at the smallest length scales. For
  incompressible turbulence, Kolmogorov scaling implies
  $\langle\sigma\rangle\langle S\rangle\sim \rho\epsilon
  (\Delta/\ell_{\mathrm{K}})^{-4/3}\sim \rho\epsilon/\Reyn(\Delta)$, where
  $\ell_{\mathrm{K}}$ is the Kolmogorov length \citep{RoepSchm09}. For high Reynolds numbers,
  the ratio $\Delta/\ell_{\mathrm{K}}$ is typically very large. As a
  result, $\langle\sigma\rangle\langle S\rangle$ is negligible
  compared to $\rho\epsilon\simeq\langle\sigma S\rangle$. From the
  same estimates follows $\tau \sim (\Delta/\ell_{\mathrm{K}})^{4/3}\langle\sigma\rangle\sim
  \Reyn(\Delta)\langle\sigma\rangle$, which explains why
  the viscous stress term in Eq.~(\ref{eq:momt_flt_high_Re}) can be neglected.}
	\begin{equation}
		\langle\rho\rangle\epsilon = 
		\langle\sigma_{ij}S_{ij}\rangle - \langle\sigma_{ij}\rangle\tilde{S}_{ij}
		\simeq \langle\sigma_{ij}S_{ij}\rangle =
		\langle\eta|S^{\ast}|^{2}+\zeta d^2\rangle\,,
		\label{eq:sgs_diss}
	\end{equation}
	where $S_{ij}$ is defined by Eq.~(\ref{eq:strain}), $|S^{\ast}|^{2} = 2S_{ij}^{\ast}S_{ij}^{\ast}$ 
	is the squared norm of the trace-free rate-of-strain tensor $S_{ij}^{\ast}=S_{ij}-\frac{1}{3}d\delta_{ij}$ 
	and $d=S_{ii}$. Although the viscous stresses can be neglected in the filtered momentum equation,
	viscous dissipation is crucial for the energy balance of turbulent flows. 
\item Rate of subgrid-scale pressure dilatation:
	\begin{equation}
		\langle\rho\rangle\lambda = -\langle d P\rangle + \tilde{d}\langle P\rangle, 
		\label{eq:sgs_press_dilt}
	\end{equation}
	where $\tilde{d}=\tilde{S}_{ii}=\partial\tilde{u}_i/\partial x_i$.
\item Convective internal energy flux on sub-grid scales:%
\footnote{In \cite{SchmNie06b}, the convective flux in Eq.~(\ref{eq:pde_energy_res}) is erroneously defined in 
	terms of the enthalpy. Generally, it is important to distinguish between 
	internal energy and pressure fluctuations because they are associated with different transport terms
	in the energy equation.}
	\begin{equation}
  		\label{eq:flux_conv}
  		\Fconv = 
		-\langle\rho\vecu e\rangle + \frho\fvecu\fe\,.
	\end{equation}
	The convective flux is closely related to the flux of a chemical species, $\Fchem_n$, defined by 
	Eq.~(\ref{eq:spec_flux}). Instead of heat fluctuations, it is the fluctuation in the concentration 
	of a species that is advected by turbulent velocity fluctuations: 
	$\Fchem_n\simeq -\rho\langle\vect{u}'X_n'\rangle$ if
	the flow is weakly compressible.
\item The flux associated with pressure fluctuations:
	\begin{equation}
		\label{eq:flux_press}
		\Fpress = -\langle\vecu P\rangle+\fvecu\langle P\rangle.
	\end{equation}
      For ideal gas with adiabatic exponent $\gamma$, we have $\Fpress=(\gamma-1)\Fconv$.
\item The diffusive flux of turbulent energy on sub-grid scales:
	\begin{equation}
	  \label{eq:flux_diff}
	  \Fdiff =
	  -\frac{1}{2}\langle\rho u^2 \vecu\rangle +
	  \frac{1}{2}\langle\rho u^2\rangle \fvecu - \fvecu\cdot\tens{\tau}\,.
	\end{equation}
      For Reynolds operators in the weakly compressible limit, \Fdiff can be expressed as
      a third-order moment of the velocity fluctuation: 
      $2\mathfrak{F}^{(\mathrm{kin})}_j\simeq -\rho\langle u_i^\prime u_i^\prime u_j^\prime\rangle$
      \citep{Germano92,Canuto97}.
\item There is also a viscous flux, which can be neglected relative to other flux terms if the
      Reynolds number is sufficiently high.
\end{itemize}
By adding the Eqs..~(\ref{eq:pde_energy_res}) and (\ref{eq:pde_energy_sgs}), we obtain an equation
for the filtered total energy:
\begin{equation}
	\begin{split}
	\label{eq:pde_energy_tot}
	\frac{\partial}{\partial t}\langle\rho\rangle(\fE + K)\, 
	+& \vecnab\cdot\langle\rho\rangle\fvecu(\fE + K + \fP) =
    		\langle\rho\rangle\fvecu(-\vecnab\langle\phi\rangle + \vecf) \\
	&+ \Psi + \fvecu\cdot\vect{\psi} 
		+ \vecnab\cdot\left[\fvecu\cdot\tens{\tau} + \Fdiff + \Fconv + \Fpress\right]\,.
	\end{split}
\end{equation}
All SGS terms are collected in the second line. Except for the source terms related to self-gravity, production and dissipation rates cancel out. Thus, the filtered total energy is conserved in the absence of external forces and gravity. The rightmost fluxes are related to turbulent transport processes below the filter length. In particular, the sum of $\Fconv$ and $\Fpress$ can be expressed as
convective enthalpy flux,
\begin{equation}
 	\Fconv + \Fpress= 
	-\langle\rho\vecu h\rangle + \frho\fvecu\fh\,,
\end{equation}
where $\rho h=\rho e+P$, which corresponds to $-\rho\langle\vect{u}'h'\rangle$ in the weakly compressible limit.
For a closed system of PDEs, it is necessary to compute all terms defined above in terms of known
quantities. A rigorous calculation requires further PDEs, which involve higher-order moments and so on ad infinitum.
This is known as the closure problem. A subgrid-scale model truncates the closure problem by approximating
moments above a given order by lower-order moments.
  
\subsection{Cosmological fluid dynamics}
\label{sc:cosmo}

In cosmological simulations, the equations of fluid dynamics are solved in a co-moving coordinate system. Coordinates of observers that are stationary relative to the Hubble expansion of the Universe are constant in this system. The expansion
is characterized by the scale factor $a(t)$, which is determined by the Friedmann equations for a homogeneous
and isotropic cosmology.\footnote{See, for example, \citep{Peacock}, Chapter 3.} 
If the proper coordinates, which include changes of position due to the expansion of the Universe, 
are denoted by $\vecx_{\mathrm{proper}}$ and $t_{\mathrm{proper}}$, the corresponding co-moving coordinates 
are $\vecx=\vecx_{\mathrm{proper}}/a$ and $t=t_{\mathrm{proper}}$. Derivative operators transform as
\[
	\left.\frac{\partial}{\partial t}\right|_{\mathrm{proper}}=\frac{\partial}{\partial t}-\frac{\adot}{a}\vecx\cdot\vecnab
	\quad\text{and}\quad
	\vecnab_{\mathrm{proper}}=\frac{1}{a}\vecnab\,.
\]
Furthermore, the invariance of mass implies that the co-moving baryonic density $\rho$ is related to 
the proper density by $\rho = a^3 \rho_{\mathrm{proper}}$. It can then be shown that the
continuity equation for $\rho$ in co-moving coordinates assumes exactly the same form as 
Eq.~(\ref{eq:navier_rho}):
\[
	\frac{\partial \rho}{\partial t} + \nabla \cdot (\rho \vecu) = 0\,.	
\]
Here, $\vecu$ is the so-called peculiar velocity, which is defined as
\begin{equation}
	\vecu = \dot{\vecx} = \frac{1}{a}\vecu_{\mathrm{proper}}-H\vecx,
\end{equation}
where $\vecu_{\mathrm{proper}}=\dot{\vecx}_{\mathrm{proper}}$ is the proper velocity and $H=\adot/a$ the Hubble constant.
This means that, in the co-moving coordinate system, matter moves with velocity $\vecu$ relative to the Hubble flow $H\vecx$.
With some algebra also the momentum and energy equations can be transformed to co-moving coordinates. The resulting
equations do not have the same form as Eqs..~(\ref{eq:navier_momt}) and~(\ref{eq:navier_energy}), but include additional terms 
with prefactors $H$. However, a particularly simple representation of the momentum and energy equations is obtained 
if the proper peculiar velocity
\begin{equation}
	\label{eq:prop_peculiar}
	\vecU = a\vecu = \vecu_{\mathrm{proper}}-\adot\vecx
\end{equation}
is used in place of $\vecu$.

Filtered dynamical equations for cosmological fluids were first derived by \cite{MaierIap09}
and presented in an alternative formulation in \cite{SchmAlm13}. The applied filter kernel is static
in co-moving coordinates, i.e., the filter length increases proportional to the cosmological
scale factor $a$. Consequently, commutation of the filter with time derivatives is unaffected by the
cosmological expansion and equations for filtered dynamical variables follow completely analogous to 
Sect.~\ref{sc:decomp_navier}. By neglecting gravitational terms associated with fluctuations below the
filter length, the following equations for the filtered mass density 
$\langle\rho\rangle$, the filtered momentum density $\langle\rho\vecU\rangle=\langle\rho\rangle\fvecU$, and 
the energy density $\langle\rho\rangle\fE$, where $\fE = \fe + \frac{1}{2}\fU^2$, are obtained:
\begin{align}
	\label{eq:cosmo_mass_les}
	\frac{\partial \langle\rho\rangle}{\partial t} 
		+ \frac{1}{a} \vecnab \cdot [\langle\rho\rangle\fvecU] =&\, 0\, , \\
	\label{eq:cosmo_momt_les}
	\frac{\partial}{\partial t}a \langle\rho\rangle \fvecU  
	 + \vecnab \cdot [\langle\rho\rangle \fvecU\otimes\fvecU] =&\,
	 - \langle\rho\rangle\vecnab\langle\phi\rangle
	 - \vecnab \fP + \vecnab\cdot\tens{\tau} + \vect{\psi}\, , \\
	\label{eq:cosmo_energy_les}
	\begin{split}
	\frac{\partial}{\partial t}a^2\langle\rho\rangle\fE + a \vecnab\cdot[\langle\rho\rangle\fvecU\fE] =\,
    &- a\langle\rho\rangle\fvecU\cdot\vecnab\langle\phi\rangle 
     + a \vecnab\cdot\left[-\fvecU\fP + \fvecU\cdot\tens{\tau} + \Fconv\right]\\
	&- a \dot{a} [2 - 3 (\gamma - 1)] \langle\rho\rangle\fe 
     - a \left[\Sigma + \langle\rho\rangle(\epsilon + \lambda)\right]
     + a \fvecU\cdot\vect{\psi}\,,
	\end{split}
\end{align}
Here, the filtered internal energy density is $\langle\rho\rangle\fe=\langle\rho e\rangle = \fP/(\gamma - 1)$, 
where $P=a^3 P_{\mathrm{proper}}$, 
and the gravitational potential $\langle\phi\rangle$ of baryonic and dark matter density fluctuations
is given by the cosmological Poisson equation
\begin{equation}
	\nabla^2\langle\phi\rangle = \frac{3H_0^2\Omega_{\mathrm{m}}(t_0)}{2a}\langle\delta_{\mathrm{m}}\rangle,
\end{equation}
where $H_0=\adot(t_0)$ is the Hubble constant and $\Omega_{\mathrm{m}}(t_0)$ the density parameter of matter
at redshift zero. 
Since the mean matter density $\rho_{\mathrm{m, 0}}$ is constant in co-moving coordinates, 
the source term of the Poisson equation can expressed in terms of the
density fluctuation $\delta_{\mathrm{m}}=(\langle\rho_{\mathrm{dm}}+\rho\rangle-\rho_{\mathrm{m, 0}})/\rho_{\mathrm{m, 0}}$ for 
the local dark matter density $\rho_{\mathrm{dm}}$ and baryonic mass density $\rho$. 
The density parameter is defined by
$\Omega_{\mathrm{m}}(t_0)=\rho_{\mathrm{m, 0}}/\rho_{\mathrm{crit,0}}$, where $\rho_{\mathrm{crit,0}}=3H_0^2/(8\pi G)$ is the
critical density at time $t=t_0$.

The kinetic energy associated with peculiar velocity fluctuations below the filter length,
\[
  \langle\rho\rangle K = 
  \frac{1}{2}\langle\rho U^2\rangle - \frac{1}{2}\langle\rho\rangle\fU^2\,,
\]
is given by the dynamical equation
\begin{equation}
	\label{eq:cosmo_k_les}
	\frac{\partial}{\partial t}a^2\langle\rho\rangle K + a\vecnab\cdot[\langle\rho\rangle\fvecU K] =\,
	a\left[\Psi + \Sigma - \langle\rho\rangle(\epsilon + \lambda)\right] + a\vecnab\cdot\left[\Fdiff+\Fpress\right]\,.
\end{equation}
Source and transport terms have prefactors of unity in the momentum equation and $a$ in energy equations. As a result, the definitions of all terms resulting from scale separation in Eqs..~(\ref{eq:cosmo_momt_les}), (\ref{eq:cosmo_energy_les}) and (\ref{eq:cosmo_k_les}) are analogous to the definitions given in Sect.~\ref{sc:decomp_navier}, with $u_i$ being replaced by $U_i$. This suggests that closures for turbulence in a
static space are applicable to cosmological fluids as well. In principle, cosmological expansion causes
a dampening of the kinetic energy \citep{SchmAlm13}. However, this effect is subdominant for 
turbulent eddies even on the largest scales in galaxy clusters because turbulence is driven in gravitationally
bound gas on time scales shorter than the current Hubble time $1/H_0$. 

\subsection{Magnetohydrodynamics}
\label{sc:mhd}

The compressible Navier-Stokes~(\ref{eq:navier_rho}-\ref{eq:navier_energy}) equations can be extended to magnetohydrodynamics. In the following, we focus on the magnetic field and do not consider gravity. Gravitational terms on resolved scales can be added just as in Section~\ref{sc:decomp_navier}. While the continuity equation remains unchanged, the momentum and energy equations for a conducting fluid in the presence of a magnetic field $\vecB$ can be written in Gaussian units as \citep{Spruit16,Shukurov_Subramanian_2021}
\begin{align}
 \label{eq:mhd_momt}
  \frac{\partial}{\partial t}\rho\vecu + 
  \vecnab\cdot\left(\rho\vecu\otimes\vecu - \frac{1}{4\uppi}\vecB\otimes\vecB\right) =\,&
   \rho\vect{f} -\vecnab \left(P + \frac{B^2}{8\uppi}\right) + \vecnab\cdot\tens{\sigma}\,, \\
 \label{eq:mhd_energy}
   \frac{\partial}{\partial t} \rho E + \vecnab\cdot\left[\vecu\left(\frac{1}{2}\rho u^{2} + \rho e + P\right) + \frac{c}{4\uppi}\vecE\times\vecB\right] =\,&
   \rho\vect{u}\cdot\vect{f} + \vecnab\cdot\left(\vecu\cdot\tens{\sigma}\right)\,,
\end{align}
where
\begin{equation}
  \label{eq:mhd_e_tot}
    \rho E = \rho e + \frac{1}{2}\rho u^{2} + \frac{B^2}{8\uppi}
\end{equation}
is the total energy density (take care not to confuse $E$ with the absolute value of the electric field in the following). The last term on the left-hand side of Eq.~(\ref{eq:mhd_energy}) is the Poynting flux. The evolution of the magnetic field is governed by the induction equation:
\begin{equation}
	\frac{\partial\vecB}{\partial t} = \vecnab\times\left(\vecu\times\vecB + \eta\vecnab\times\vecB\right)\,, 
\end{equation}
where $\eta$ is the resistivity (also called magnetic diffusivity). In terms of the conductivity $\sigma$ of the fluid, it can be expressed as $\eta= c^2/(4\uppi\,\sigma)$.

Since $B^2/8\uppi$ is the magnetic energy density, the magnetic field is filtered without mass-weighing and a factor of $1/\sqrt{4\uppi}$ is included for convenience:\footnote{This convention is customarily used for code units in astrophysical MHD codes.}
\begin{equation}
	\label{eq:b_flt}
		\fvecB = \frac{\langle\vecB\rangle}{\sqrt{4\uppi}}\,.
\end{equation}
In the limit of high Reynolds numbers, the following filtered momentum equation for MHD can be derived \citep{VlayGrete16}:
\begin{equation}
  \label{eq:mhd_momt_flt_high_Re}
  \begin{split}
  \frac{\partial}{\partial t}\frho\fvecu + 
  \vecnab\cdot\left(\langle\rho\rangle\fvecu\otimes\fvecu - \fvecB\otimes\fvecB\right)
  =\, & \langle\rho\rangle\vect{f} - \vecnab\!\left(\fP + \frac{1}{2}\fB^2 + \langle\rho\rangle K^\mathrm{b}\right) \\
  &+ \vecnab\cdot\left(\tens{\tau}^\mathrm{u} - \tens{\tau}^\mathrm{b}\right)\,.
  \end{split}
\end{equation}
where $\tens{\tau}^\mathrm{u}$ is the SGS turbulence stress tensor defined by Eq.~(\ref{eq:turb_stress}) and the SGS magnetic stress tensor $\tens{\tau}^\mathrm{b}$ is defined by an expression that is analogous to the incompressible Reynolds stress tensor:
\begin{equation}
	\label{eq:turb_mag_stress}
	\tens{\tau}^\mathrm{b} := -\frac{1}{4\uppi}\langle\vecB\otimes\vecB\rangle + \fvecB\otimes\fvecB\,.
\end{equation}
Similar to equation~(\ref{eq:sgs_energy}), the SGS magnetic energy is related to the trace of this tensor:
\begin{equation}
	\label{eq:sgs_mag_energy}
	\langle\rho\rangle K^\mathrm{b} := -\frac{1}{2}\tau_{ii}^\mathrm{b} = 
	\frac{1}{8\uppi}\langle|\vecB|^2\rangle - \frac{1}{2}\fB^2\,.
\end{equation}
In contrast to Eq.~(\ref{eq:momt_flt_high_Re}), the stress tensors on the right-hand side of the filtered
momentum equation for MHD are not trace-free. The magnetic pressure is a consequence of splitting the full Maxwell stress tensor into a symmetric tensor and a diagonal tensor \citep[see][Section 1.3.2]{Spruit16}. The quantity $K^\mathrm{b}$ can be interpreted as contribution to the magnetic pressure caused by subgrid-scale fluctuations of the magnetic field.

From equation~(\ref{eq:mhd_momt_flt_high_Re}) follows an equation for the kinetic energy $\fEu: = \frac{1}{2}\left|\fvecu\right|^2$ of the filtered
flow:\footnote{The time derivative and advection term can be written in this form by using the continuity equation.}
\begin{equation}
  \label{eq:mhd_energy_kin_flt}
  \begin{split}
	\frac{\partial}{\partial t}\langle\rho\rangle\fEu 
	+& \vecnab\cdot\langle\rho\rangle\fvecu\fEu 
		- \fvecu\cdot\left[\vecnab\cdot\left(\fvecB\otimes\fvecB\right)\right] = \\
	& \langle\rho\rangle\fvecu\cdot\vecf 
		- \fvecu\cdot\vecnab\!\left(\fP + \frac{1}{2}\fB^2 + \langle\rho\rangle K^\mathrm{b}\right)
  		+ \fvecu\cdot\left[\vecnab\cdot\left(\tens{\tau}^\mathrm{u} - \tens{\tau}^\mathrm{b}\right)\right]\,,
	\end{split}
\end{equation}

Similar to the viscous stresses in the momentum equation, we can neglect resistive terms if the magnetic diffusion scale is small compared to the filter scale. Thus, the filtered induction equation for sufficiently high magnetic Reynolds numbers, $\Reyn_\mathrm{m} = VL/\eta$, can be written as
\begin{equation}
	\label{eq:induction_flt}
	\frac{\partial\fvecB}{\partial t} = \vecnab\times\left(\fvecu\times\fvecB + \vect{\mathcal{E}}\right), 
\end{equation}
where the SGS electromotive force (EMF) is defined by
\begin{equation}
	\label{eq:emf_sgs}
	\vect{\mathcal{E}} := \frac{1}{\sqrt{4\uppi}}\langle\vecu\times\vecB\rangle - \fvecu\times\fvecB\,. 
\end{equation}
Eq.~(\ref{eq:induction_flt}) implies
\begin{equation}
	\label{eq:energy_mag_flt}
	\frac{\partial}{\partial t}\langle\rho\rangle\fEb = 
	\fvecB\cdot\left[\vecnab\times\left(\fvecu\times\fvecB + \vect{\mathcal{E}}\right)\right]
\end{equation}
where $\langle\rho\rangle\fEb: = \frac{1}{2}\fB^2$ is both the magnetic energy density and the magnetic pressure associated with the filtered magnetic field. 

For a physical interpretation, it is instructive to derive the equation for $\fEb$ along the lines of \cite{Shukurov_Subramanian_2021}, Section 2.4.
Starting from Faraday's law for the filtered electric and magnetic fields and the filtered current density\footnote{Ampere's law in the MHD approximation.}
\begin{equation}
	\fvecJ := \sqrt{4\uppi}\langle\vecJ\rangle = c\vecnab\times\fvecB\,,
\end{equation}
substitution of Ohm's law gives
\begin{equation}
	\label{eq:energy_mag_flt2}
	\frac{1}{2}\frac{\partial}{\partial t}\fB^2 = 
	-c\vecnab\cdot\left(\fvecE\times\fvecB\right)
	+\frac{1}{c\sqrt{4\uppi}}\langle\vecu\times\vecB\rangle \cdot \fvecJ - \frac{\fJ^2}{4\uppi\,\sigma}\,.
\end{equation}
The first term on the right-hand side is the Poynting flux expressed in terms of $\fvecB$ and the filtered electric field $\fvecE = \langle\vecE\rangle/\sqrt{4\uppi}$. The filtered product $\langle\vecu\times\vecB\rangle$ originates from the Lorentz force. We can use Eq.~(\ref{eq:emf_sgs})
to substitute $\langle\vecu\times\vecB\rangle$ and neglect Ohmic dissipation $\fJ^2/4\uppi\,\sigma = \eta\fJ^2/c^2$ if losses due to microscopic magnetic reconnection on length scales above the filter length are negligible compared to turbulent dissipation of magnetic energy. This corresponds to the ideal MHD limit $\eta\rightarrow 0$
for the filtered magnetic field:
\begin{equation}
  \begin{split}
	\label{eq:energy_mag_flt_high_Re}
	\frac{\partial}{\partial t}\langle\rho\rangle\fEb  
	+ c\vecnab\cdot\left(\fvecE\times\fvecB\right) &= 
	\frac{1}{c}\left(\fvecu\times\fvecB + \vect{\mathcal{E}}\right) \cdot \fvecJ \\
	&= \fvecu\cdot\left[-\vecnab\cdot\left(\fvecB\otimes\fvecB\right) + \vecnab\left(\frac{1}{2}\fB^2\right)\right]
		+ \frac{1}{c}\vect{\mathcal{E}}\cdot\fvecJ \,.
  \end{split}
\end{equation}
The first two terms on the right-hand side (second line) occur with opposite signs in Eq.~(\ref{eq:mhd_energy_kin_flt}) and express the conversion of magnetic energy $\fEb$ into kinetic energy $\fEu$ and vice versa due to tension and pressure of the magnetic field. By applying the ideal MHD relation $c\langle\vecE\rangle = -\langle\vecu\times\vecB\rangle$ in combination with Eq.~(\ref{eq:emf_sgs}) for the EMF, it follows that
\begin{equation}
	\label{eq:poynt_flt}
	\left(c\,\fvecE + \vect{\mathcal{E}}\right)\times\fvecB = \fvecu\,\fB^2 - (\fvecu\cdot\fvecB)\fvecB\,
	= \fvecu_\perp\left(\langle\rho\rangle\fEb + \frac{1}{2}\fB^2\right)\,.
\end{equation}	
\medskip
The last expression can be interpreted as magnetic enthalpy flux in the plane perpendicular to $\fvecB$ \citep{Spruit16}. 

The definition of for the total energy density~(\ref{eq:energy_total}) can be generalized by adding the magnetic energy density associated with the filtered magnetic field:
\begin{equation}
	\label{eq:mhd_energy_total_flt}
	\begin{split}
		\langle\rho\rangle\fE 
	 	:= & \;\langle\rho\rangle\left(\tilde{e} + \frac{1}{2}\tilde{u}^2\right) + \frac{1}{2}\fB^2 \\
	 	= & \;\langle\rho\rangle\left(\tilde{e} + \fEu + \fEb\right)
		= \langle\rho E\rangle - \langle\rho\rangle\left(K^\mathrm{u} + K^\mathrm{b}\right)\,,
	\end{split}
\end{equation}
where the SGS energies $K^\mathrm{u}$ and $K^\mathrm{b}$ are defined by Eqs.~(\ref{eq:sgs_energy}) and~(\ref{eq:sgs_mag_energy}), respectively.
The PDE for $\langle\rho\rangle\fE$ is obtained by adding Eqs.~(\ref{eq:mhd_energy_kin_flt}) and (\ref{eq:energy_mag_flt_high_Re}) and the equation for the filtered internal energy, $\langle\rho\rangle\tilde{e}$. If
Eq.~(\ref{eq:poynt_flt}) is used and the SGS convective flux is neglected, the
resulting equation is
\begin{equation}
	\label{eq:pde_mhd_energy_res}
	\begin{split}
	\frac{\partial}{\partial t}\langle\rho\rangle\fE\,
		+& \vecnab\cdot\left[\fvecu\left(\langle\rho\rangle\fE + \fP + \frac{1}{2}\fB^2\right) 
		- (\fvecu\cdot\fvecB)\fvecB\right] =\,
    	\langle\rho\rangle\fvecu\cdot\vecf\\
	    & + \fvecu\cdot\left[\vecnab\cdot\left(\tens{\tau}^\mathrm{u} - \tens{\tau}^\mathrm{b}\right) 
	      - \vecnab\langle\rho\rangle K^\mathrm{b}\right] + \fvecB\cdot\left(\vecnab\times\vect{\mathcal{E}}\right)
	  	  + \langle\rho\rangle(\epsilon + \lambda)\,,
	\end{split}
\end{equation}
where the terms in the second line are all associated with SGS processes. This equation is non-conservative with respect to the exchange of energy across the filter scale. In principle, one could split the SGS stress terms into conservative transport terms and terms that account for energy transfer across the filter scale, such as the negative rate of SGS turbulence energy production $-\Sigma$ in Eq.~(\ref{eq:pde_energy_res}). The SGS EMF term accounts for the amplification of the magnetic field by a turbulent small-scale dynamo \citep[see][Chapter~6]{Shukurov_Subramanian_2021}. While $-\Sigma$ is mostly dissipative with respect to $\fEu$, the filtered magnetic energy reservoir $\fEb$ can gain energy from the SGS energy reservoir $K^\mathrm{b}$ through an inverse cascade and from $K^\mathrm{u}$ due to small-scale dynamo action. The rate of microscopic energy dissipation (conversion of SGS kinetic and magnetic energy into internal energy) in the limit of high Reynolds numbers is given by
\begin{equation}
	\langle\rho\rangle\epsilon = \langle\rho\rangle\left(\epsilon^\mathrm{u} + \epsilon^\mathrm{b}\right)
	\simeq \langle\sigma_{ij}S_{ij}\rangle + \frac{\langle J^2\rangle}{\sigma}\,.
	\label{eq:sgs_diss_mhd}
\end{equation}
Analogous to viscous dissipation (see Sect.~\ref{sc:decomp_navier}), the second-order moment $\langle J^2\rangle$ corresponds to the filtered Ohmic dissipation on microscopic scales, while $\fJ^2$ is negligible. The pressure dilation $\lambda$ is defined by equation~(\ref{eq:sgs_press_dilt}). 

In addition, one would have to solve PDEs for the SGS energies $K^\mathrm{u}$ and $K^\mathrm{b}$ just like in the hydrodynamical case. However, such a two-equation model would be highly complex and has not been applied in astrophysics so far. An alternative approach that is solely based on the momentum equation~(\ref{eq:mhd_momt_flt_high_Re}), the induction equation~(\ref{eq:induction_flt}), and Eq.~(\ref{eq:pde_mhd_energy_res}) will be discussed in Sect.~\ref{sec:struc}. 



\section{Subgrid-scale models}
\label{sec:subgrid}

There is a beautiful correspondence between finite-volume discretization and filtering. Finite-volume
methods solve an equation for the cell averages of some dynamical variable q(\vect{x}):
\[
	Q_{ijk}=
	\int_{z_k-\Delta/2}^{z_k+\Delta/2}\int_{y_j-\Delta/2}^{y_j+\Delta/2}\int_{x_i-\Delta/2}^{x_i+\Delta/2}
	 q(x,y,z)\,\dd x\,\dd y\,\dd z
\]
Here, $(x_i,y_j,z_k)$ are the cell-centred coordinates and $\Delta$ is the linear size of a grid cell. It is not
difficult to see that $Q_{ijk}$ equals the box-filtered variable $\langle q\rangle_G$ for a box filter $G$ with
filter length $\Delta$ at the discrete points $(x_i,y_j,z_k)$.%
\footnote{An \emph{effective} filter length for isotropic
  turbulence simulations can be calculated from the second moment of
  the compensated energy spectrum as shown in \cite{SchmHille06}.}
Also finite differences correspond to low-pass filters. Thus, numerical
discretization can be interpreted as \emph{implicit filtering}.
The numerical errors in the approximations to $Q_{ijk}$
can be characterized by the truncation error of the finite-volume method. For stable schemes with
flux limiters, these errors are usually associated with terms of the diffusion type, i.e.\
proportional to $\nabla^2 q$.%
\footnote{For higher-order methods such as PPM \citep{ColWood84},
  the leading-order truncation errors correspond to hyper-viscosity
  terms proportional to $\nabla^4 q$.}
In ILES, this is what causes the dissipation of kinetic energy
into heat (see \citealp{Garnier} for further discussion). 
As we shall see, a similar expression follows if the Boussinesq expression for the turbulent
stresses is used as an explicit SGS model for the interaction between numerically resolved and
unresolved turbulent eddies. However, an important difference is that the
turbulent viscosity in the resulting diffusion terms in the momentum and energy
equations is controlled by the dynamical variable $K$, which is the kinetic energy associated with turbulent 
velocity fluctuations below the grid scale. In this section, we first discuss the computation of $K$ in LES.
The aim of this approach, which is known as \emph{functional} modelling, is to model 
only statistical effects of SGS turbulence on the dynamics of the filtered fields,
which are identified with the numerical solution. An alternative strategy is \emph{structural} modelling 
(see Chapter~5 in \citealt{Garnier}), which will be covered in the remainder of this section.

\subsection{Closures for the turbulence stress tensor}
\label{sec:turb_stress}

The SGS turbulence stress tensor, which is associated with the non-linear energy transfer between
large and small scales, is the central quantity that has to be modelled in hydrodynamical LES. The most commonly
used closure is the eddy-viscosity closure. The underlying assumption is that the form of the
trace-free part $\tens{\tau}^\ast$ is analogous to the anisotropic
viscous stress tensor $\tens{\sigma}^{\ast}$, with the correspondence%
\footnote{This idea was originally proposed by Boussinesq in the
  19th century \citep{Boussin77}.}
\begin{align}
	S_{ij} &\longleftrightarrow \tilde{S}_{ij},\\
	\eta   &\longleftrightarrow \langle\rho\rangle\nu_{\mathrm{sgs}}\,,
\end{align}
where $S_{ij}$ and $\tilde{S}_{ij}$ are defined by Eqs..~(\ref{eq:strain}) and (\ref{eq:strain_flt}), respectively. 
The turbulent viscosity $\nu_{\mathrm{sgs}}$ is assumed to depend on the grid scale and the unresolved turbulent velocity fluctuation (see, for example, Sect.~4.3 in \citealp{Sagaut}):
\begin{equation}
	\label{eq:visc_sgs}
	\nu_{\mathrm{sgs}} = C_{\nu}\Delta \sqrt{K}\,.
\end{equation}
Hence,
\begin{equation}
	\label{eq:tau_eddy_flt}
	\tau_{ij}^{(\mathrm{eddy})} = 2\langle\rho\rangle\left(\nu_{\mathrm{sgs}}\tilde{S}_{ij}^\ast-\frac{1}{3}K\delta_{ij}\right)\,.
\end{equation}
For brevity, we omit brackets and tildes that indicate filtered and Favre-filtered quantities so that we can write the turbulent stresses as
\begin{equation}
	\label{eq:tau_eddy}
	\tau_{ij}^{(\mathrm{eddy})} = 2\rho\left(\nu_{\mathrm{sgs}}S_{ij}^{\,\ast}-\frac{1}{3}K\delta_{ij}\right)\,.
\end{equation}
In the following, it is understood that all quantities are either numerically resolved variables or 
modelled in terms of these variables. The production rate (turbulence energy flux) corresponding to the eddy-viscosity
closure is
\begin{equation}
	\label{eq:flux_eddy}
	\Sigma^{(\mathrm{eddy})} = C_{\nu}\rho\Delta K^{1/2}|S^\ast|^2-\frac{2}{3}\rho K d\,,
\end{equation}
where $d=S_{ii}$ is the divergence. The eddy-viscosity coefficient $C_{\nu}$ is typically in the range
from $0.05$ and $0.1$ \citep{Sagaut,SchmNie06b,SchmFeder11}.

In the incompressible case $(d = 0)$, the eddy-viscosity closure admits only positive energy flux.
However, data from numerical simulations show that there is a certain amount of backscattering
from smaller to larger scales, corresponding to a negative energy flux \citep{SchmNie06b,SchmFeder11}. 
An expression for the SGS turbulence stress tensor that is non-linear in terms of the Jacobian matrix $u_{i,k}$ 
of the velocity follows from a Taylor series expansion around grid cell centers \citep{Canuto94,WoodPort06}. 
A normalization that satisfies the identity $\tau_{ii} = -2\rho K$ leads to 
\begin{equation}
	\label{eq:tau_nonlin_norm}
	\tau_{ij}^{(\mathrm{nonlin})} = -4\rho K\frac{u_{i,k}u_{j,k}}{|\vecnab\otimes\vecu|^2}\,,
\end{equation}
where $|\vecnab\otimes\vecu|=(2u_{i,k}u_{i,k})^{1/2}$ is the norm of the velocity derivative.
However, Eq.~(\ref{eq:tau_nonlin_norm}) is generally not adequate as a model 
for the turbulence stress tensor in LES. In contrast to the eddy-viscosity closure, rotation invariance 
is violated because of the antisymmetric part of $\vecnab\otimes\vecu$. This would cause spurious production of $K$ in a uniformly
rotating fluid. A further problem is that $K = 0$ would be a fixed point of Eq.~(\ref{eq:pde_energy_sgs}) if all
other sources of turbulence energy were zero. This results in unphysical behaviour. With the eddy-viscosity closure, 
on the other hand, $K$ can grow sufficiently fast from arbitrarily small initial values because $\nu_{\mathrm{sgs}}$ 
is proportional to $\sqrt{K}$ rather than $K$. For this reason, \citet{WoodPort06} proposed to use 
a linear combination of $\tau_{ij}^{(\mathrm{nonlin})}$ and $2\nu_{\mathrm{sgs}}S_{ij}^{\,\ast}$, where $\nu_{\mathrm{sgs}}$
is modelled by the determinant of $S_{ij}^{\,\ast}$. The additional term has a small coefficient and seeds the production 
of turbulence energy if the flow is not smooth, while the production rate vanishes for a uniformly rotating fluid.

With the standard turbulent viscosity defined by Eq.~(\ref{eq:visc_sgs}), the same idea leads to the
generalized two-coefficient closure \citep{SchmFeder11}:
\begin{equation}
	\label{eq:tau_mixed}
	\tau_{ij} = 2\rho\left[C_1\Delta (2K)^{1/2}S_{ij}^{\,\ast} -
	2C_2 K\frac{u_{i,k}u_{j,k}}{|\vecnab\otimes\vecu|^2} - \frac{1}{3}(1-C_2)K\delta_{ij}\right]\,.
\end{equation}
The coefficient $C_2$ determines the relative contributions of the non-linear and divergence terms
to the trace $\tau_{ii}$. The purely non-linear closure corresponds to $C_1 = 0$ and $C_2 = 1$. Equation~(\ref{eq:tau_eddy}),
on the other hand, is obtained for $C_1 = C_{\nu}/\sqrt{2}$ and $C_2 = 0$. Calibration for supersonic turbulence
shows that $C_1 = 0.02$ and $C_2 = 0.7$ are robust values for the application in LES (see Sect.~\ref{sec:closure}). 
The rate of production following from the generalized closure is
\begin{equation}
	\label{eq:flux_mixed}
	\Sigma = C_{1}\rho\Delta (2K)^{1/2}|S^\ast|^2
	-4C_2\rho K\frac{u_{i,k}u_{j,k}S_{ij}^{\,\ast}}{|\vecnab\otimes\vecu|^2} - \frac{2}{3}\rho K d\,.
\end{equation}
The first term dominates if $K^{1/2}$ is small compared to $\Delta|S^\ast|$. For strong turbulence intensity,
i.e.\ $K^{1/2}\gtrsim\Delta|\vecnab\otimes\vecu|$,
the second term contributes significantly. The transition is influenced by the ratio $C_2/C_1$.

\subsection{The Smagorinsky model for weakly compressible turbulence}
\label{sec:smag}

In the case of isotropic incompressible turbulence, the mean SGS turbulence energy $\overline{K}$ 
for a sharp cut-off at the length scale $\Delta$ is obtained by integrating the Kolmogorov spectrum 
$E(k)$ over wave numbers $k\ge\pi/\Delta$:
\begin{equation}
	\label{eq:ksgs_kolmogorov}
    \overline{K} = 
    \int_{\pi/\Delta}E(k)\dd k =
    \frac{3}{2}C\bar{\epsilon}^{2/3}
    \left(\frac{\pi}{\Delta}\right)^{-2/3}.
\end{equation}
The mean dissipation rate is therefore given by
\begin{equation}
    \bar{\epsilon} =
    \pi\left(\frac{3C}{2}\right)^{-3/2}
    \frac{\overline{K}^{3/2}}{\Delta} \approx
    0.81\frac{\overline{K}^{3/2}}{\Delta}
\end{equation}
for the Kolmogorov constant $C\approx 1.65$ \citep{Pope}. It is commonly assumed that an
expression of this form also holds for the \emph{local} dissipation rate in LES
(see, for example, \citealp{Sagaut}):
\begin{equation}
    \label{eq:diss_close}
    \epsilon = 
    C_{\epsilon}\frac{K^{3/2}}{\Delta}\,,
\end{equation}
with $C_{\epsilon}\sim 1$. This basically means that the local time scale of 
energy dissipation is given by $\tau_{\epsilon}\sim\Delta/\sqrt{K}$.

A particularly simple SGS model for nearly incompressible turbulence ($d\simeq 0$) can be formulated 
by neglecting all gravitational and transport terms associated with subgrid-scale effects 
in Eq.~(\ref{eq:pde_energy_sgs}). In this case, balance between production and dissipation, i.e.
\[
  \Sigma \simeq \rho\epsilon\,,
\]
implies
\begin{equation}
  \label{eq:sar_smag}
  C_{\nu}\Delta K^{1/2}|S^{\ast}|^2
  \simeq C_{\epsilon}\frac{K^{3/2}}{\Delta}\,.
\end{equation}
Here, the eddy-viscosity closure~(\ref{eq:tau_eddy}) is substituted for $\tau_{ij}^{\ast}$. By
solving for $K$, the Smagorinsky model is obtained \citep{Smago63}:
\begin{equation}
  \label{eq:energy_smag}
  K \simeq \frac{C_\nu}{C_\epsilon}\Delta^2|S^{\ast}|^2\qquad\text{and}\qquad
  \nu_{\mathrm{sgs}} = (C_{\mathrm{s}}\Delta)^2|S^{\ast}|\,,
\end{equation}
where $C_{\mathrm{s}} = (C_\nu^3/C_\epsilon)^{1/4}$. The PDEs~(\ref{eq:dens_flt}), (\ref{eq:momt_flt_high_Re}), 
and~(\ref{eq:pde_energy_res}) with the approximations outlined above form a closed system. 

The corresponding equilibrium dissipation rate is
\begin{equation}
  \label{eq:diss_smag}
  \epsilon \simeq (C_{\mathrm{s}}\Delta)^2|S^{\ast}|^3\,.
\end{equation}
This expression has an important implication. Let us assume that $\epsilon$ in ILES (without explicit SGS model)
can be estimated by an expression that is analogous to the viscous dissipation rate 
in the Navier-Stokes equations:
\[
  \epsilon \sim \nu_{\mathrm{eff}}|S^{\ast}|^{2}\,,
\]
where $\nu_{\mathrm{eff}}$ is the numerical viscosity. While the microscopic viscosity is a property of the
fluid, numerical viscosity is determined by numerical truncation errors. If ILES is assumed to be equivalent to
LES, where $\epsilon$ is given by Eq.~(\ref{eq:diss_smag}), it follows that
\[
	\mathrm{\nu}_{\mathrm{eff}} \simeq \Delta^2|S^{\ast}|\,.
\]
Since the rate of strain $|S^{\ast}|$ fluctuates strongly, this cannot be reconciled with a numerical viscosity that is given by a
constant effective Reynolds number \citep{PanPad09}:
\[
	\nu_{\mathrm{eff}} \sim \frac{V L}{\Reyn_{\mathrm{eff}}}\,.
\]
However, the two expressions for $\epsilon$ are in agreement if $\Reyn_{\mathrm{eff}}$ is estimated from Eq.~(\ref{eq:Re_eff}), where $v'(\Delta) \sim (\epsilon\Delta)^{1/3} \sim |S^{\ast}|\Delta$ is the local velocity fluctuation.
This is a consequence of Eq.~(\ref{eq:sgs_diss}), which implies that the dissipation rate
on the grid scale cannot be expressed as the contraction of the filtered viscous stress tensor with 
the filtered rate-of-strain tensor. The dissipation rate is instead given by the filtered contraction 
of the two tensors (see Sect.~\ref{sc:decomp_navier}). Although the Smagorinsky model is very simple, it sheds some light
on basic properties of turbulent dissipation. As shown in \cite{SchmFeder11}, the argument remains valid even if the 
dissipation rate is calculated for LES of supersonic turbulence with the advanced SGS model 
presented in the following section.
 
\subsection{The compressible subgrid-scale turbulence energy model}
\label{sec:K_eq}

To determine the SGS turbulence energy $K$, one can either invoke the equilibrium condition (\ref{eq:sar_smag}) or numerically solve the PDE~(\ref{eq:pde_energy_sgs}). The latter is called the SGS turbulence energy model \citep{Sagaut,Schu75,MoinSqui91,Yoshi91,Germano92,SchmNie06b,SchmFeder11}.
If gravitational terms are negligible, the turbulence energy equation can be explicitly written as
\begin{equation}
	\label{eq:pde_energy_close}
	\begin{split}
	\frac{\partial}{\partial t}\rho K + \vecnab\cdot\left(\rho\vecu K\right) = &\,
		C_1\rho\Delta(2K)^{1/2}|S^\ast|^2 - 4C_2\rho K\frac{u_{i,k}u_{j,k}S_{ij}^{\,\ast}}{|\vecnab\otimes\vecu|^2}
		-\frac{2}{3}\rho K d\\
		&- C_{\epsilon}\frac{K^{3/2}}{\Delta} + \vecnab\cdot\left[C_\kappa\rho\Delta K^{1/2}\vecnab K\right]\,,
	\end{split}
\end{equation}
where the closure~(\ref{eq:flux_mixed}) for the production rate $\Sigma$, the dissipation rate $\epsilon$ defined by 
Eq.~(\ref{eq:diss_close}), and the gradient-diffusion closure for $\Fdiff + \Fpress$ were substituted
into Eq.~(\ref{eq:pde_energy_sgs}). 
The gradient-diffusion hypothesis, which is also known as Kolmogorov--Prandtl relation, 
is based on the assumption that the turbulent transport of $K$ is a diffusion 
process satisfying Fick's law (see, for example, \citealp{Pope,SchmNie06b}):
\begin{equation}
	\label{eq:grad_diff}
	\Fdiff + \Fpress = \rho\kappa_{\mathrm{sgs}}\vecnab K\,,
\end{equation}
with a turbulent diffusivity
\begin{equation}
	\label{eq:sgs_diff}
	\kappa_{\mathrm{sgs}} = C_{\kappa}\Delta \sqrt{K} = \frac{C_{\kappa}}{C_{\nu}}\nu_{\mathrm{sgs}}\,.
\end{equation}
The ratio $C_{\kappa}/C_{\nu}$ is the Prandtl number of turbulent transport. The determination of the coefficients in Eq.~(\ref{eq:pde_energy_close}) is discussed in Sect.~\ref{sec:closure}.

The non-diagonal components $\tau_{ij}^{\ast}$ in the filtered momentum equation~(\ref{eq:momt_flt_high_Re}) cause a diffusion-like effect that complements the numerical diffusion. However, there is crucial distinction between these two effects. The turbulent stresses in the momentum equation convert numerically resolved kinetic energy into SGS turbulence energy, which is then transported through advection and SGS diffusion and eventually gets dissipated. In contrast, numerical diffusion directly dissipates kinetic energy into heat. This corresponds to the assumption of local equilibrium, as in the Smagorinsky model. 
For strongly diffusive numerical schemes, numerical dissipation is the dominant factor. Conversely, the explicit turbulent stresses gain importance for high-resolution schemes. Additionally, the non-linear term in Eq.~(\ref{eq:tau_mixed}) modifies the diffusive behaviour of the eddy-viscosity term, while the trace term acts as an additional turbulent pressure in Eqs.~(\ref{eq:momt_flt}) and~(\ref{eq:pde_energy_res}).
The sum of the thermal and turbulent pressures is sometimes called the effective pressure:
\begin{equation}
	\label{eq:press_eff}
	P_{\mathrm{eff}} = P + \frac{2}{3}\rho K\,.
\end{equation}
It is important to keep in mind that $P_{\mathrm{eff}}$ depends on the numerical resolution and 
$P_{\mathrm{eff}}\rightarrow P$ in the limit $\Delta\rightarrow 0$ (DNS). Although $\frac{2}{3}\rho K$ is small 
compared to the thermal pressure, the intermittency of turbulence can locally produce an effective pressure 
that exceeds the thermal pressure by one order of magnitude \citep{SchmFeder11}. Thus, the turbulent pressure
can become important for compressible turbulence, particularly if there are other sources than the turbulent cascade. 
As an example, turbulent feedback in galaxy simulations is discussed in Sect.~\ref{sec:galaxy}.

The pressure-dilatation $\lambda$ is usually neglected \citep{WoodPort06,SchmFeder11}.
At high Mach numbers, negligible pressure-dilatation is a reasonable assumption because kinetic energy is large
compared to internal energy and non-linear interactions between turbulent velocity fluctuations should be the dominant mode
of energy transfer. Both theoretical and numerical studies support this conjecture. For instance, \citet{Aluie11,Aluie13} argues that a range of length scales exists where the kinetic and internal energies decouple, and the flux through the kinetic energy cascade becomes asymptotically constant while $\rho\lambda$ becomes subdominant. Furthermore, the computation of the different contributions to the total energy flux at varying length scales from supersonic turbulence data corroborates this conclusion in \citep{KritWag13}.
Based on an analytical theory for the two-point correlations of compressible turbulence, \citet{GalBan11} show that 
the main contribution to the energy flux is
\begin{equation}
	F_{\parallel}(r) = \langle \delta(\rho\vecu)\cdot\delta\vecu\,\delta u_{\parallel} \rangle\,,
\end{equation}
where $\delta\vecu$ is the velocity difference between two points separated by a distance $r$, $\delta u_{\parallel}$
is the longitudinal component of the velocity difference in the direction of $\vect{r}$, and the brackets
denote the ensemble average. Equation~(\ref{eq:flux_mixed}) for $\Sigma$ has a similar structure,
with factors of $\rho K^{1/2}$ and derivatives of $\vecu$ corresponding to fluctuations on the grid scale. 

A subtlety emerges when considering discontinuities. As highlighted in Section 2.5.2 of \cite{Garnier}, the Rankine-Hugeniot conditions governing jumps across shock fronts should be filtered rather than the PDEs. This entails SGS terms that differ from those in the filtered PDEs. However, it remains uncertain whether attempting to model these terms would be beneficial. The assumption that the unmodified jump conditions apply to the numerical solution essentially represents a fallback from LES with an explicit SGS model to ILES. Since shock-capturing schemes, such as PPM, revert to stronger diffusion in the vicinity of shocks, this approach is likely the most practical. Nevertheless, the closure equation (\ref{eq:flux_mixed}) accurately captures the non-linear inter-scale transfer of energy due to supersonic turbulent velocity fluctuations. The SGS model outlined above accounts for the statistical impact of shocks and vortices interacting across the grid scale, while SGS terms in the jump conditions primarily correct geometric differences between smoothed shock fronts and the corresponding unfiltered fronts with substructure on smaller scales (similar to the turbulent flame fronts discussed in Sect.~\ref{sec:SN_Ia}).

\begin{figure}[htbp]
    \centerline{\includegraphics[width=0.9\textwidth]{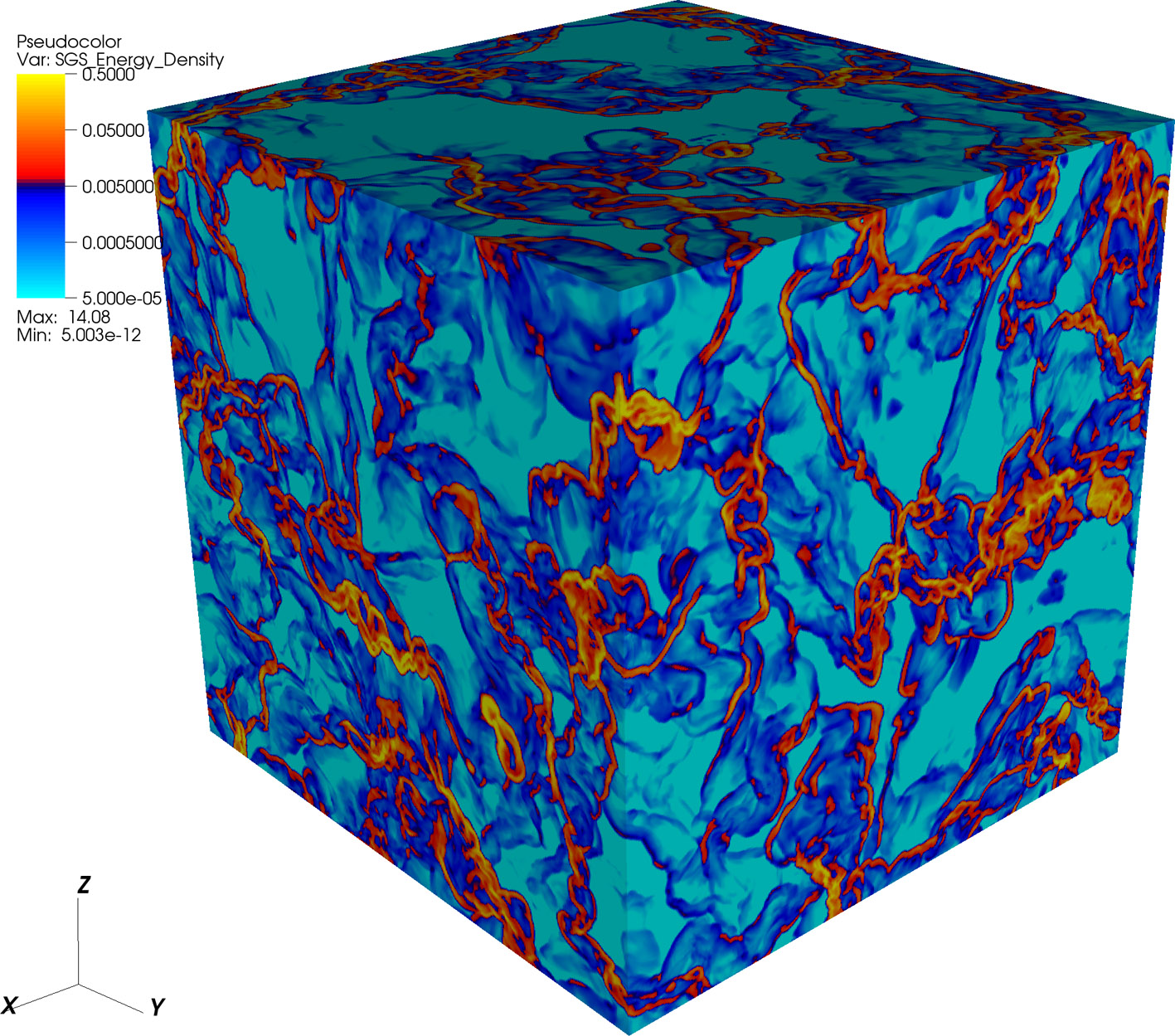}}
    \caption{Visualization of the SGS turbulence energy density $\rho
      K$ in a $512^{3}$ LES with solenoidal forcing.
      Image reproduced with permission from \cite{SchmFeder11}, copyright by ESO.}
    \label{fig:ksgs_soln}
\end{figure}

\begin{figure}[htbp]
    \centerline{\includegraphics[width=0.9\textwidth]{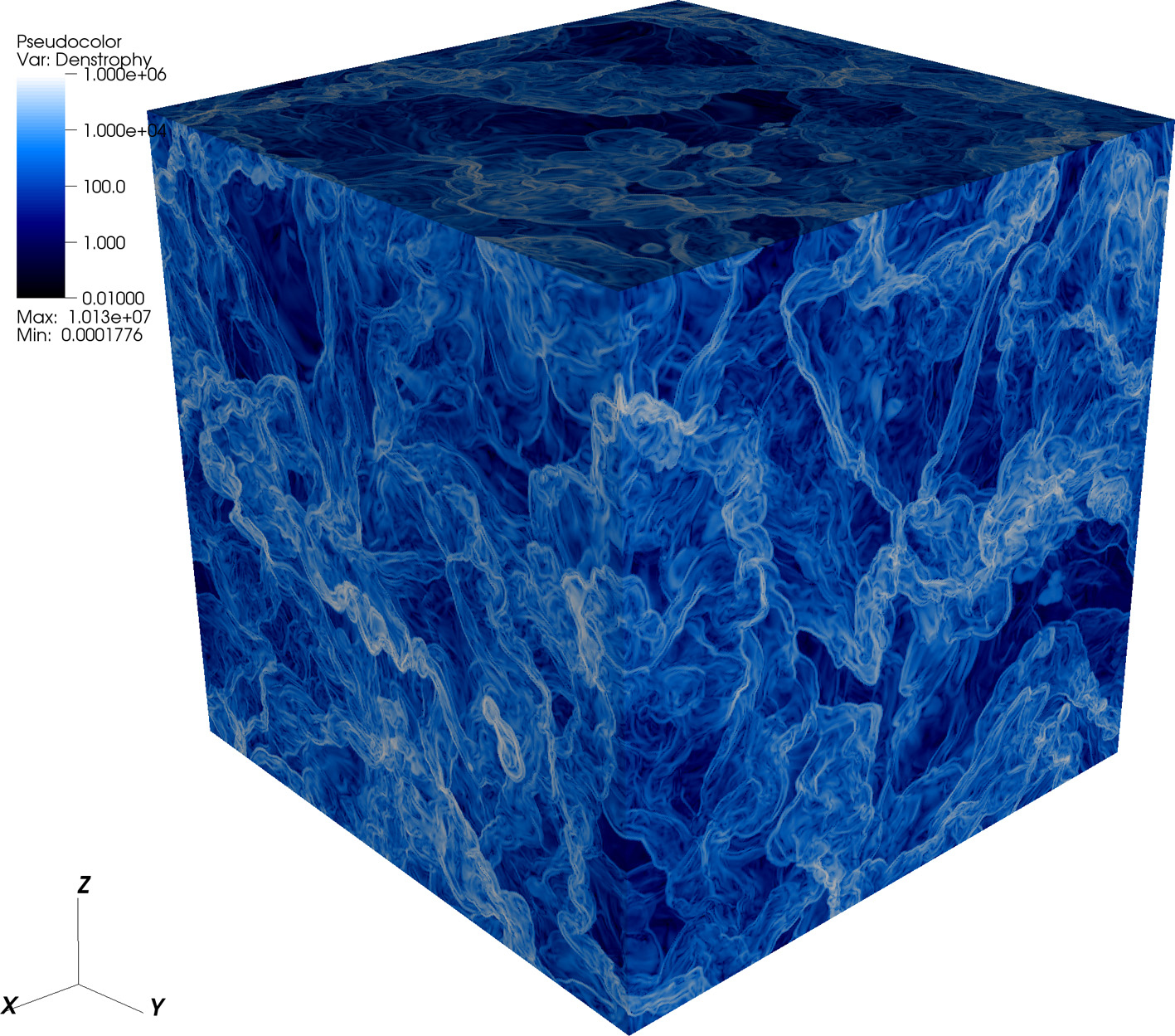}}
    \caption{Visualization of the the denstrophy $\Omega_{1/2}$ for
      the same LES as in Fig.~\ref{fig:ksgs_soln}.
      Image reproduced with permission from \cite{SchmFeder11}, copyright by ESO.}
    \label{fig:denstr_soln}
\end{figure}

\citet{SchmFeder11} demonstrated that Eq.~(\ref{eq:pde_energy_close}) for $\rho K$ works very well in the highly compressible regime. As an example, Fig.~\ref{fig:ksgs_soln} shows a visualization of $\rho K$ from an LES of isotropic supersonic turbulence, where solenoidal stochastic forcing maintains a root mean square Mach number of about 5 in the statistically stationary regime.%
\footnote{In this simulation, a quasi-isothermal equation of state is applied with an adiabatic exponent $\gamma=1.001$.}
The numerical resolution is $512^3$. In the reddish regions of the plot, $K_{\mathrm{sgs}}$ is higher than the spatial average, while it is lower in the bluish regions. In Fig.~\ref{fig:denstr_soln}, the numerically resolved turbulent flow is illustrated by the so-called denstrophy,
\[
	\Omega_{1/2}=\frac{1}{2}\left|\vecnab\times\left(\rho^{1/2}\vecu\right)\right|^2\,.
\] 
Since $\Omega_{1/2}$ combines density fluctuations and the rotation of the velocity, $\vecnab\times\vecu$, it highlights both compression and eddy-like motion on small scales \citep{KritNor07}. There is evidently a correlation between $\Omega_{1/2}$ and $\rho K$, which reflects the local interaction between the smallest resolved scales and subgrid and scales, as expressed by the production terms in Eq.~(\ref{eq:pde_energy_close}). This correlation is akin to the equilibrium  condition~(\ref{eq:energy_smag}) following from the Smagorinsky model for incompressible turbulence. However, due to the non-local effects inherent in the PDE~(\ref{eq:pde_energy_close}), the SGS turbulence energy cannot be reliably estimated from local quantities such as $\Omega_{1/2}$. In particular, turbulent diffusion smears out $\rho K$ in comparison to $\Omega_{1/2}$. 

\begin{figure}[tbp]
    \centerline{
      \includegraphics[width=0.49\textwidth]{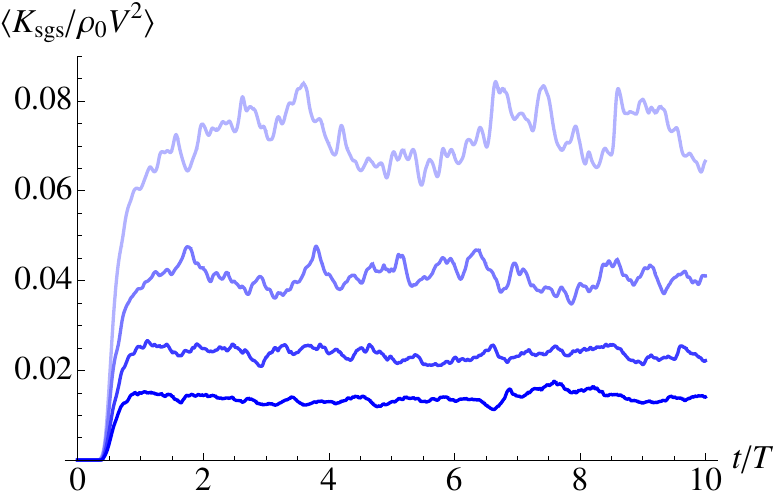} 
      \includegraphics[width=0.49\textwidth]{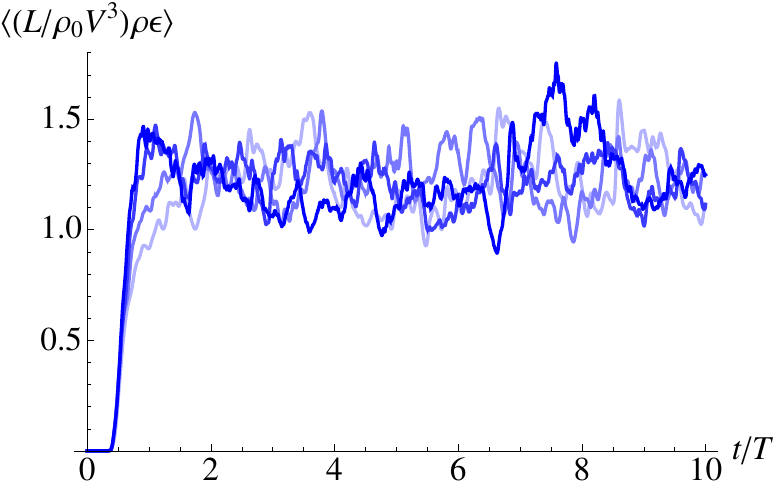}
    }
    \caption{Temporal evolution of the spatially averaged SGS
      turbulence energy (left) and the dissipation rate (right) for
      forced supersonic turbulence. The grid scale $\Delta$ decreases
      from $L/32$ (light colour) to $L/256$ (full colour).
      Image reproduced with permission from \cite{SchmFeder11}, copyright by ESO.}
    \label{fig:statistics_soln}
\end{figure}

\begin{figure}[ttbp]
    \centerline{\includegraphics[width=0.6\textwidth]{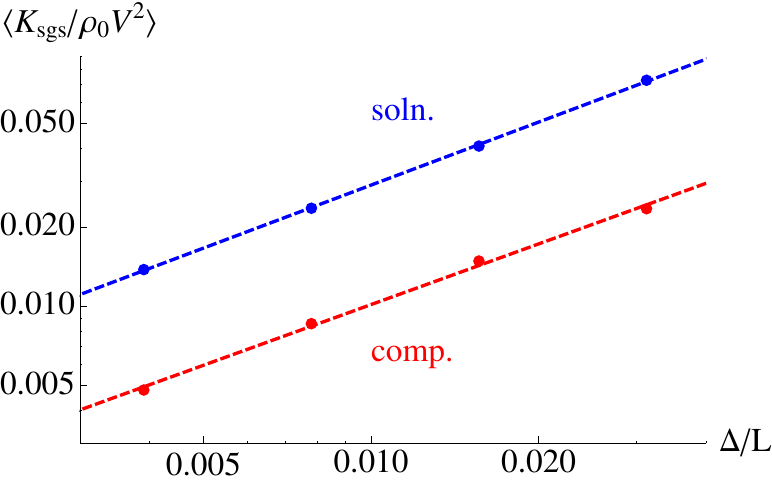}}
    \caption{Time-averaged mean values of the SGS turbulence energy in
      LES with different resolutions (dots) and power-law fits (dashed
      lines) for solenoidal and compressive forcing.
      Image reproduced with permission from \cite{SchmFeder11}, copyright by ESO.}
    \label{fig:K_power_law}
\end{figure}

A critical property is the scaling behaviour of the SGS turbulence energy. For statistically stationary homogeneous turbulence,
the mean value of $\rho K$ should scale as a power of the grid resolution because the fraction of unresolved kinetic energy
changes as the the cut-off of the energy spectrum is shifted (see Eq.~\ref{eq:ksgs_kolmogorov}). This was verified in 
\cite{SchmFeder11} by running LES with different grid scales $\Delta$ and fixed forcing length $L$. The global spatial averages
$\langle\rho K\rangle$ in these simulations are plotted in Fig.~\ref{fig:statistics_soln} (left plot) for $\Delta$ ranging 
from $L/256$ to $L/32$, where the case $\Delta=L/256$ corresponds to the $512^3$ simulation depicted in Figs.~\ref{fig:ksgs_soln} and \ref{fig:denstr_soln}. Although there are substantial fluctuations, one can see that $\langle\rho K\rangle$ decreases with $\Delta$. Time-averaging over the statistically stationary regime yields mean values that are close to the power law 
\begin{equation}
	\label{eq:sgs_scaling}
	\langle \rho K\rangle \propto \Delta^{\alpha},
\end{equation}
with $\alpha\approx 0.799\pm0.009$ (power-law fits are shown in Fig.~\ref{fig:K_power_law}). 
The scaling exponent falls between the Kolmogorov and
Burgers exponents and roughly comparable to the slope of the second-order structure functions with fractional mass-weighing 
reported in \cite{SchmFeder08}. The mean dissipation rate, on the other hand, does not change with resolution (see Fig.~\ref{fig:statistics_soln}, right plot). This is a necessary property for energy dissipation because it must balance the energy injection by the forcing on large scales.

\subsection{Energy and momentum conservation in AMR simulations}
\label{sec:consrv}

The scale-dependence of SGS turbulence energy illustrated in Fig.~\ref{fig:K_power_law} bears a special significance in adaptive mesh refinement (AMR) simulations. AMR \citep{BergOli84,BergCol89} is commonly applied in finite-volume codes to follow gravitational collapse and to locally resolve structures at higher resolution than the surrounding regions. In AMR simulations, data have to be transferred between different refinement levels by conservative interpolation or averaging. For example, when a region is refined, data from coarser grids are interpolated to finer grids. The same operation is used to fill ghost cells at the boundaries between a finer and a coarser level, which is required to compute fluxes through the faces of adjacent finer and coarser cells. Moreover, block-structured AMR codes usually average down the data from the highest-level grid to coarser levels. This operation is called restriction. 

The standard method to restore energy consistency between different refinement levels in AMR simulations leads to
artificial changes in the internal energy. The mass density, momentum, and energy variables at two levels, say, $l$ and $l+1$, 
are in the simplest case related by
\begin{align}
	\label{eq:mass_crs}	
	\rho_{\mathrm{crs}} :=&\,\overline{\rho} = \frac{1}{N}\sum_n\rho_{n}\,,\\
	\label{eq:vel_crs}	
	(\rho\vecU)_{\mathrm{crs}} :=&\,\overline{\rho\vecU} = \frac{1}{N}\sum_n(\rho\vecU)_{n}\,,\\
	\label{eq:energy_crs}	
	(\rho E)_{\mathrm{crs}} :=&\,\overline{\rho E} =  \frac{1}{N}\sum_n (\rho E)_{n} =
		\frac{1}{N}\sum_n\left[(\rho e)_n + \frac{1}{2}\frac{(\rho U)_{n}^2}{\rho_{n}}\right]\,,
\end{align}
where $N$ grid cells at level $l+1$ are summed up to a single value in a coarse cell at level $l$. 
Obviously, these relations guarantee mass, momentum, and energy conservation.
For more sophisticated interpolation schemes, the fine-grid values have different weights $w_n$ in the
above sums. Without loss of generality, we assume $w_n=1$ in the following. The internal energy at the coarser level 
is given by
\begin{equation}
	(\rho e)_{\mathrm{crs}}
	= (\rho E)_{\mathrm{crs}} - \frac{1}{2}\frac{(\rho U)_{\mathrm{crs}}^2}{\rho_{\mathrm{crs}}}\\
	= \overline{\rho e} + \Delta(\overline{\rho K})\,,
\end{equation}
where the energy difference  
\begin{equation}
	\label{eq:delta_energy}
	\Delta(\overline{\rho K}) :=
	\frac{1}{N}\sum_{n}\frac{1}{2}\frac{(\rho U)_{n}^2}{\rho_{n}} -
	\frac{1}{2}\frac{(\rho U)_{\mathrm{crs}}^2}{\rho_{\mathrm{crs}}}
\end{equation}
is generally non-zero. This implies
\[
	(\rho e)_{\mathrm{crs}} \ne \overline{\rho e} = \frac{1}{N}\sum_n(\rho e)_{n}\,.
\]
To put it another way, the kinetic energy differences between refinement levels results in numerical cooling or heating.

This can be mitigated by combining LES and AMR \citep{MaierIap09,SchmAlm13}. In principle, energy conservation can be met without altering the internal energy if the SGS turbulence energy $\rho K$ serves as an additional energy reservoir. 
(see Sect.~\ref{sec:K_eq}). By setting
\begin{equation}
	\label{eq:rho_K_balance}
	(\rho K)_{\mathrm{crs}} = \overline{\rho K} + \Delta(\overline{\rho K}),\quad
	\text{where}\quad
	\overline{\rho K} = \frac{1}{N}\sum_{n}(\rho K)_n
\end{equation}
energy is exchanged between the resolved and SGS components of the turbulence energy when regions are refined or de-refined. By
substituting Eq.~(\ref{eq:delta_energy}) for $\Delta(\overline{\rho K})$, it can be seen that the total (resolved plus unresolved) kinetic energies at the coarser ($l$) and finer ($l+1$) levels are consistent:
\begin{equation}
	\label{eq:energy_balance_ftec}
	\frac{1}{2}\frac{(\rho U)_{\mathrm{crs}}^2}{\rho_{\mathrm{crs}}} + (\rho K)_{\mathrm{crs}} = 
	\frac{1}{N}\sum_n\left[\frac{1}{2}\frac{(\rho U)_{n}^2}{\rho_{n}} + (\rho K)_n\right]\,,
\end{equation}
In this case, $(\rho e)_{\mathrm{crs}}=\overline{\rho e}$.

As demonstrated in \cite{SchmAlm13}, Eq.~(\ref{eq:energy_balance_ftec}) results in the expected scaling of the SGS turbulence energy
in multi-level simulations of homogeneously forced compressible turbulence. If the flow structure is strongly inhomogeneous, however, 
this method of energy compensation can lead to unphysical restrictions of the SGS turbulence energy to coarser grids. A particular problem is posed by non-turbulent bulk flows, such as gas accretion into the gravitational wells of clusters. In this case, the increment $\Delta(\overline{\rho K})$ by Eq.~(\ref{eq:delta_energy}) tends to greatly overestimate the energy difference associated with turbulent velocity fluctuations. In addition, fallbacks are necessary to avoid negative energy values resulting from conservative interpolation. To resolve these problems and to preserve global energy conservation, a combination of changes in the SGS turbulence energy, internal energy, and momentum is applied in the \textsc{Nyx} code \citep{AlmBell13}. Apart from cut-offs ensuring positivity, the power-law relation $(\rho K)_{\mathrm{crs}}=r^{2\eta}\overline{\rho K}$ ($\eta$ = 1/3 for Kolmogorov scaling) is used as an estimator to constrain differences of the SGS turbulence energy between two levels with refinement ratio $r=\Delta_{l+1}/\Delta_l$ \citep[see][for the detailed algorithm]{SchmAlm13}. 

\subsection{The structural subgrid-scale model for compressible MHD}
\label{sec:struc}

A non-linear closure similar to Eq.~(\ref{eq:tau_nonlin_norm}) follows from a Yeo-Bedford expansion of the
inverse filter kernel in Fourier space, $\FT{G}^{-1}(\veck)$, in terms of the filter scale $\Delta$ 
(see \citealt{Garnier}, Section 5.2). For an isotropic Gaussian filter, 
the transfer function defined by Eq.~(\ref{eq:gauss_filter}) can be expressed as a Taylor series expansion:
\begin{equation} 
	\FT{G}(k) = \sum_{n=0}^\infty\frac{(-1)^n}{n!}\left(\frac{\Delta^2 k^2}{4\gamma}\right)^n\,.
\end{equation}
The inverse is given by
\begin{equation}
	\FT{G}^{-1}(k) = \sum_{n=0}^\infty\frac{1}{n!}\left(\frac{\Delta^2 k^2}{4\gamma}\right)^n\,.
\end{equation}
Hence, a generic field $f$ can be formally reconstructed from the filtered field $\langle f\rangle$. Computing the 
inverse Fourier transform of $\FT{G}^{-1}(k)\langle\FT{f}\rangle$ gives
\begin{equation}
	f = \sum_{n=0}^\infty\frac{(-1)^n}{n!}\left(\frac{\Delta^2}{4\gamma}\nabla^2\right)^n\langle f\rangle\,.
\end{equation}
For $\gamma = 6$, one obtains the approximate deconvolution
\begin{equation}
	f \simeq \langle f\rangle - \frac{\Delta^2}{24}\nabla^2 f \pm \mathcal{O}(\Delta^4)\,.
\end{equation}
Similarly, second-order moments can be expanded. This leads to the non-linear structural closure
\begin{equation}
	\label{eq:tau_mhd_u_struc}
	\tau_{ij}^\mathrm{u} = -\frac{\Delta^2}{12}\,u_{i,k}u_{j,k}
\end{equation}
for the SGS turbulence stress tensor (tildes indicating Favre filtering are omitted). In contrast to Eq.~(\ref{eq:tau_nonlin_norm}), it has no tuneable coefficient apart from the width parameter $\gamma$ and does not depend on additional variables such as $K$. 

The structural approach was generalized to compressible MHD by \citet{VlayGrete16}. 
The closure for the SGS magnetic stress tensor defined by Eq.~(\ref{eq:turb_mag_stress}) 
is given by\footnote{
	Here it is understood that $\vecB$ is the filtered magnetic field in code units in accordance with the notation used
	for the velocity and density fields in this section.}
\begin{equation}
	\label{eq:tau_mhd_b_struc}
	\tau_{ij}^\mathrm{b} = -\frac{\Delta^2}{12}\,B_{i,k}B_{j,k}\,,
\end{equation}
and for the SGS EMF defined by Eq.~(\ref{eq:emf_sgs}) one obtains
\begin{equation}
	\label{eq:emf_struc}
	\mathcal{E}_{i} = \frac{\Delta^2}{12}\varepsilon_{ijk}u_{j,l}\left[B_{k,l} - (\ln\rho)_{,l}B_{k}\right]\,.
\end{equation}
The first term on the left-hand side is analogous to the stresses defined above with a cross product instead of
a tensor product. The second term follows from the compressibility extension proposed by \citet{VlayGrete16}.
As shown by \citet{GreteVlay16}, these closures perform extremely well in a priori tests compared to other MHD SGS models (see Sect.~\ref{sec:least_squares}). So far, it is the only subgrid-scale model applied to astrophysical flows with magnetic fields.

In contrast to the turbulence energy equation model, there is no need to solve additional PDEs for SGS quantities.
The stress tensor and EMF can be computed directly from the Jacobian matrices of the velocity and magnetic field. 
The dynamical equations that have to be integrated numerically are the filtered momentum equation~(\ref{eq:mhd_momt_flt_high_Re}), induction equation~(\ref{eq:induction_flt}), and energy equations (see Sect.~\ref{sc:mhd}), where SGS stresses and the SGS EMF are substituted by the closures~(\ref{eq:tau_mhd_u_struc}--\ref{eq:emf_struc}). Usually, Eq.~(\ref{eq:pde_mhd_energy_res}) for the total energy density $\rho E$ is solved, where the SGS dissipation term is not explicitly modelled.
In this case, the internal energy $e$ is determined by the identity~(\ref{eq:mhd_energy_total_flt}). However, this can result in negative internal energies for high Mach numbers. To alleviate this problem, pressure or temperature floors are applied, although this in turn violates energy conservation.

The structural model does not provide closures for viscous and Ohmic dissipation on microscopic scales, which explicitly occur in the hydrodynamical equations~(\ref{eq:pde_energy_res}) and~(\ref{eq:pde_energy_sgs}). Similar to the Smagorinsky model, energy that is transferred from resolved to unresolved scales is assumed to be instantaneously dissipated, i.e.\ on time scales comparable to the SGS eddy-turnover time. Without an SGS energy reservoir, there is no buffering between resolved kinetic or magnetic and internal energy (see also the discussion in Sect.~\ref{sec:K_eq}). In principle, the turbulence energy equation model could be generalised to MHD turbulence by deriving and closing PDEs for the SGS energies $K^\mathrm{u}$ and $K^\mathrm{b}$. However, this would substantially increase the model complexity and has not been attempted yet.

The remaining SGS terms on the right-hand side of equation~(\ref{eq:pde_mhd_energy_res}) can be implemented as additional source terms in customary ideal MHD solvers. These terms treat numerically unresolved transport of kinetic and magnetic energy and transfer between resolved and unresolved energy reservoirs. However, as mentioned at the end of Sect.~\ref{sc:mhd}, the structural model does not split the term associated with SGS stresses into cascade and conservative transport terms. As a result, any non-conservative contribution acts as dissipation (loss of resolved energy) or backscatter (gain of resolved energy). Positive contributions from the EMF term can be interpreted as production of magnetic energy on resolved scales due to numerically unresolved dynamo processes. The various energy exchanging processes are discussed in some detail in \citet{VlayGrete16}. 

\citet{grete_comparative_2017} applied the non-linear structural model in LES of driven MHD turbulence with a shock-capturing finite-volume scheme of second order. Compared to an equivalent ILES at doubled resolution, the kurtosis and skewness of probability density functions of the current density, dilatation, and vorticity are better reproduced if an explicit filter with filter widths of 2.71 or 4.75 times the grid scale is used. Without additional filtering of the numerically resolved velocity and magnetic fields, no improvement compared to eddy-viscosity models was found.\footnote{The application of a second filter (also known as test filter) in 	addition to the implicit filter corresponding to numerical discretization is a common technique. Examples are scale-similarity 	models (see \citealt{Garnier}, chapter 5) and the dynamic procedures introduced in Sect.~\ref{sec:dyn_proc}. } 
The temporal evolution of mean quantities, energy spectra, and structure functions do not show any significant differences compared to other SGS models or ILES at the same resolution.
It is also demonstrated that energy dissipation is dominated by numerical dissipation. The effective numerical viscosity and resistivity in ILES of MHD turbulence are analysed further in \citet{grete_dynamical_range_2023}. In summary, it can be stated that all investigated SGS models have only a very small impact on the numerically computed flow and magnetic field. However, it is established that quantities related to processes on numerically unresolved scales can be computed quite accurately. This can be utilised in some astrophysical applications.

As shown by \citet{carrasco_gradient_2020}, it is possible to obtain structural closures for a general set of conservation laws by grouping conservative and primitive variables in state vectors, decomposing the transport terms, and expanding the resulting SGS terms. Favre filtering simply follows from expressing the primitive state vector as inverse function of the conservative state vector. They derive the SGS stress tensors~(\ref{eq:tau_mhd_u_struc}, \ref{eq:tau_mhd_b_struc}) and EMF~(\ref{eq:emf_struc}) for MHD and generalize the filtered equations to relativistic fluids. In \citet{vigano_general_2020}, the formalism is applied to general relativistic MHD (GRMHD). The resulting SGS model is tested in 2D and 3D simulations of the relativistic Kelvin-Helmholtz instability in a curved background, where a high-resolution run is compared against LES of lower resolution. Since the equations get quite complicated in the relativistic case, we refer the interested reader to the original papers for details. 
 
\subsection{Turbulent mixing models and regularization}
\label{sec:mixing}

Turbulent mixing refers to the diffusion-like transport of quantities such as heat or metals due to turbulence. Numerically unresolved mixing can be modelled by conservative transport terms. For instance, the term $\vecnab\cdot\Fdiff$ in Eq.~(\ref{eq:pde_energy_sgs}) accounts for the diffusion of SGS turbulence energy when applying the gradient-diffusion hypothesis:
\begin{equation}
	\label{eq:div_grad_diff}
	\vecnab\cdot\Fdiff = \vecnab\cdot\left(\rho\kappa_{\mathrm{sgs}}\vecnab K\right)\,.
\end{equation}
In a similar way, the convective flux $\Fconv$ in Eq.~(\ref{eq:pde_energy_res}) describes the transport of internal energy due to SGS turbulence. These are examples of diffusive processes that tend to flatten gradients by redistributing energy without converting energy from one form to another. On numerically resolved scales, turbulent transport is explicitly treated by the advection terms in the PDEs.

Assuming a similarity of SGS transport processes, \citet{engels_modelling_2019} propose to model turbulent mixing by Kolmogorov-Prandtl-relations, as expressed in Eq.~(\ref{eq:grad_diff}), with a universal turbulent diffusivity: 
\begin{align}
	\label{eq:heat_diff}
	\Fconv &= \rho\kappa_{\mathrm{sgs}}\vecnab e\,,\\
	\Fchem_n &=\rho\kappa_{\mathrm{sgs}}\vecnab X_n\,.
\end{align}
Independently, \citet{rennehan_dynamic_2019-1} put forward a similar gradient-diffusion model where they assumed an equivalence between turbulent viscosity and diffusivity and set $\kappa_{\mathrm{sgs}} = \nu_{\mathrm{sgs}}$. This implies that the Prandtl number in Eq.~(\ref{eq:sgs_diff}) is $C_\kappa/C_\nu = 1$. However, as demonstrated in \citet{SchmNie06b}, this assumption is not supported by data obtained from turbulence simulations. Instead, by testing correlations of the diffusive flux rather than the divergence of the flux, higher Prandtl numbers of the order $10$ are obtained (see also Section~\ref{sec:hierarch_flt}).

As argued by \citet{Pope}, turbulent transport is not aligned with the gradient of the transported quantity. As a result, the gradient-diffusion
hypothesis is of limited validity. This problem can be alleviated by using structural closures. \citet{rennehan_mixing_2021-1} considers an anisotropic tensor diffusivity of the form\footnote{\citet{rennehan_mixing_2021-1} uses negative sign convention for diffusive fluxes with a diffusivity tensor $\tens{K}=-\rho\tens{\upkappa}_\mathrm{sgs}$ on the right-hand side of the PDEs.}
\begin{equation}
	\label{eq:diff_anisotropic}
	\tens{\upkappa}_\mathrm{sgs} = -2 C\Delta^2\,\vecnab\otimes\vecu
\end{equation}
such that
\begin{equation}
	\Fchem_n = \rho\tens{\upkappa}_\mathrm{sgs}\cdot\vecnab X_n\,.
\end{equation}
This is a special instance of the structural closures introduced by \citet{VlayGrete16}. From the expansion of Eq.~(\ref{eq:spec_flux}) follows the lowest-order approximation
\begin{equation}
	\left(\mathfrak{F}^\mathrm{(chem)}_n\right)_i = -\frac{\Delta^2}{12}\,\rho u_{i,k}(X_n)_{,k}
\end{equation}
and a similar closure is obtained for the convective flux. Comparison with Eq.~(\ref{eq:diff_anisotropic}) yields $C=1/24$.
\citet{rennehan_mixing_2021-1} employs a dynamical procedure to determine $C$ \emph{in situ}.\footnote{For comparison, also a fixed coefficient is used that is larger than the coefficient of the structural closure. However, the choice of the coefficient also depends on the definition of the resolution scale $\Delta$, which is more ambiguous for particle codes than for grid-based codes.} Moreover, the tensor diffusivity is modified to include only negative eigenvalues of the rate-of-strain tensor $S_{ij}$, corresponding to positive diffusivities \citep{balarac_dynamic_2013}. This technique is called regularization and ensures that the action of anisotropic diffusion in particle codes is repulsive rather than attractive. Regularization is also applied to the SGS turbulence stress tensor to prevent numerical instabilities and to address issues with the antisymmetric part (rate of rotation) of the velocity derivative \citep{vollant_dynamic_2016}. However, it was demonstrated that the unmodified structural closures for SGS turbulence and magnetic stresses can be used in grid-based codes without encountering any problems \citep{grete_comparative_2017}. A possible explanation is that the backscattering associated with positive eigenvalues is partially compensated by numerical diffusion, while particle codes have no intrinsic diffusion. Currently, the exact conditions under which regularization is necessary remain elusive.



\section{Determination of closure coefficients}
\label{sec:closure}

One of the basic assumptions of the Kolmogorov theory is that
turbulence is statistically self-similar in the inertial subrange
(see, for example, \citealp{Frisch}). With regard to subgrid-scale
closures, the self-similarity of turbulence implies that
dimensionless coefficients such as $C_{\nu}$ in Eq.~(\ref{eq:visc_sgs})
should be independent of the chosen filter scale. This is not only a
necessary condition for the feasibility of LES, but it also allows for 
the calibration of closure coefficients by explicitly filtering turbulence
data. Since closures do not exactly match SGS terms, an improved approximation
can be achieved by so-called dynamic procedures, which estimate coefficients
from properties of the numerically resolved flow under the assumption of local
self-similarity.

\subsection{Hierarchical filtering}
\label{sec:hierarch_flt}

As a formal framework, let us consider an infinite series of
isotropic and time-independent filter operators
$\langle\ \rangle_{n}$. Each filter is defined by a
kernel $G_n(r)$ with filter length $\Delta_n$ (see Sect.~\ref{sec:separation}). 
We shall assume that $\Delta_0\sim L$, where $L$ is the integral length scale of the flow,
and 
\begin{equation}
  \label{eq:self_siml}
  \forall n\in\mathbb{N}_{0}: 
  G_{n}(\vect{x}) = \lambda^{3}G_{n-1}(\lambda\vect{x}),\quad\text{where}\ \lambda > 1\,,
\end{equation}
i.e., $\langle\ \rangle_{n}$ for $n=0,1,2,\ldots$ is a self-similar hierarchy of filters
and the filter lengths are given by $\Delta_{n}=\Delta_{0}/\lambda^n$. Since $G_{n}(r)\rightarrow\delta(r)$
in the limit $n\rightarrow\infty$, $\langle\ \rangle_{\infty}$ is the identity operator.
Intuitively, one expects the following asymptotic equation to hold:
\begin{equation}
  \label{eq:flt_twice}
  \langle\langle q\rangle_{m}\rangle_{n}\simeq\langle q\rangle_{n}\quad
  \text{if}\ \Delta_{n}\gg\Delta_{m}\,.
\end{equation}
Because filtering in physical space corresponds to a multiplication with the transfer function
of the filter in Fourier space, the validity of this approximation becomes evident
by calculating the product of the transfer functions in the case of Gaussian filters:
\begin{equation}
  \widehat{G}_{m}(k)\widehat{G}_{n}(k) = 
  \exp\left[-\frac{k^{2}(\Delta_{m}^{2}+\Delta_{n}^{2})}{24}\right] \simeq 
  \exp\left[-\frac{k^{2}\Delta_{n}^{2}}{24}\right] = \widehat{G}_{n}(k)\,.
\end{equation}

We can now apply the scale separation of the Navier--Stokes equations introduced 
in Sect.~\ref{sec:separation} at different levels of the filter hierarchy. In particular, the
filtered density field at the $n$-th level is $\langle\rho\rangle_n$, and the 
Favre-filtered velocity is given by
\begin{equation}
  \fvecu^{[n]} = \frac{\langle\rho\vecu\rangle_n}{\langle\rho\rangle_n}\,.
\end{equation}
By filtering twice at levels $m$ and $n$, we obtain 
\begin{equation}
  \fvecu^{[m][n]}\langle\langle\rho\rangle_m\rangle_{n} = 
  \langle\langle\rho\rangle_m\vecu^{[m]}\rangle_{n} =
  \langle\langle\rho\vecu\rangle_{m}\rangle_{n}\,.
\end{equation}
If the $n$-th level is much coarser than the $m$-th level, 
the asymptotic relation~(\ref{eq:flt_twice}) for $\Delta_{n}\gg\Delta_{m}$ implies
\begin{equation}
  \label{eq:vel_flt_twice}
  \fvecu^{[m][n]}\langle\langle\rho\rangle_m\rangle_{n} =
  \langle\langle\rho\vecu\rangle_{m}\rangle_{n} \simeq
  \langle\rho\vecu\rangle_n =
  \fvecu^{[n]}\langle\rho\rangle_n\,.
\end{equation}

The turbulence stress tensor on the length scale $\Delta_n$ of the $n$-th filter is defined by
\begin{equation}
    \label{eq:turb_stress_n}
    \tens{\tau}^{[n]} =
    -\langle\rho\vecu\otimes\vecu\rangle_{n} + \langle\rho\rangle_n \fvecu^{[n]}\fvecu^{[n]}.
\end{equation}
The stress tensors for two filter levels $m$ and $n$, where $\Delta_n>\Delta_m$, are
related by the Germano identity  
(see Sect.~3.3.3 in \citealp{Sagaut} and \citealp{Germano92,SchmidtPhD}):
\begin{equation}
  \label{eq:germano}
  \tens{\tau}^{[m][n]} = 
  \langle\tens{\tau}^{[m]}\rangle_{n} + \tens{\tau}^{[m,n]}\,.
\end{equation}
The stress tensor associated with the double-filtered variables
is defined by
\begin{equation}
  \begin{split}
  \tens{\tau}^{[m][n]} =
  &-\langle\langle\rho \vecu\otimes \vecu\rangle_{m}\rangle_{n}
  + \langle\langle\rho\rangle_{m}\rangle_{n}\fvecu^{[m][n]}\fvecu^{[m][n]}= \\
  &-\langle\langle\rho \vecu\otimes \vecu\rangle_{m}\rangle_{n}
  + \frac{\langle\langle\rho \vecu\rangle_{m}\rangle_{n}\otimes
          \langle\langle\rho \vecu\rangle_{m}\rangle_{n}}
         {\langle\langle\rho\rangle_{m}\rangle_{n}}
  \end{split}
\end{equation}
and
\begin{equation}
  \label{eq:leonard_stress_general}	
  \begin{split}
  \tens{\tau}^{[m,n]} =
  &-\langle\langle\rho\rangle_m\fvecu^{[m]}\otimes\fvecu^{[m]}\rangle_{n}
  + \frac{\langle\langle\rho\rangle_m\fvecu^{[m]}\rangle_{n}\otimes
        \langle\langle\rho\rangle_m \fvecu^{[m]}\rangle_{n}}
       {\langle\langle\rho\rangle_m\rangle_{n}} = \\
  &-\langle\langle\rho\rangle_m\fvecu^{[m]}\otimes\fvecu^{[m]}\rangle_{n} 
  + \frac{\langle\langle\rho \vecu\rangle_{m}\rangle_{n}\otimes
          \langle\langle\rho \vecu\rangle_{m}\rangle_{n}}
         {\langle\langle\rho\rangle_{m}\rangle_{n}}
  \end{split}
\end{equation}
is the Leonard stress tensor, which is associated with velocity fluctuations 
in the intermediate range of length scales $\Delta_m \le \ell \le \Delta_{n}$. 
The Germano identity also holds for two arbitrary filters
in the hierarchy. In the limit $\Delta_{n}\gg\Delta_{m}$, the
contribution from $\langle\tens{\tau}^{[m]}\rangle_{n}$ becomes
negligible and
\begin{equation}
  \label{eq:turb_stress_asympt}
  \tens{\tau}^{[m][n]} \simeq \tens{\tau}^{[m,n]} \simeq \tens{\tau}^{[n]}\,,
\end{equation}
where the second relation follows from Eq.~(\ref{eq:vel_flt_twice}).
As a result, the turbulent stresses associated with the scale $\Delta_{n}$
are not sensitive to the flow structure on much smaller scales. In particular,
it follows that $K^{[n]}\simeq K^{[m,n]}$ if $\Delta_m\ll\Delta_n$.

In \cite{SchmNie06b,SchmFeder11,GreteVlay16}, Gaussian filters are applied to 
data from ILES of forced compressible turbulence to validate closures \emph{a priori}.
A crucial aspect of estimating closure coefficients from finite-resolution data lies in the following reasoning.
Let us assume that $\rho$ and $\vecu$ are the physical density and velocity fields. Let us further assume that
the implicit filter of an ILES corresponds to the filter level $m=I$, meaning that
$\langle\rho\rangle_I$ and $\fvecu^{[I]}$ represent the numerically computed density and velocity fields.
Now, if the numerical data are coarse-grained by an explicit filter $\langle\ \rangle_n$ within the inertial subrange, the turbulence stress tensor $\tens{\tau}^{[I,n]}$ can be calculated using Eq.~(\ref{eq:leonard_stress_general}). By definition, a closure for the turbulent stresses on the length scale $\Delta_n$ applies to $\tens{\tau}^{[n]}$.
However, if $\Delta_n$ is sufficiently large compared to $\Delta_{I}$, the distinction between physical densities and velocities (or DNS data) and ILES data becomes irrelevant and we can make use of the approximation $\tens{\tau}^{[n]}\simeq\tens{\tau}^{[I,n]}$ (see Eq.~\ref{eq:turb_stress_asympt} for $m=I$). 
Consequently, it is possible to calculate coarse-grained eddy-viscosity coefficients as follows:
\[
  C_\nu \simeq
  \frac{\tau_{ij}^{[n]\ast}S_{ij}^{[n]}}
       {\langle\rho\rangle_n\Delta_{n}\left(K^{[n]}\right)^{1/2}\left|S^{[n]\ast}\right|^2}\,.
\]
Of course, since the eddy-viscosity closure is not exact, the value of $C_\nu$ varies. 
But the mean value turns out to be roughly $0.05$ for different
filter lengths and simulation parameters \citep{SchmNie06b}, in good agreement with other estimates in the literature.
The same method was used to determine the coefficient $C_{\kappa}$ for the gradient-diffusion closure~(\ref{eq:grad_diff}).
The result $C_{\kappa}\approx 0.4$ implies a Prandtl number $C_{\kappa}/C_{\nu}$
around $10$, contrary to the common assumption that the kinetic Prandtl number is of the order unity
\citep{Sagaut}.  
 
\subsection{Dynamic procedures}
\label{sec:dyn_proc}

Subgrid-scale models in their standard form apply to statistically
stationary and isotropic turbulence. But turbulent flows in nature
often deviate from this idealization:
In terrestrial applications, flow inhomogeneities are inevitably caused by
boundary conditions (``walls''). In astrophysics, one of the major 
energy sources is gravity. It causes matter to clump (galaxies and clusters)
or to move under the action of central gravitational fields (stars),
which produces inherently inhomogeneous and anisotropic flows.
For example, turbulent convection in stars introduces a vertical anisotropy
of the flow. Turbulence driven by violent local energy release, such as supernova
explosions in the interstellar medium, can also be highly inhomogeneous.

One of the solutions to this problem is to \emph{localize} closures, i.e.,
to calculate local closure coefficients. This requires local estimators
that take properties of the flow in some small region as input.
Obviously, this works only if the size of this region is not significantly
affected by the flow inhomogeneity on larger scales. In other words,
the flow must be asymptotically homogeneous and isotropic at least
on length scales of the order of the grid scale. 
In this case, a test filter, denoted by $\langle\ \rangle_{\mathrm{T}}$, can be applied in LES. 
The filter length, $\Delta_{\mathrm{T}}$, is a small multiple of the grid scale, $\Delta$.
Test filters are usually implemented as discrete filters over several grid cells
(see Sect.~2.3.2 in \citealp{Garnier}). A multi-dimensional test filter can be composed as a 
succession of one-dimensional filters.\footnote{For filters with large stencils, test filtering can nevertheless become too inefficient because of the access to remote blocks of memory.}
The test filter length $\Delta_{\mathrm{T}}$ can be adjusted
by varying the weights of the cells. An optimal ratio $\gamma_{\mathrm{T}}=\Delta_{\mathrm{T}}/\Delta$
is determined by the closest match between the transfer functions of the discrete and analytical
box filters with filter length $\Delta_{\mathrm{T}}$ \citep{VasilLund98,SchmidtPhD,grete_comparative_2017}. 
For instance, a test filter with $\gamma_{\mathrm{T}}=2.711$ is optimal when a five-point stencil is used 
in each spatial dimension.

By identifying $\Delta_m$ with $\Delta$ and $\Delta_{n}$ with $\Delta_{\mathrm{T}}$,
the Germano identity~(\ref{eq:germano}) allows us to express the
turbulence stress tensor on the length scale of the test filter as the sum of
the test-filtered SGS turbulence stress tensor and the Leonard
tensor for the intermediate velocity fluctuations (see also Sect.~4.3 in
\citealt{Sagaut}):
\begin{equation}
  \label{eq:germano_test}
  \tens{T} = 
  \langle\tens{\tau}\rangle_{\mathrm{T}} + \tens{L}\,.
\end{equation}
Here, the Leonard tensor $\tens{L}$ associated with the
test filter is defined by
\begin{equation}
	\label{eq:leonard_test}
  \tens{L} =
  -\langle\rho\vecu\otimes\vecu\rangle_{\mathrm{T}} + 
  \frac{\langle\rho\vecu\rangle_{\mathrm{T}}\otimes\langle\rho\vecu\rangle_{\mathrm{T}}}
       {\langle\rho\rangle_{\mathrm{T}}}\,,
\end{equation}
where we use the simplified notation $\rho$ and $\vecu$ for the density
and velocity on the grid scale, as in Sect.~\ref{sec:turb_stress}.
Because of the scale-invariance of turbulence, \cite{GerPio91} proposed that
the eddy-viscosity closure~(\ref{eq:tau_eddy}) holds for both $\tens{T}$
and $\tens{\tau}$. In the case of the Smagorinsky model 
(see Eq.~\ref{eq:energy_smag} for $\nu_{\mathrm{sgs}}$), the corresponding tensors are:
\begin{align}
  \label{eq:prod_germ_lilly}
  \tau_{ij}^{\ast} &=
    2\rho (C_{\mathrm{S}}\Delta)^{2}|S|S_{ij} =: C_{\mathrm{S}}^2\beta_{ij}\,, \\
  T_{ij}^{\ast} &\simeq
    2\rho_{\mathrm{T}}(C_{\mathrm{S}}\Delta_{\mathrm{T}})^{2}|S_{\mathrm{T}}|(S_{\mathrm{T}})_{ij} =:
    C_{\mathrm{S}}^2\alpha_{ij}\,.
\end{align}
The rate-of-strain tensor $(S_{\mathrm{T}})_{ij}$ at the test filter level is
given by the symmetrized derivative of the test-filtered numerically resolved velocity field, 
$\partial_i\langle u_j\rangle_{\mathrm{T}}$, analogous to Eq.~(\ref{eq:strain_flt}). 
The variable $C_{\mathrm{S}}(\vecx,t)$ needs to be determined. This can be achieved by substituting
the above expressions for $\tau_{ij}^{\ast}$ and $T_{ij}^{\ast}$ into the trace-free part of the
Germano identity~(\ref{eq:germano_test}), 
which implies
\begin{equation}
  \label{eq:leonard_GL}
  L_{ij}^{\ast} \simeq C_{\mathrm{s}}^2\alpha_{ij}-\langle C_{\mathrm{s}}^2\beta_{ij}\rangle_{\mathrm{T}}\,.
\end{equation}
Under the assumption that $C_{\mathrm{s}}$ varies only little over the smoothing length of the test filter,
one can set
$\langle C_{\mathrm{s}}^2\beta_{ij}\rangle_{\mathrm{T}}\simeq C_{\mathrm{s}}^2\langle\beta_{ij}\rangle_{\mathrm{T}}$.
Since $L_{ij}$ can be evaluated from Eq.~(\ref{eq:leonard_test}),
minimization of the residual error between $L_{ij}^{\ast}$ and the expression on the right-hand side
of Eq.~(\ref{eq:leonard_GL}) yields
\begin{equation}
  C_{\mathrm{S}}^2 = \frac{m_{ij}L_{ij}^{\ast}}{m_{ij}m_{ij}},
\end{equation}
where $m_{ij}=\alpha_{ij}-\langle\beta_{ij}\rangle_{\mathrm{T}}$. This is the 
Germano--Lilly dynamic procedure, which was applied, for example, in LES of turbulent channel flows
\citep{Pio93}.
In principle, this procedure could also be applied to the non-equilibrium model with
the turbulent viscosity defined by Eq.~(\ref{eq:visc_sgs}). In this case, the
turbulence energy associated with $\tens{T}$ is given by the contracted Germano
identity, 
\[
	-\frac{1}{2}T_{ii}=\langle\rho K\rangle_{\mathrm{T}}+\rho_{\mathrm{T}}K_{\mathrm{T}}\,,
\]
where $\rho K=-\tau_{ii}/2$ and $\rho_{\mathrm{T}}K_{\mathrm{T}} = -L_{ii}/2$.

However, the dynamic procedure as outlined above has several caveats.
In particular, the assumption of negligible variation of $C_{\mathrm{S}}$ over the the 
test filter length is found to be violated significantly. Moreover,
$C_{\mathrm{S}}$ diverges if $m_{ij}$ vanishes. Consequently, several
attempts were made to improve the dynamic procedure \citep{LiuMen94,PioLiu95,GhoLund95}.
A particularly simple modification was found by analyzing experimental measurements 
of turbulent velocity fluctuations in consecutive wave number bands $[k_{n-1},k_{n}]$, corresponding 
to a hierarchy of filters. By explicitly evaluating the turbulent stresses $\tens{\tau}^{[n]}$
associated with the wave numbers $k_n=\pi/\Delta_n$, the correlations with localized
closures were verified. Although some correlation between the turbulent
stresses at different filter levels was found, the correlation of $\tens{\tau}^{[n][n-1]}$
with the Leonard stresses $\tens{L}^{[n,n-1]}$ turned out to be significantly better.
This observation can be understood as a consequence of the locality of the energy
transfer \citep{Kraich76,Sagaut}, i.e., the energy transfer across a certain wave number $k$ is mainly
caused by interactions in the narrow spectral band $[\frac{1}{2}k,2k]$. 
In the context of test filtering in Large Eddy Simulations (LES), this implies that the eddy-viscosity closure should be applied to $\tens{L}$ instead of $\tens{T}$. Consequently, the localized coefficient $C_{\nu}(\vecx,t)$
of the turbulent viscosity is then given by \citep{KimMen99,SchmNie06b,RoepSchm09}
\begin{equation}
  \label{eq:C_nu_test}
  C_{\nu} \simeq
  \frac{L_{ij}^{\ast}(S_{\mathrm{T}})_{ij}}
       {\rho_{\mathrm{T}}\Delta_{\mathrm{T}}K_{\mathrm{T}}^{1/2}|S_{\mathrm{T}}^{\ast}|^2}\,,
\end{equation}
where $K_{\mathrm{T}}=-L_{ii}/(2\rho_{\mathrm{T}})$ is the resolved kinetic energy on length scales 
$\Delta\le\ell\le\Delta_{T}$. By substituting the expression for $C_{\nu}$ into Eq.~(\ref{eq:flux_eddy}) 
for the localized rate of production, we obtain
\begin{equation}
  \label{eq:prod_locl}
  \Sigma=\tau_{ij}S_{ij} \simeq
  \frac{\rho\Delta}{\rho_{\mathrm{T}}\Delta_{\mathrm{T}}}  
  \left(\frac{K}{K_{\mathrm{T}}}\right)^{1/2}
  \left(\frac{|S^\ast|}{|S_{\mathrm{T}}^{\ast}|}\right)^2
  L_{ij}^{\ast}(S_{\mathrm{T}})_{ij} - \frac{2}{3}\rho K d\,.
\end{equation}
The above formula was used for simulations of turbulent thermonuclear combustion in white dwarfs
(see Sect.~\ref{sec:SN_Ia}). A straightforward generalization of the dynamic procedure would involve 
localizing both $C_1$ and $C_2$ in the closure~(\ref{eq:tau_mixed}). In this case, a linear system 
in the coefficients $C_1$ and $C_2$ needs to be solved to minimize the residual. 
However, this approach has yet to be applied. 

\begin{figure}[htbp]
    \centerline{\includegraphics[width=0.5\textwidth]{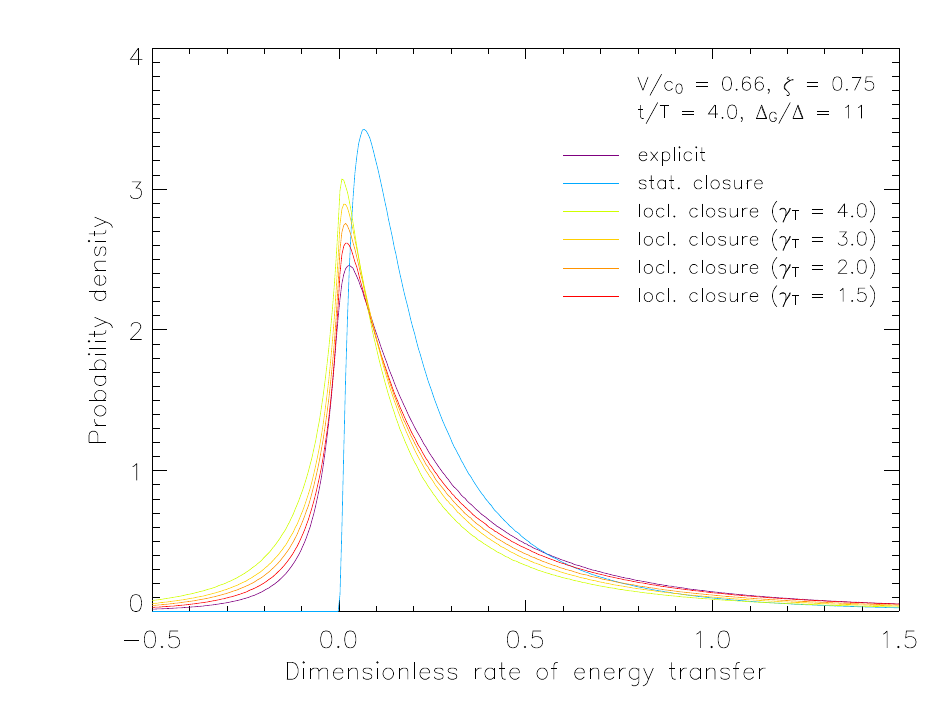} 
    	        \includegraphics[width=0.5\textwidth]{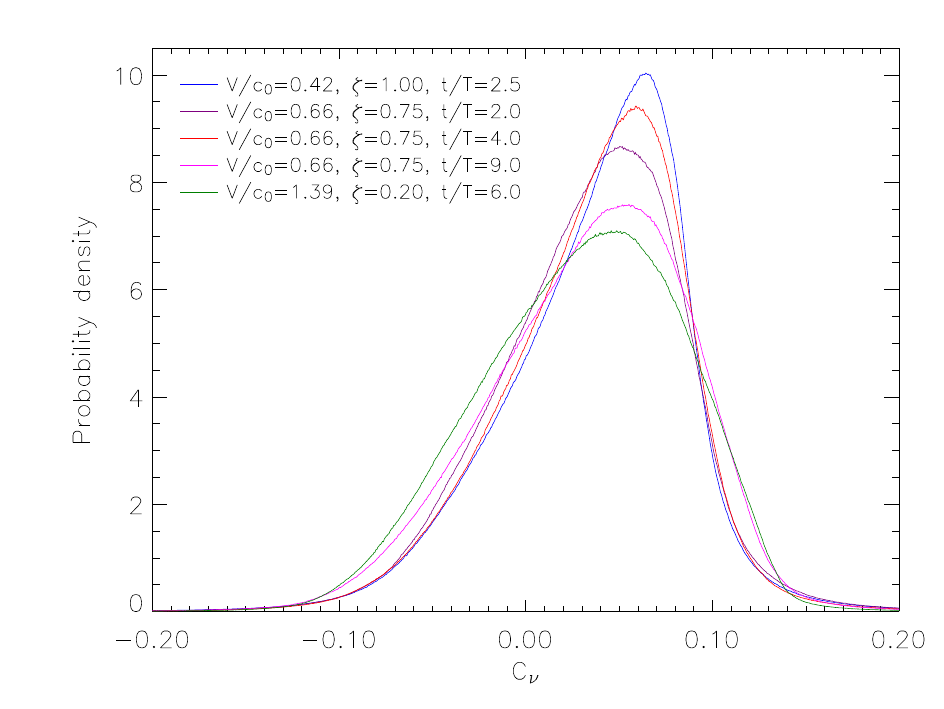}}
    \caption{\emph{Left plot:} comparison of the probability density functions of the coarse-grained turbulence energy flux
		due to anisotropic shear with different closures. \emph{Right plot:} probability density functions of the
		localized closure coefficient $C_{\nu}$ obtained by test-filtering different coarse-grained ILES. 
		The ratio of the test filter length to the coarse-graining length is $\gamma_{\mathrm{T}}$. The static
		closure refers to the case with a constant coefficient that is given by a spatial average. The parameters of the random 
		forcing are the characteristic Mach number $V/c_0$ and the weight $\zeta$ of the 
		Helmholtz decomposition into solenoidal and compressive modes. 
		Image reproduced with permission from \citet{SchmNie06b}, copyright by ESO.}
    \label{fig:pdfs_eddy_visc}
\end{figure}

Closures can be tested by coarse-graining numerical data with hierarchical Gaussian filters $\langle\ \rangle_n$ as explained in Sect.~\ref{sec:hierarch_flt}. In the functional modelling approach, the coefficient $C_\nu$ is determined
by matching the rate of energy transfer across the length scale $\Delta_n$ with the eddy-viscosity closure in terms of the filtered fields at level $n$:
\begin{equation}
	\label{eq:flux_eddy_coarse}
	\Sigma^{[n]} + \frac{2}{3}\langle\rho\rangle_n K^{[n]} d^{[n]} = 
	C_\nu\langle\rho\rangle_n\Delta_n\sqrt{K^{[n]}}\,|S^{[n]\ast}|^2\,.
\end{equation}
The quantity $\Sigma^{[n]}=\tau_{ij}^{[n]}S_{ij}^{[n]}$ can be interpreted as rate of production of turbulence energy, $K^{[n]}$, below the length scale $\Delta_n$.
For instance, Fig.~\ref{fig:pdfs_eddy_visc} shows results for ILES of subsonic and transonic turbulence generated by stochastic forcing in periodic boxes \citep{SchmHille06}. By averaging the coefficient obtained from Eq.~(\ref{eq:flux_eddy_coarse}) over the whole simulation box and evaluating the production rate for the eddy-viscosity closure with the average coefficient, $\langle C_\nu\rangle$, the light blue line is obtained. It is evident that this line does not accurately reproduce the explicitly calculated energy transfer $\Sigma^{[n]}$ (purple line).
One can see that $\Sigma^{[n]}$ is negative in about 20\% of all grid cells, 
which corresponds to backscattering from smaller to larger scales.

The bias toward positive transfer can be mitigated by localizing the eddy-viscosity closure. In \citet{SchmNie06b}, 
test filters of varying filter length $\Delta_{\mathrm{T}} = \gamma_{\mathrm{T}}\Delta_n$ were employed to compute
$C_\nu$ locally. As demonstrated by the lines ranging from green to red in Fig.~\ref{fig:pdfs_eddy_visc} (left plot), a significantly improved match is obtained. In particular, the localized closure reproduces negative energy transfer quite well, indicating that even homogeneous turbulence simulations benefit from the dynamic procedure. Moreover, it seems that the distributions of $C_\nu$ are relatively insensitive to the Mach number and forcing parameter (see right plot in Fig.~\ref{fig:pdfs_eddy_visc}). 

\subsection{Global least squares method}
\label{sec:least_squares}

Closures can also be tested by analysing correlations. This allows for the calibration of
the closure coefficients by least squares minimization of the integrated residual \citep{SchmFeder11}.
For example, let us consider a generic closure with a single coefficient $C_1$ at the $n$-th filter
level:
\begin{equation}
	\label{eq:trace_free_fit}
	C_1 f^{[n](\mathrm{cls})} =
	\Sigma^{[n](\mathrm{cls})} + \frac{2}{3}\langle\rho\rangle_n K^{[n]} d^{[n]}\,,
\end{equation}
The global residual of the explicitly computed turbulence energy flux $\Sigma^{[n]}=\tau_{ij}^{[n]}S_{ij}^{[n]}$,
where $\tau_{ij}^{[n]}$ is defined by Eq.~(\ref{eq:turb_stress_n}), can be quantified 
by the squared error integrated over the whole domain $\mathcal{V}$ of the turbulent flow:
\begin{equation}
	\mathrm{err}^{2}(C_1) =
	\int_{\mathcal{V}} \left|\Sigma^{[n]} + \frac{2}{3}\langle\rho\rangle_n K^{[n]} d^{[n]} -
		C_1 f^{[n](\mathrm{cls})}\right|^{2}\dd^{3}x\,.
\end{equation}
The minimum of $\mathrm{err}^{2}(C_1)$ is obtained by setting the derivative with respect to $C_1$ equal to zero:
\begin{equation}
	\label{eq:coeff_lse}
	C_1=\frac{\int_{\mathcal{V}} f^{[n](\mathrm{cls})}
	\left[\Sigma^{[n](\mathrm{cls})} + \frac{2}{3}\langle\rho\rangle_n K^{[n]} d^{[n]}\right]\,\dd^{3}x}
		{\int_{\mathcal{V}} |f^{[n](\mathrm{cls})}|^2\dd^{3}x}\,.
\end{equation}
In contrast to the dynamic procedure, the resulting closure coefficients are constants.

The method of least squares described above is applied in \cite{SchmFeder11} to various ILES of supersonic
isothermal turbulence produced by stochastic forcing \citep{SchmFeder09,FederRom10}. To coarse-grain the data, 
a Gaussian filter with a smoothing length $\Delta_4=L/16=32\Delta_{I}$ is used, where $\Delta_{I}=\Delta_9\equiv\Delta$
is the grid resolution of the ILES.\footnote{The simulation from \cite{SchmFeder09} was performed on a 
	$768^3$ grid. In this case, the filter length is $\Delta_4=L/16=24\Delta$.} 
Since the filter length is large compared to the grid resolution in this case,
it is advantageous to apply the filter operation in Fourier space.
For the eddy-viscosity closure, we have
\[
	f^{[n](\mathrm{cls})} = \Delta_{n}\langle\rho\rangle_n\sqrt{2K^{[n]}}|S^{[n]\ast}|^2\,.
\]
The closure coefficients following from Eq.~(\ref{eq:coeff_lse}) are, for instance,
$C_1\approx 0.102$ for the $1024^3$ ILES with purely solenoidal (divergence-free) forcing, and 
$C_1\approx 0.092$ in the case of compressive (rotation-free) forcing. The corresponding value
of $C_{\nu}=\sqrt{2}C_1$ is about 0.14. When comparing this value 
to the results in \cite{SchmNie06b}, one has to bear in mind not only that a different method is applied to determine the coefficients, but also that the Mach numbers differ substantially.

\begin{figure}[tbp]
    \centerline{
      \includegraphics[width=0.5\textwidth]{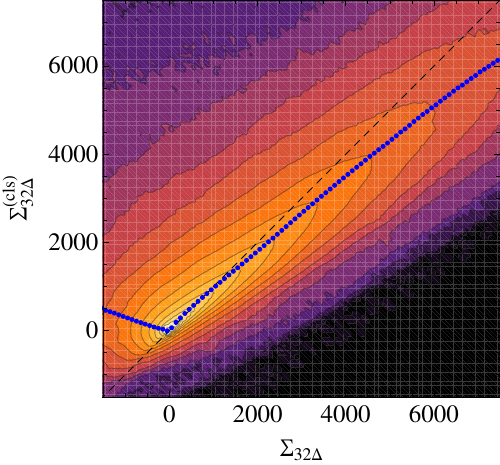} 
      \includegraphics[width=0.5\textwidth]{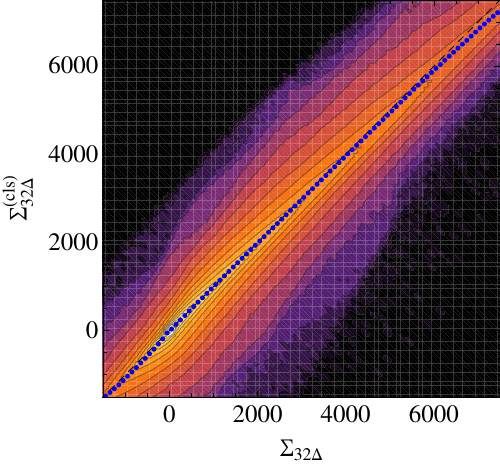}
    }
    \caption{Correlations of the coarse-grained turbulence energy flux
      with the	eddy-viscosity (\emph{left plot}) and non-linear (\emph{right plot}) closures
      for supersonic isothermal turbulence produced by solenoidal
      forcing. The applied filter length is $32\Delta$, where $\Delta$ is the grid resolution. 
      The average prediction of the closure for each bin is indicated by the blue
      dots.
      Image reproduced with permission from \cite{SchmFeder11}, copyright by ESO.}
    \label{fig:corrl_sol1024}
\end{figure}

The correlation diagram for $\Sigma^{[4](\mathrm{cls})}$, with the least-squares coefficient $C_1$, 
versus $\Sigma^{[4]}$ in the case of solenoidal forcing is shown in Fig.~\ref{fig:corrl_sol1024} (left plot). 
The overall correlation is actually quite good. A quantitative measure is the correlation coefficient
\begin{equation}
\begin{split}
   \mathrm{corr} &[\Sigma^{[n]},\Sigma^{[n](\mathrm{cls})}] =\\
   &\frac{\int_{\mathcal{V}} \left(\Sigma^{[n]}-\langle\Sigma^{[n]}\rangle\right)
   \left(\Sigma^{[n](\mathrm{cls})}-\langle\Sigma^{[n](\mathrm{cls})}\rangle\right)\,\dd^{3}x}
   		{\mathrm{std}(\Sigma^{[n]})\,\mathrm{std}(\Sigma^{[n](\mathrm{cls})})},
   \end{split}
\end{equation}
where $\mathrm{std}$ denotes the standard deviation and the brackets indicate an average over the whole domain.
The correlation coefficients of the eddy-viscosity closure are found to be 0.95 and 0.93 for solenoidal and
compressive forcing, respectively \citep{SchmFeder11}. However, it becomes apparent that the closure breaks
down for negative fluxes. This corresponds to the bias of the probability density function for 
an averaged coefficient in Fig.~\ref{fig:pdfs_eddy_visc} (left plot). This is remedied by the non-linear closure~(\ref{eq:tau_nonlin_norm}) for the turbulence stress tensor, which yields an excellent correlation 
between $\Sigma^{[n](\mathrm{cls})}$ and $\Sigma^{[n]}$ (see Fig.~\ref{fig:corrl_sol1024}, right plot).
The correlation coefficients are 0.991 for both solenoidal and compressive forcing.

However, as explained in Sect.~\ref{sec:turb_stress}, one faces potential problems when using the energy-scaled non-linear closure in LES. For this reason, \citet{SchmFeder11} applied the least squares method to the generalized closure with two coefficients, $C_1$ and $C_2$:
\begin{equation}
	C_{1}f^{[n](\mathrm{cls})} +  C_{2}g^{([n]\mathrm{cls})} =
	\Sigma^{[n](\mathrm{cls})} + \frac{2}{3}\langle\rho\rangle_n K^{[n]} d^{[n]}\,.
\end{equation}
The closure coefficients are given by the linear system of equations
\begin{align}
	\label{eq:coeff_lse_2}
	\begin{split}
	\left(\int_{\mathcal{V}} |f^{[n](\mathrm{cls})}|^2\dd^{3}x\right)\,C_1 + 
	&\left(\int_{\mathcal{V}} f^{[n](\mathrm{cls})}g^{[n](\mathrm{cls})}\dd^{3}x\right)\,C_2
		=\\
		&\int_{\mathcal{V}} f^{[n](\mathrm{cls})}
		\left(\Sigma^{[n](\mathrm{cls})} + \frac{2}{3}\langle\rho\rangle_n K^{[n]} d^{[n]}\right)\,\dd^{3}x\,, 
	\end{split}\\
	\begin{split}
	\left(\int_{\mathcal{V}}f^{[n](\mathrm{cls})}g^{[n](\mathrm{cls})} \dd^{3}x\right)\,C_1 + 
	&\left(\int_{\mathcal{V}} |g^{[n](\mathrm{cls})}|^{2}\dd^{3}x\right)\,C_{2}
	=\\
	&\int_{\mathcal{V}} g^{[n](\mathrm{cls})}
	\left(\Sigma^{[n](\mathrm{cls})} + \frac{2}{3}\langle\rho\rangle_n K^{[n]} d^{[n]}\right)\,\dd^{3}x\,,
	\end{split}
\end{align}
where 
\begin{equation}
	f^{[n](\mathrm{cls})} =
	\Delta_{n}\langle\rho\rangle_n\sqrt{2K^{[n]}}|S^{[n]\ast}|^2
	\quad\text{and}\quad
	g^{[n](\mathrm{cls})}  =-4\langle\rho\rangle_n K^{[n]}
	\frac{u_{i,k}^{[n]}u_{j,k}^{[n]}S_{ij}^{[n]\ast}}
	{|\vecnab\otimes\vecu^{[n]}|^{2}}\,.
\end{equation}
The resulting coefficients are $C_{1}\approx 0.02$ and $C_{2}\approx 0.7$, with a correlation of $0.990$.
These coefficients provide good approximations also for isothermal and adiabatic turbulence simulations 
at lower Mach numbers. Additionally, the coefficients appear to vary minimally with the filter length, 
particularly in the range from $\Delta_3=64\Delta$ to $\Delta_5=16\Delta$.
Choosing longer or shorter filter lengths is not advisable because of the influence of the forcing ($\Delta_n$ must be small
compared to $\Delta_0=L$) and numerical dissipation ($\Delta_n\gg\Delta$).

\begin{figure}[htbp]
\centering
\subfigure[Correlations of different SGS energy closures (isotropic component of the SGS stress tensors). ]{
\includegraphics[width=0.9\linewidth]{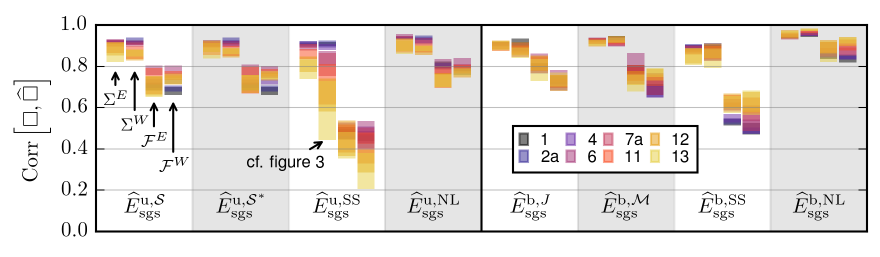}
\label{fig:EnergyOverview}
}

\subfigure[Correlations of different traceless SGS Reynolds stress $\tau^{\mathrm{u}*}$ closures. ]{
\includegraphics[width=0.9\linewidth]{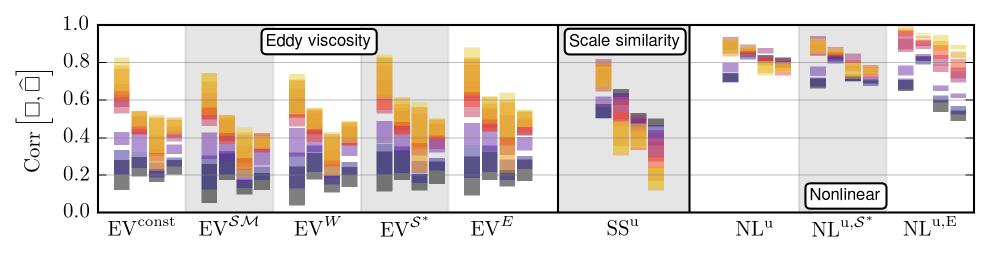}
\label{fig:tauUStarOverview}
}

\subfigure[Correlations of different traceless SGS Maxwell stress $\tau^{\mathrm{b}*}$ closures.]{
\includegraphics[width=0.9\linewidth]{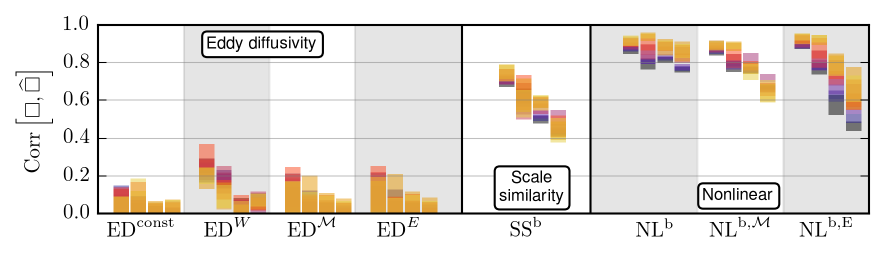}
\label{fig:tauBStarOverview}
}

\subfigure[Correlations of different electromotive force $\mathcal{E}$ closures.]{
\includegraphics[width=0.9\linewidth]{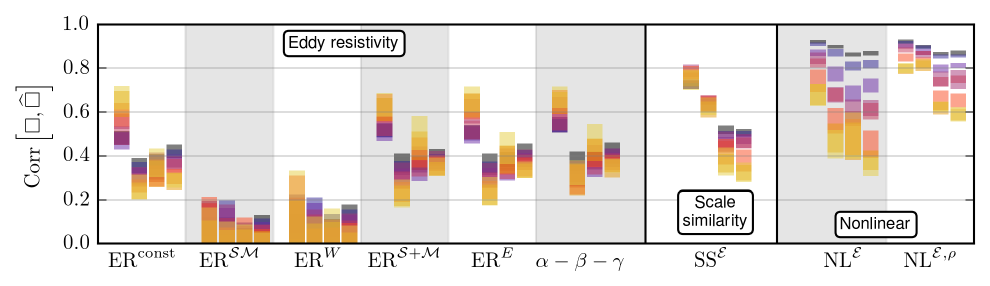}
\label{fig:EMFOverview}
}

\caption{
Correlation coefficients of closures for SGS energies (a), SGS kinetic and magnetic stresses (b,c), and the SGS electromotive force (d). The coloured bars indicate the range of correlation over all snapshots from different MHD turbulence simulations (color-coded). Each color corresponds to a particular combination of sonic and Alfv\'{e}nic Mach numbers (subsonic runs are towards the dark end and supersonic at the bright end of the palette). The bottom labels indicate different closures as detailed in \citet{GreteVlay16}. The four bars for each closure show the results from fits to cascade-related ($\Sigma$) and total ($\mathcal{F}$) fluxes of turbulence energy ($E$) and cross helicity ($W$). The variable $E_\mathrm{sgs}$ corresponds to $K$ in the notation used in this review. Reprinted from \citet{GreteVlay16}, with the permission of AIP Publishing.
\label{fig:ModelOverview}
}
\end{figure}

The least squares method was extended to a large suite of MHD turbulence simulations with Mach numbers reaching from the subsonic to the supersonic regime by \citet{GreteVlay16}. They consider not only the cascade-related flux term, $\Sigma$, but also the total flux including conservative transport terms (see Sect.~\ref{sc:mhd}). While in Eq.~(\ref{eq:trace_free_fit}) only the trace-free contribution to the turbulence energy flux is evaluated and no assumption about the SGS turbulence energy is made, \citet{GreteVlay16} focus on local equilibrium models for SGS energies. For contributions related to the anisotropic, trace-free parts of the SGS stresses, various flavours of linear eddy-diffusivity closures and non-linear closures (here, diffusivity generically refers to viscosity, resistivity, or some other kind of diffusivity) are investigated. Moreover, scale-similarity models were investigated. They utilize the similarity between the stresses computed with a test filter and the SGS stresses. In contrast to dynamic procedures, a constant coefficient is determined in this case. The results are presented in Fig.~\ref{fig:ModelOverview}. These plots contain a wealth of information. The most significant conclusions that can be drawn from the correlations between model and data are:
\begin{itemize}
\item For all tested quantities, closures based on isotropic diffusivities (e.g.\ eddy-viscosity closure for kinetic turbulence stress tensor) have poor to moderate correlations. In particular, magnetic stresses cannot be approximated by eddy diffusivities at all, as the sign (direction of the cascade) matches only in approximately $50\,\%$ of the volume.
\item Scale-similarity closures reach acceptable correlation coefficients for the kinetic turbulence energy flux $\Sigma$, but fail to provide satisfactory results for cross helicity.
\item High correlations (close to unity) are obtained only for the non-linear closures defined by Eqs.~(\ref{eq:tau_mhd_u_struc}), (\ref{eq:tau_mhd_u_struc}), and (\ref{eq:emf_struc}). Variants of these closures with normalizations as in Eq.~(\ref{eq:tau_nonlin_norm}) degrade correlation. However, this does not necessarily imply that energy-scaled non-linear closures are intrinsically less accurate because they are only tested under the assumption of local equilibrium, i.e., the SGS turbulence energies are simply defined by the traces of $\tau_{ij}^\mathrm{u}$ and $\tau_{ij}^\mathrm{b}$.
\item The eddy-viscosity closures shown in Fig.~\ref{fig:ModelOverview}(b) tend to improve with Mach number. This finding is surprising, as this type of closure was originally applied to incompressible or weakly compressible turbulence. A similar trend is observed for non-linear closures, although the difference between the lowest and highest Mach numbers is much smaller.
\item As demonstrated by Fig.~\ref{fig:ModelOverview}~(d), modelling the SGS EMF is notoriously difficult. Only the non-linear closure that incorporates the compressibility correction reproduces the cascade-related energy fluxes with sufficient accuracy. Without this correction, correlations are substantially reduced in the supersonic regime.
\end{itemize}
Therefore, the preferred model encompasses the non-linear closures with compressibility correction outlined in Section~\ref{sec:struc}. The closure coefficients for the SGS stresses and EMF are equal to unity within the error bars in agreement with the structural modelling approach.

\section{Astrophysical applications}
\label{sec:astro}

\subsection{Thermonuclear combustion in white dwarfs}
\label{sec:SN_Ia}

One possible scenario for supernovae of type Ia is the
thermonuclear explosion of a Chandrasekhar-mass white dwarf that accretes in a 
close binary system \citep{HilleKrom13}. 
If the mass of a white dwarf approaches the Chandrasekhar limit, 
explosive carbon and oxygen burning is ignited \citep{NonaAsp12}.
Owing to the degeneracy of white dwarf matter, the thermonuclear 
reaction zones propagate as thin flame fronts, whose thickness
$\delta_{\mathrm{f}}$ and propagation speed $s_{\mathrm{f}}$ are determined by 
the very high thermal conductivity of the fuel and the nuclear reaction rates. 
This mode of burning is called deflagration. 
Since the burned material has lower density than the fuel, 
it rises because of its buoyancy. Consequently, the energy released by 
thermonuclear deflagration drives convection. Since eddies
exert strong shear on rising bubbles of burning material, 
they are deformed into mushroom-like shapes and Kelvin-Helmholtz instabilities 
at the surface are rapidly producing turbulence \citep{MalNona13}.
Eventually, this results in a very complex flame front with a fractal structure that
cannot be resolved in numerical simulations over the full range of length scales.

This problem was addressed by performing LES with
an effective propagation speed of the turbulent flame front (\citealp{ReinHille02,SchmNie06a,
RoepHille07}, to mention just a few examples). In these simulations, the flame front 
is tracked by means of the level set method \citep{OshSeth88,ReinHille99}, which is able to follow
complex topological changes by determining the interface between fuel and burned material
as the spatial surface for which a signed distance function is zero.
The evolution of the distance function from given initial conditions is
given by the advection velocity relative to the fluid plus the flame propagation
speed. In a fully resolved simulation, the flame propagation speed
would be given by the microscopic flame speed $s_{\mathrm{f}}$ (also called laminar
flame speed). If the flame front is underresolved, however, the wrinkling and folding 
of the front by turbulent eddies below the grid scale leads to an enhanced rate
of energy release. In this case, the relevant time scale is not the
microscopic diffusion time scale but the turn-over time of numerically
unresolved eddies. Simple dimensional reasoning implies that
$s_{\mathrm{f}}$ has to be replaced by the turbulent flame speed
$s_{\mathrm{t}}=\sqrt{2K}$ \citep{NieHille95,NieKer97,Peters99}, where $K$ is the SGS turbulence
energy given by Eq.~(\ref{eq:pde_energy_close}).
In \cite{SchmNie06c}, the formula of \citet{Poch94},
\begin{equation}
  \label{eq:flame_speed_turb}
  s_{\mathrm{t}}= s_{\mathrm{f}} 
  \sqrt{1 + C_{\mathrm{t}}\frac{2K}{s_{\mathrm{f}}^2}}\,,
\end{equation}
is adopted for a smooth transition between laminar and turbulent flame propagation. They
set $C_\mathrm{t}=4/3$ to get the expected flame speed in the highly turbulent regime.
A further complication comes from the pronounced anisotropy of turbulence
at the flame front \citep{CirSchm08,MalNona13}. While the burned material inside the flame 
is highly turbulent, there is little or no turbulence in the fuel just outside the flame. In a way,
this is similar to walls in terrestrial flows. For this reason, it is important to apply
the dynamic procedure explained in Sect.~\ref{sec:dyn_proc} to locally calculate the 
eddy-viscosity coefficient, which determines the rate of production of $K$.

The minimal length scale for which the flame front is affected by 
turbulence is given by the Gibson scale $\ell_{\mathrm{G}}$.
If $v^{\prime}(\ell)$ is the mean turbulent velocity fluctuation
on the length scale $\ell$, then $\ell_{\mathrm{G}}$ is implicitly
given by the condition 
\begin{equation}
  v^{\prime}(\ell_{\mathrm{G}}) = s_{\mathrm{f}}\,,
\end{equation}
i.e., the turbulent velocity fluctuation on the Gibson scale
equals the microscopic flame speed. For $\ell\ll\ell_{\mathrm{G}}$,
the turn-over time associated with $v^{\prime}(\ell_{\mathrm{G}})$
is much longer than the crossing time of the
flame front through an eddy of size $\ell$. On these scales, the flame front is
virtually unaffected by turbulence. If $\ell\gg\ell_{\mathrm{G}}$,
on the other hand, eddies will significantly distort the flame front 
on length scales between $\ell_{\mathrm{G}}$ and $\ell$.
In LES of a whole white dwarf, the Gibson scale $\ell_{\mathrm{G}}$ is much
smaller than computationally feasible grid resolutions $\Delta$ during most
of the deflagration phase and, consequently, a turbulent flame speed model has
to be applied.%
\footnote{To be more precise, the notion of a flame front propagating
	with the turbulent flame speed $s_{\mathrm{t}}$ applies to the so-called flamelet regime,
	for which $\ell_{\mathrm{f}}\ll\ell_{\mathrm{G}}$. 
	When $\ell_{\mathrm{G}}$ becomes comparable to $\ell_{\mathrm{f}}$,
	turbulence affects the internal structure of the flame and distributed
	burning sets in \cite{NieKer97,RoepHille05,Schm07}.}

\begin{figure}[htbp]
    \centerline{\includegraphics[width=\textwidth]{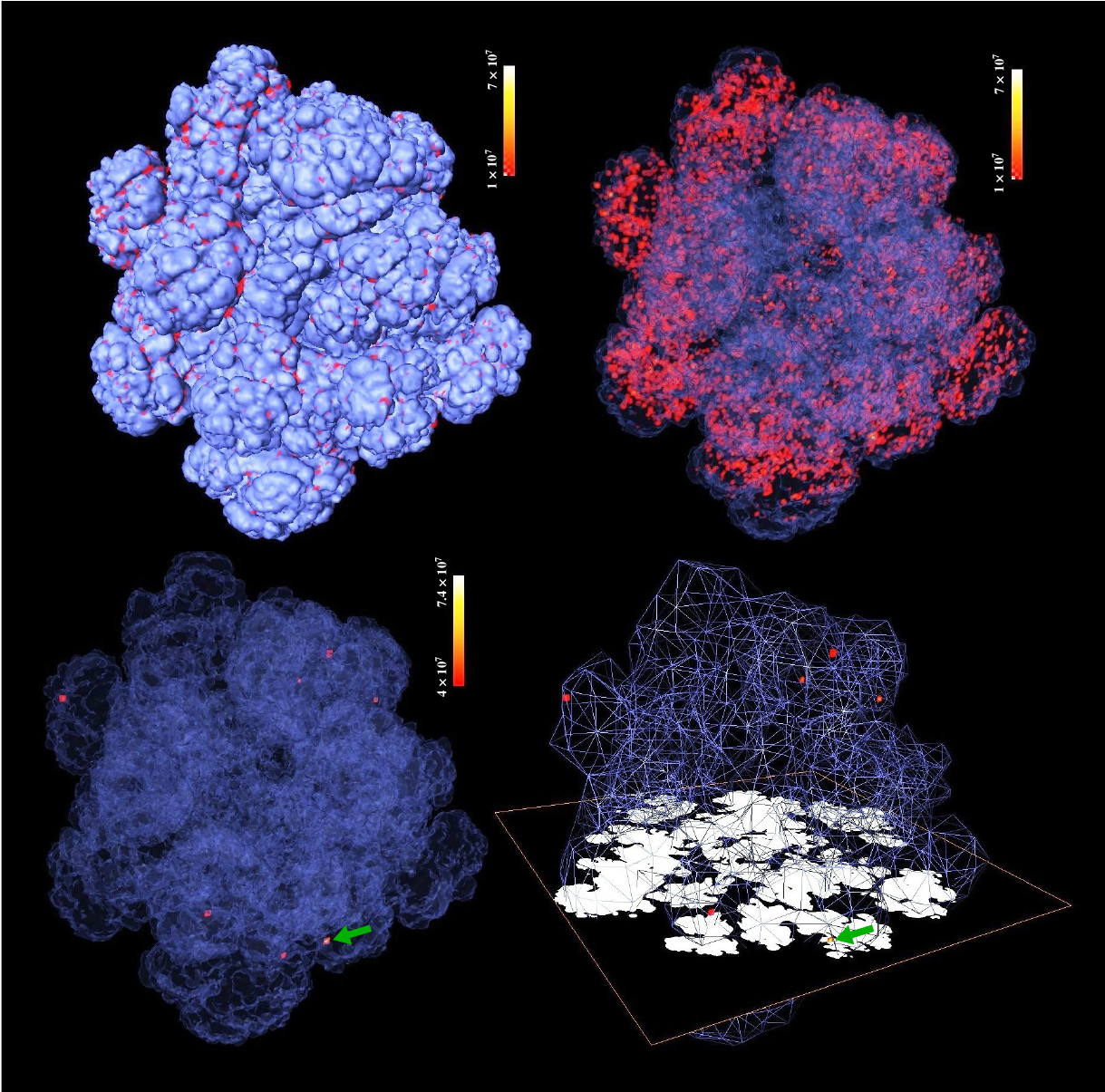}}
    \caption{Flame front in a high-resolution simulation of a
      thermonuclear supernova. High turbulent velocity fluctuations
      occur in the reddish regions. Extremely strong fluctuations that
      could trigger a transition to a detonation are indicated by the
      green arrows. 
      Image reproduced with permission from \cite{Roepke07},
      copyright by AAS.}
    \label{fig:flame_turb}
\end{figure}

As an illustration of the level set method, Fig.~\ref{fig:flame_turb} shows a flame front evolved with the turbulent flame speed defined by Eq.~(\ref{eq:flame_speed_turb}) from \citet{Roepke07}. However, the predictions from pure deflagration models are not consistent with observational properties of the majority of type Ia supernovae \citep{Taubenberger2017}.
While pure deflagrations could explain subluminous events in the subclass SN Iax \citep{Jha2017}, models that yield the energies and luminosities of typical type Ia supernovae necessitate a detonation \citep{Roepke18}. 
Among the proposed mechanisms that cause a detonation is the transition from the subsonic flame front to a supersonic shock front in a Chandrasekhar-mass white dwarf. However, a theoretical explanation for this transition remains an enduring challenge. One plausible scenario is that very strong local velocity fluctuations at the onset of distributed burning might trigger the transition to a detonation \citep{NieKer97,KhokhOran97,poludnenko_detonation_19}. As depicted in Fig.~\ref{fig:flame_turb}, the strong intermittency of turbulent velocity fluctuations following from the SGS turbulence energy $K$ in high-resolution simulations suggests that such events might indeed occur \citep{Roepke07,Seitenzahl12,CiaSeit13}.

\subsection{Star formation in galaxy simulations} 
\label{sec:galaxy}

Numerical simulations of galaxies, especially from cosmological initial conditions, cannot fully resolve processes
such as star formation and feedback from supernova (SN) explosions. As suggested in \cite{JoungLow09}, \cite{ScanBruegg10}, and \cite{BraunSchm12}, 
a production term $\Sigma_{\star}$ due to feedback can be incorporated as an additional source term in the SGS 
turbulence energy equation~(\ref{eq:pde_energy_sgs}). The simplest model for this term is
\begin{equation}
	\label{eq:sigma_SN}
	\Sigma_{\star}= C_{\star}\frac{\rho e_{\mathrm{SN}}}{\tau_{\mathrm{ff}}}\,,
\end{equation}
where $e_{\mathrm{SN}}$ is the typical energy released by
a core-collapse supernova and $\tau_{\mathrm{ff}}=\sqrt{3\uppi/32 G\rho}$ is the free-fall time scale. The parameter $C_{\star}$ controls the effective feedback time scale. The above expression for $\Sigma_{\star}$ follows from the simple Kennicutt--Schmidt relation $\dot{\rho}_{\star}\propto\rho^{-3/2}$ for the star formation rate, 
where the factor $C_{\star}\rho/\tau_{\mathrm{ff}}$ is some fraction of $\dot{\rho}_{\star}$. Apart from stellar feedback, turbulence is produced by instabilities in the ISM, such as gravitational, thermal, and hydrodynamical instabilities. Typically, energy is injected on numerically resolved length scales by these instabilities and transferred to smaller scales by the turbulent cascade.
As a very crude model, let us consider the local equilibrium between production and dissipation, i.e., $\Sigma+\Sigma_{\star}\sim\epsilon$. By substituting $\Sigma = \rho K/\tau$, where $\tau$ is the dynamical time scale associated with energy transfer through the cascade, the energy injection due to feedback given by Eq.~(\ref{eq:sigma_SN}), and
the dissipation rate $\epsilon\sim\rho K^{3/2}/\Delta$, the equilibrium condition reads
\[
	\frac{K}{\tau} + C_{\star}\frac{e_{\mathrm{SN}}}{\tau_{\mathrm{ff}}} \sim \frac{K^{3/2}}{\Delta}\,.
\]
Depending on the energy ratio $K/e_{\mathrm{SN}}$ and 
the ratio of the dynamical and feedback time scales, $\tau/\tau_{\mathrm{ff}}$, the production of SGS turbulence energy
is dominated by the turbulent cascade or by supernovae. The quantity $\frac{2}{3}\rho K$ can be interpreted as numerically unresolved contribution to the pressure of the turbulent interstellar medium (see Sect.~\ref{sec:K_eq}).

The star formation rate, which in turn determines the feedback rate, can be parametrized by the
gas density, temperature, and turbulence intensity \citep{KrumMcKee05,Padoan11,HenneChab11}.
The two main parameters appearing in these parametrizations are the turbulent Mach number $\mathcal{M}_{\star}$ and 
the virial parameter $\alpha_{\star}$ \citep{BertMcKee92}, which correspond to the ratios of the turbulence energy to the internal and gravitational energies, respectively:
\begin{equation}
  \mathcal{M}_{\star} = \sqrt{3}\,\frac{\sigma(\ell_{\mathrm{c}})}{c_{\mathrm{s}}}
  \quad\text{and}\quad
  \alpha_{\star} = \frac{15\sigma^2(\ell_{\mathrm{c}})}{\pi G\rho\ell_{\mathrm{c}}^2}\,.
\end{equation}
Here, $G$ is Newton's constant, $c_{\mathrm{s}}$ is the speed of sound, and $\sigma(\ell_{\mathrm{c}})$ is the 
turbulent velocity dispersion on a length scale, $\ell_{\mathrm{c}}$, that is characteristic for cold, dense clumps.
The star formation rate is then given by
\begin{equation}
	\label{eq:sfr}
	\dot{\rho}_{\star} = \epsilon_\mathrm{ff}(\alpha_{\star},\mathcal{M}_{\star})\frac{\rho}{\tau_{\mathrm{ff}}}\,,
\end{equation}
where star formation efficiency $\epsilon_\mathrm{ff}(\alpha_{\star},\mathcal{M}_{\star})$ specifies the mass fraction that is converted into stellar mass per free-fall time. This factor, which accounts for the gravo-turbulent fragmentation of star-forming clouds \citep{HenneFalg12}, depends on the adopted star formation law. For example, power-law functions based on data from numerical simulations are proposed in \citet{KrumMcKee05} and \citet{PadHaug12}. \citet{Padoan11} suggest that $\epsilon_\mathrm{ff}(\alpha_{\star},\mathcal{M}_{\star})$ is determined by the log-normal probability density function (PDF), whose width depends on the turbulent Mach number $\mathcal{M}_{\star}$ and the forcing \citep{Padoan_2002,KritNor07,FederRom10}. In the model proposed by \citet{HenneChab11}, the star formation efficiency is calculated by integrating the mass spectrum of substructures within a cloud, weighted by their respective free-fall time factors. These different parametrizations are compared and calibrated using numerical data in \cite{FederKless12}. 

A star formation rate proportional to the density of molecular hydrogen, $f_{\mathrm{H_2}}\rho$, was suggested on grounds of the observed tight correlation between the star formation rate and the column density of molecular hydrogen \citep{KrumMcKee09,GnedTass09}. Since molecular hydrogen forms only in the cold phase of the interstellar medium, \citet{BraunSchm12} combined the star formation and feedback framework outlined above with a two-phase model based on \citet{SpringHern03} to approximate the turbulent multi-phase structure of the ISM. The model was applied in AMR simulations of isolated disk galaxies \citep{Braun14}, starting with a rotating gaseous disk of uniform temperature in hydrostatic equilibrium and a spherically symmetric gravitational potential representing the dark matter halo \citep[see also][]{wang_equilibrium_2010}. In addition, the SGS turbulence energy was computed using the treatment of energy and momentum conservation outlined in Sect.~\ref{sec:consrv}. The initial disk rapidly collapses into a thin cold disk because it radiates away energy and loses pressure support against its own gravity. Due to various instabilities, the disk does not uniformly collapse but fragments and develops an intricate clumpy structure. After the disk has collapsed, the gas in the clumps is cold and dense enough to form stars. In the star-forming disk, a turbulent velocity dispersion around $10\;\mathrm{km\,s^{-1}}$ is maintained by the turbulent cascade, while the internal driving caused by supernovae excites much stronger turbulent velocity fluctuations. 

\begin{figure}[htp]
\centering
\includegraphics[width=0.8\linewidth]{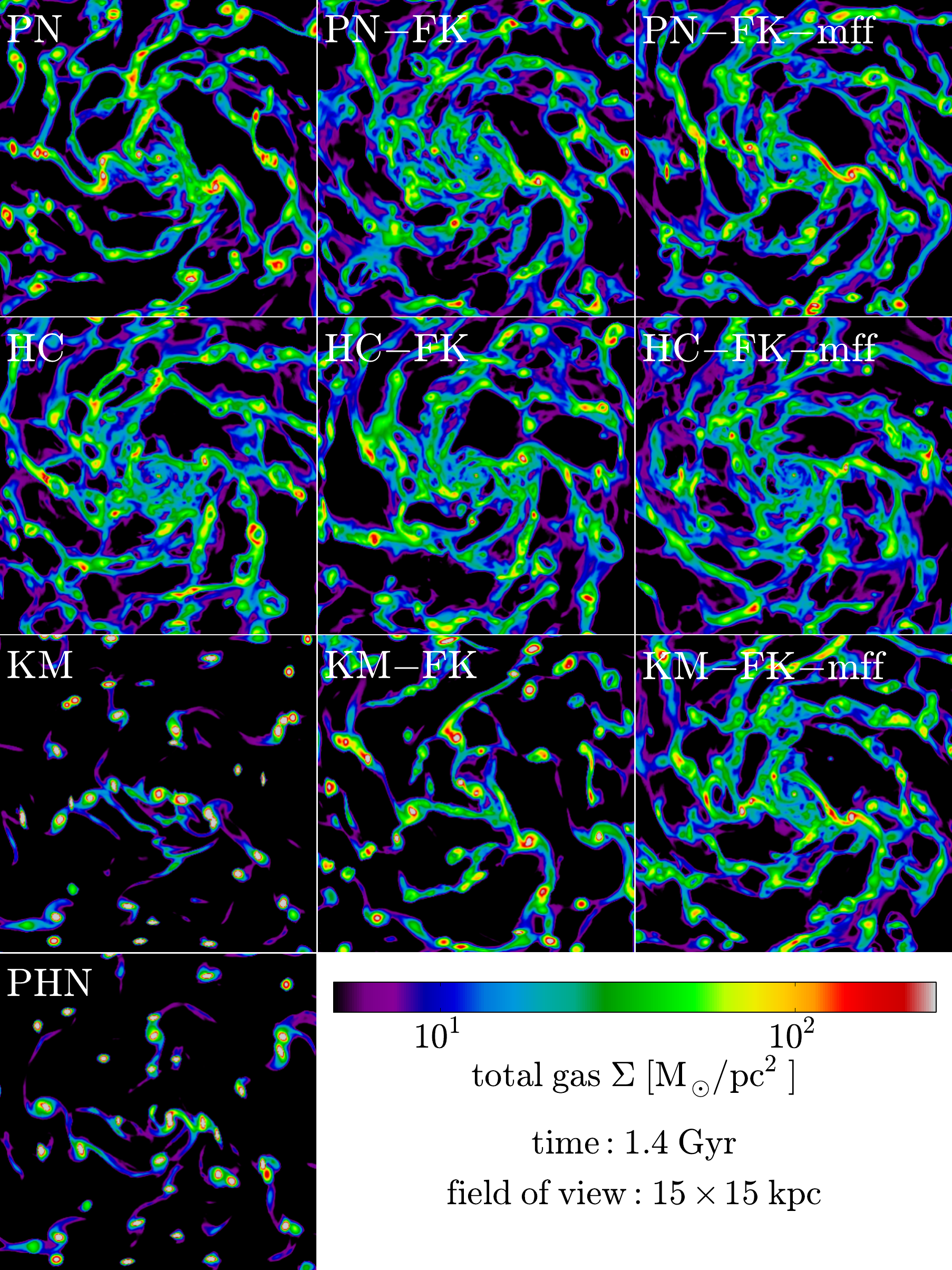}
\caption{Gas column densities in isolated disk galaxy simulations using the multiphase star formation and feedback model of \cite{Braun14} with various flavours of the star formation laws of \citet{Padoan11}, \citet{HenneChab11}, and \citet{KrumMcKee05}, which are labelled with the prefixes PN, HC, and KM, respectively (see table 1 in \citealt{Braun15} for details). The simple star formation law of \citet{PadHaug12} is abbreviated by PHN. Image reproduced with permission from \cite{Braun15}, copyright by OUP.
}
\label{fig:discs_sf_law}
\end{figure}

\begin{figure}[tp]
\centering
\includegraphics[width=0.7\linewidth]{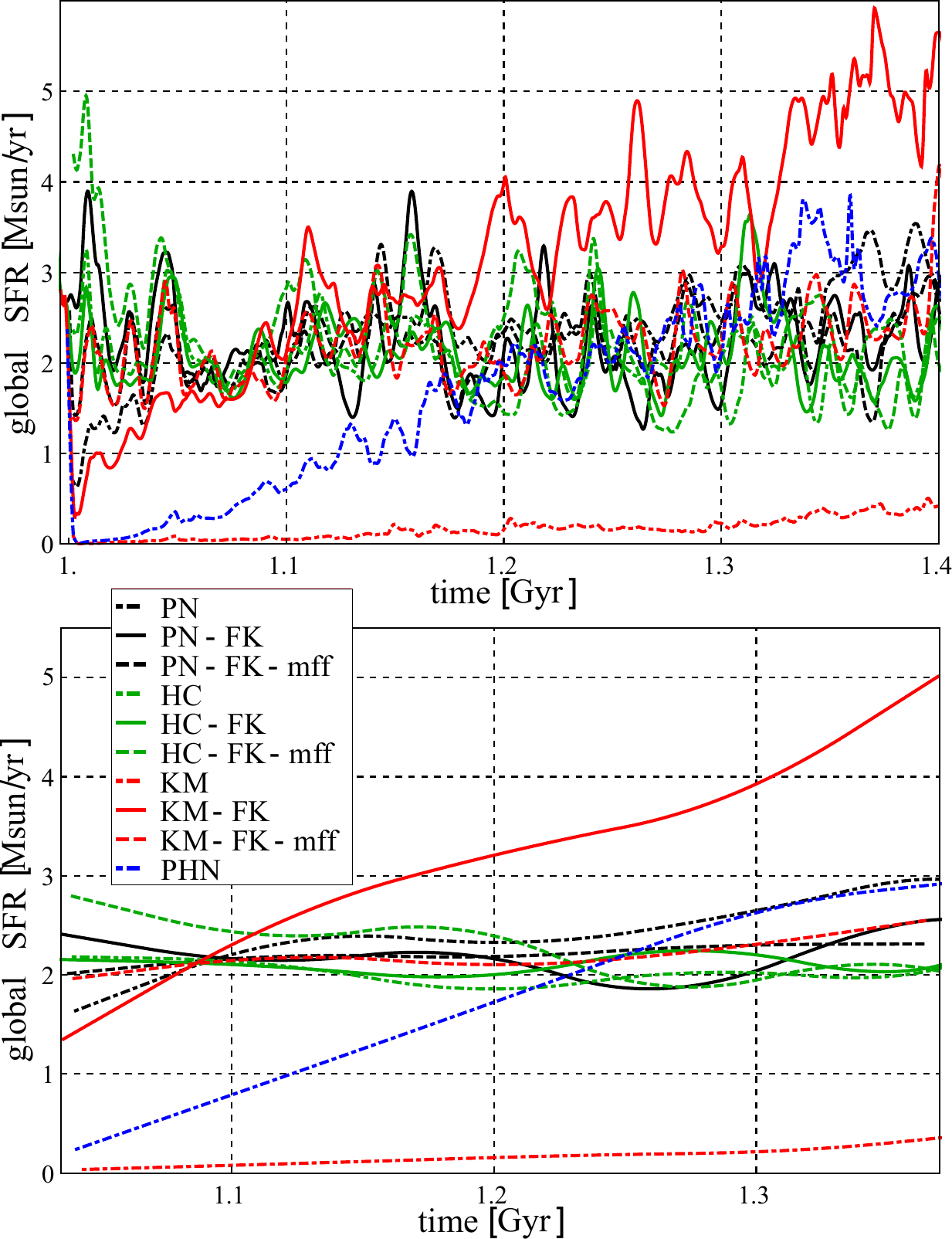}
\caption{\emph{Upper plot:} Global star formation rate for the same cases as in Fig~\ref{fig:discs_sf_law}. \emph{Lower plot:} To highlight the overall changes induced by different star formation laws, the star formation rate is smoothed over a time of $80\;{\rm Myr}$. Image reproduced with permission from \cite{Braun15}, copyright by OUP.
}
\label{fig:mean_sf}
\end{figure}

A Kennicutt-Schmidt relation between the star formation rate and the column density of molecular hydrogen in agreement with observations is not assumed as input but emerges from the simulations of \citet{Braun14}. While constant star formation efficiencies are assumed as  parameter in many isolated disk galaxy simulations \citep[e.g.][]{Renaud13,Murante14,kim_agora_2016,Su17,bland-hawthorn_turbulent_2024}, an average total star formation efficiency of $0.01$ is predicted by the turbulence-regulated star formation model. In dense molecular regions the efficiency is locally enhanced. In a comparative study, \citet{Braun15} demonstrated that the employed star formation model significantly impacts the disk structure. For some models, particularly the so-called simple star formation law proposed by \citet{PadHaug12}, an extremely clumpy disk is obtained (see Fig.~\ref{fig:discs_sf_law}). In this case, the star formation efficiency is too low and feedback is not sufficient to stop the accumulation of cold gas in very massive high-density clumps. More complex models, such as \citet{Padoan11} and \citet{HenneChab11}, allow for stronger fragmentation of the disk and star formation efficiencies in agreement with observations. As shown in Fig.~\ref{fig:mean_sf}, this results in a global star formation rate around a few solar masses per year for a Milky Way-like galaxy. The local quenching of star formation by supernova explosions causes the fluctuations of the star formation rate on time scales comparable to the typical lifetime of star-forming clouds in the upper plot in Fig.~\ref{fig:mean_sf}. An open question is how the star formation rate is affected by magnetic fields. Small-scale simulations of the turbulent ISM suggest a reduction of the star formation rate due to the stabilizing effect of magnetic fields in turbulent gas clouds \citep{Federrath15}.

\begin{figure}[tp]
\centering
\includegraphics[width=0.6\linewidth]{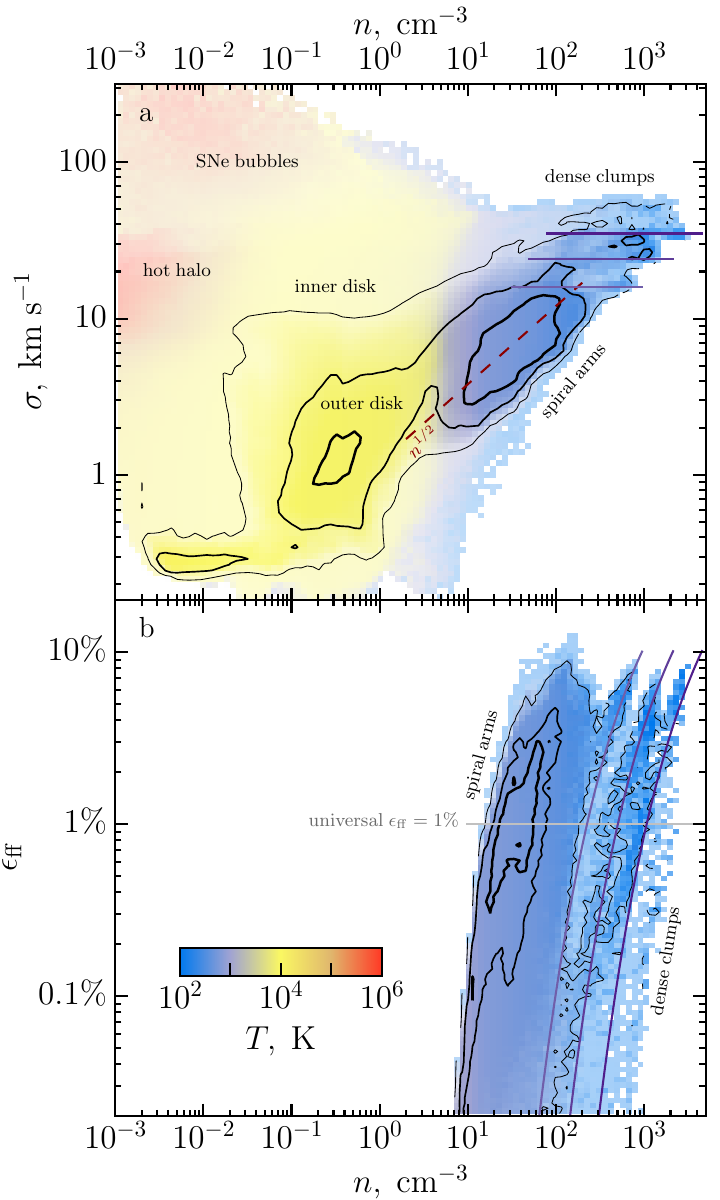}
\caption{Distribution of the SGS turbulent velocity dispersion $\sigma$ (\emph{upper plot}) and the resulting star formation efficiencies (\emph{lower plot}) at different densities in a simulation of an isolated disk galaxy by \citet{Semenov16}. Colors show the mass-weighted average temperature in the bin, and its intensity indicates the total mass in the bin. Black contours enclose 25$\,\%$, 68$\,\%$, 95$\,\%$ (\emph{upper plot}) and 5$\,\%$, 15$\,\%$, 30$\,\%$ (\emph{lower plot}) of the total gas mass. Image reproduced with permission from \citet{Semenov16}, copyright by IOP.
}
\label{fig:sigma_eps}
\end{figure}

A major problem in galaxy simulations is the choice of a characteristic scale $\ell_{\mathrm{c}}$ of star formation. Since the SGS turbulence energy model outlined in Sect.~\ref{sec:K_eq} predicts the specific turbulence energy $K=3\sigma^2(\Delta)/2$ on the grid scale $\Delta$, it is possible to estimate the turbulent velocity dispersion in star-forming clumps as 
\begin{equation}
    \label{eq:sigma_c}
	\sigma(\ell_{\mathrm{c}})^2 = \frac{2}{3}K\left(\frac{\ell_{\mathrm{c}}}{\Delta}\right)^{2\eta}\,,
\end{equation}
where the scaling exponent $\eta$ is in the range between $1/3$ and $1/2$, depending on the compressibility
of the gas and the intermittency of turbulence \citep{SchmFeder08,FederRom10,HenneFalg12}. The turbulent velocity dispersion in turn determines the magnitude of density fluctuation inside a clump.
In \citet{Braun14}, clumps are treated as unresolved substructure of the ISM and it is assumed that $\ell_{\mathrm{c}}$ is given by the local Jeans length for gravitational instability. \citet{Semenov16} use the star formation recipe of \citet{PadHaug12} and assume $\ell_{\mathrm{c}}=\Delta$ for numerical resolution scales $\Delta$ between $10$ and $40\;$pc. This model implies that $\epsilon_\mathrm{ff}(\alpha_{\star},\mathcal{M}_{\star})$ is asymptotically determined by the virial parameter $\alpha_{\star}$ if $\mathcal{M}_{\star}\gg 1$. Such conditions are found in highly turbulent cold clumps. Otherwise the star formation efficiency is limited by the thermal energy of the gas. The distribution of the turbulent velocity dispersion in one of their simulations is shown in Fig.~\ref{fig:sigma_eps} (upper plot). Particularly high values of $\sigma$ are associated with supernova bubbles and dense gas clumps, in which $\sigma$ tends to be independent of density. On the other hand, $\sigma\propto n^{1/2}$ in the spiral arms, where $n$ is the number density, with a characteristic value of $10\;\mathrm{km/s}$. Similar to \citet{Braun14}, they find that the star formation efficiency varies in a wide range from $10^{-3}$ to $0.1$ in most of the gas (see lower plot in Fig.~\ref{fig:sigma_eps}). The great appeal of the approach put forward by \citet{Semenov16} is that the  model is very simple to implement in comparison to the complex system of equations used in \citet{Braun14}, which relies on numerous assumptions about the sub-resolution multi-phase structure of the ISM.

\citet{polzin_universality_2024} show that star formation efficiencies computed with the model of \citet{Semenov16} are not sensitive to metallicity and UV flux. Although $\epsilon_\mathrm{ff}$ undergoes local variation, the mean value resulting from gas compression and ensuing stellar feedback in isolated gas disks is universal. However, compared to a model with a constant efficiency, the model based on SGS turbulence produces a significantly earlier disk formation and bursty star formation. This was recently demonstrated by re-simulating a  Milky-Way-like disk from the TNG50 simulation \citep{semenov_how_2024}. The variation of the star formation rate agrees with observed scatter in early disk galaxies. 
Moreover, \citet{semenov_uv-bright_2024} identify two regimes: Before the disk forms, $\epsilon_\mathrm{ff}$ varies strongly due to gas accretion in cold streams and mergers. This produces star-forming gas with efficiencies above $0.1$. Once the a disk has formed, star formation settles into a more steady regime with lower $\epsilon_\mathrm{ff}$, which is regulated by the equilibrium between turbulence production due to disk instabilities and dissipation. Simulations with fixed efficiencies fail to reproduce both regimes, resulting in disks that either lack episodes of bursty star formation or prevent long-lived disks.

\subsection{Turbulent velocity dispersion in cosmological simulations} 
\label{sec:clusters}

Since structure formation produces a strongly clumped medium through gravitational contraction or
collapse, the numerical simulation of turbulence on cosmological scales is particularly challenging. Potential production mechanisms of turbulent flows in the baryonic gas are mergers between dark-matter haloes and the accretion of gas into haloes, 
but also feedback from active galactic nuclei and winds produced by strongly star-forming galaxies \citep{DoBy08,KravtBor12}.
A largely open question concerns whether the gaseous component of haloes is in a state of developed turbulence. 
While there is no doubt about turbulence in the interior of galaxies, there is no direct observational evidence yet for
turbulence on larger scales, particularly in the intracluster medium (ICM). Theoretical and numerical studies
suggest that magnetic fields play a key role in the physical dissipation mechanism and, possibly, the onset 
of instabilities, but the associated length scales are highly uncertain \citep{FerrGov08,Brueggen13}.
Despite these uncertainties, turbulent flows resulting from cosmological structure formation 
are numerically investigated by computing the evolution of an ideal fluid subject to the gravitational potential of dark matter, which is modelled as a collisionless $N$-body system \citep{BorKravt11}. 

\begin{figure}[htbp]
  	\centerline{
          \includegraphics[width=0.53\textwidth]{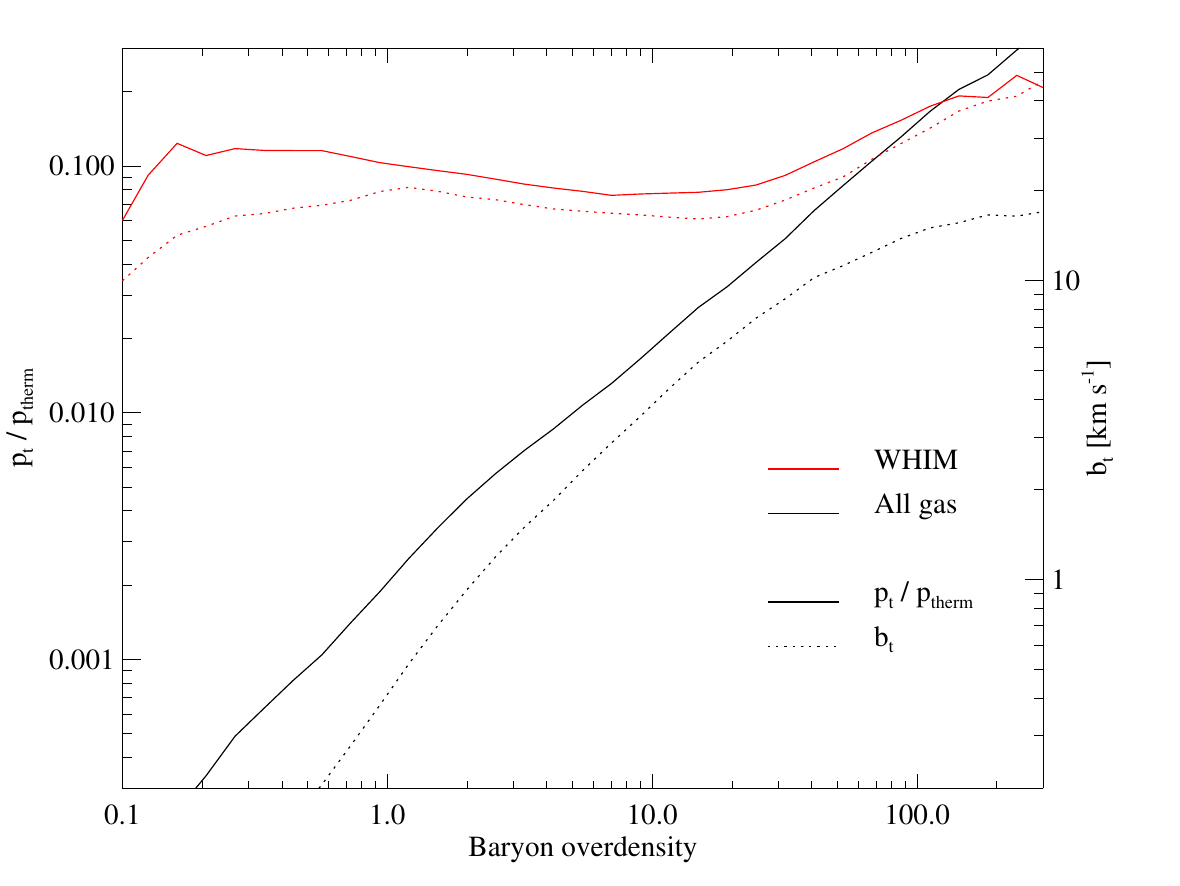}\quad
          \includegraphics[width=0.45\textwidth]{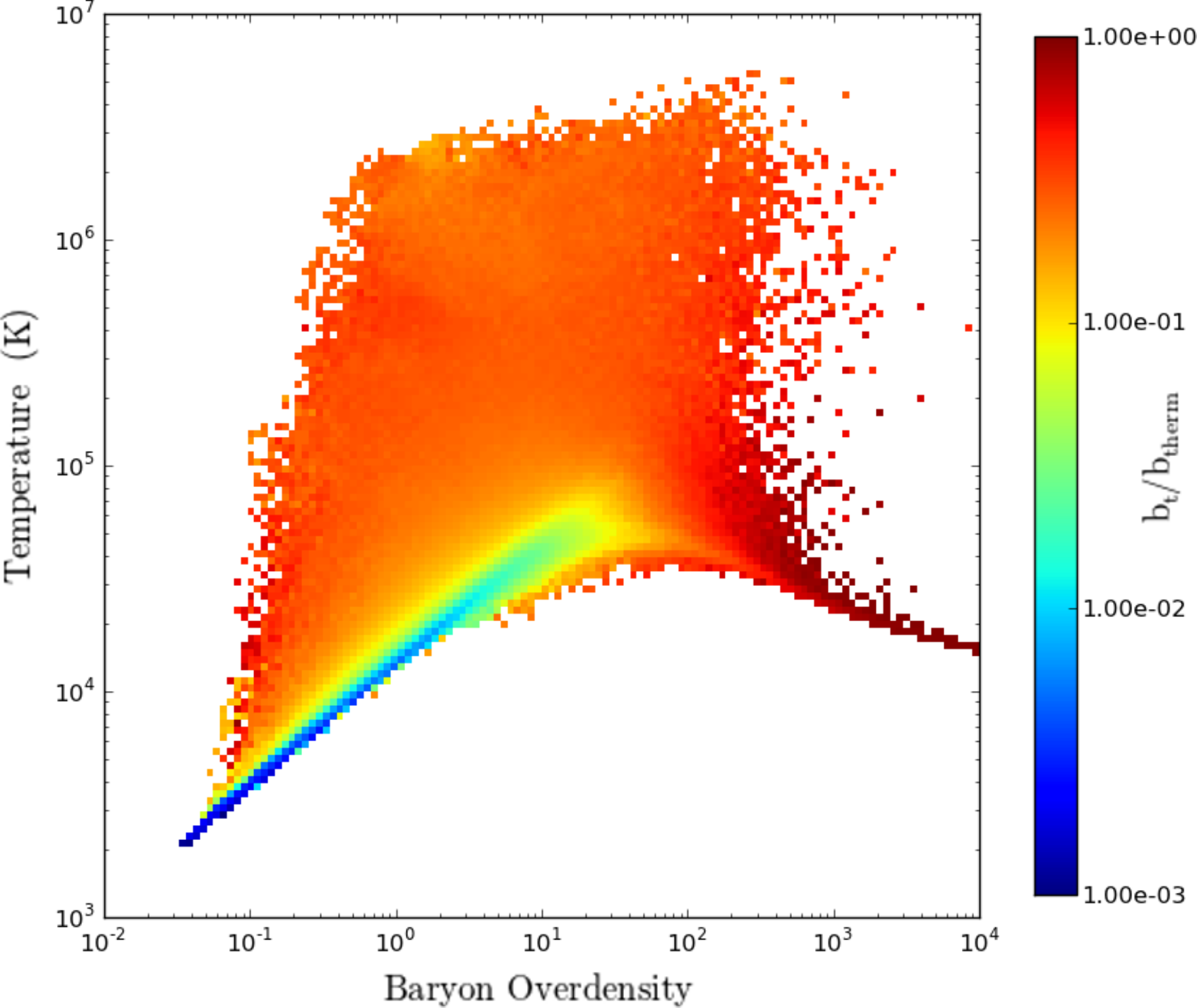}
        }
    \caption{\emph{Left plot:} Average ratios of turbulent and thermal
      pressures vs.\ baryonic overdensity in a $10\mathrm{\ Mpc}\
      h^{-1}$ box with heating and cooling at $z=2.0$. \emph{Right
        plot:} Ratio of turbulent to thermal Doppler broadening for
      bins of temperature and baryonic overdensity. 
      Image reproduced with permission from \cite{IapViel13},
      copyright by the authors.}
    \label{fig:doppler_broad}
\end{figure}

A crucial question in analysing turbulence production during cosmological structure formation concerns the computation of the turbulent velocity dispersion, denoted as $\sigma_{\mathrm{turb}}$. \citet{IapSchm11,IapViel13} adopt the SGS turbulence energy model based on \cite{MaierIap09} and define $\sigma_{\mathrm{turb}}=\sqrt{2K}$, where $K$ represents the turbulent kinetic energy. This implies that $\sigma_{\mathrm{turb}}$ corresponds to the magnitude of turbulent velocity fluctuations on the grid scale of the simulation. 
Although this variable obviously depends on numerical resolution, it is nevertheless useful to infer the dependence on various factors that influence the production of turbulence. 
For example, Fig.~\ref{fig:doppler_broad} (left plot) shows the ratio of the
turbulent and thermal pressures, where $P_{\mathrm{t}}=P_{\mathrm{sgs}}=\frac{2}{3}\rho K$, and the
Doppler broadening parameter $b_{\mathrm{t}}=\sigma_{\mathrm{turb}}/\sqrt{3}=\sqrt{2K/3}$ \citep{EvoFerr11}
for a cosmological simulation with radiative background and cooling in a box of $10\mathrm{\ Mpc}\ h^{-1}$
co-moving size. The analysis is carried out for data cubes at redshift $z=2.0$. 
For the intergalactic medium (IGM), typically characterized by moderate baryonic overdensities $\rho/\rho_0$ ranging from 1 to approximately 50, $b_{\mathrm{t}}$ is found to increase with density 
from roughly $1$ to $10\mathrm{\ km/s}$. The warm-hot intergalactic medium (WHIM) 
is associated with gas that is heated by accretion shocks to temperatures between $10^5$ and $10^7\mathrm{\ K}$, 
resulting in a flat turbulent velocity dispersion, characterized by $P_{\mathrm{t}}/P\sim 0.1$. The phase diagram in 
Fig.~\ref{fig:doppler_broad} shows that the ratio $b_{\mathrm{t}}/b$  varies
over several orders of magnitude for different densities and temperatures and reaches
peak values $\sim 1$. Although the distinction between gas in the IGM ($1\le \rho/\rho_0\le 50$) and the WHIM ($10^5\mathrm{\ K} \le T\le 10^7\mathrm{\ K}$) is not mutually exclusive and somewhat arbitrary, the phase diagram suggests that the WHIM tends to be more turbulent then the diffuse gas in the IGM. 

A caveat of the SGS model in \cite{MaierIap09} is the assumption of constant closure coefficients. Strictly speaking, constant-coefficient models are only applicable to statistically homogeneous and stationary turbulence. Turbulence produced by cosmological structure formation and feedback processes, however, is highly inhomogeneous and non-stationary. 
For this type of turbulent flow, dynamic procedures can be applied to adjust SGS model coefficients to local
flow conditions (see Sect.~\ref{sec:dyn_proc}). A completely different idea was motivated by the problem of
calculating the turbulent stresses in LES of wall-bounded turbulence. A crucial requirement for
a consistent SGS model is that the turbulent stresses should vanish in non-turbulent regions of a flow. This
is not generally the case for simple eddy-viscosity closures. Instead of adjusting the eddy-viscosity coefficient $C_\nu$, 
\citet{LevTosch07} decompose the numerically resolved velocity field into a mean flow $[\vecu]$ and turbulent fluctuations $\vect{u}'$. 
Theoretical arguments suggest that the turbulence energy flux for the Smagorinsky model should linearly depend on the shear associated with the fluctuating component, i.e.,
\[
	\Sigma=(C_{\mathrm{s}}\Delta|S|)^2\left(|S|-|[S]|\right).
\]
where $|[S]|$ is the rate of strain of the mean flow. If the flow is laminar, $[\vecu] \simeq \vect{u}$ and turbulence production vanishes because $|S|-|[S]|\simeq 0$. For developed isotropic turbulence, on the other hand, $u'\gg [u]$. In this case, the standard Smagorinsky model with $\Sigma\propto|S|^3$ is applicable.
For intermediate cases, the model corrects the turbulence energy flux $\Sigma$ by taking into
account interactions of the grid-scale fluctuations with the mean shear. This is why it is called the shear-improved model.

As proposed in \cite{SchmAlm13}, this idea can be carried over to the SGS turbulence energy model (see Sect.~\ref{sec:K_eq})
by defining the shear-improved eddy-viscosity closure as
\begin{equation}
	\label{eq:tau_si}
	\tau_{ij} = 2\rho\left(\nu_{\mathrm{sgs}}S_{ij}^{\,\prime\ast}-\frac{1}{3}K\delta_{ij}\right)\,,
\end{equation}
where $S_{ij}^{\,\prime} = S_{ij}-[S_{ij}]$ is the rate-of-strain tensor associated with
the fluctuating component of the flow. The SGS turbulence energy flux is then given by
\begin{equation}
	\label{eq:flux_si}
	\Sigma = \tau_{ij}S_{ij} =
	 C_{\nu}\rho\Delta K^{1/2}\left(|S^{\ast}|^2-2[S_{ij}^{\ast}]S_{ij}\right)-\frac{2}{3}\rho K d\,.
\end{equation}
Apart from the divergence term ($d=S_{ii}$), the expression on the right-hand side is motivated by the generalized K\'{a}rm\'{a}n--Howarth equation in \cite{LevTosch07}.

A challenge in implementing the shear-improved model is the computation of the mean flow $[\vecu]$, which is an ensemble average
over the flow velocity $\vect{u}$. A practical solution is to find an approximation to the mean flow by smoothing $\vect{u}$
with a temporal low-pass filter. For statistically stationary turbulence, it is possible to use an exponentially weighted
recursive average. In component notation, an estimate 
for the mean velocity at time $t_{n+1}$ is calculated as weighted sum of the
estimate at the previous time step and the local velocity $u_i^{(n+1)}$ for each grid cell
(cell indices are omitted):
\begin{equation}
	\label{eq:filter_recursive}
	[u_i]^{(n+1)} = \left(1-\alpha_i^{(n+1)}\right)[u_i]^{(n)} + \alpha_i^{(n+1)}u_i^{(n+1)},	
\end{equation}
The weighting coefficients are defined by 
\[
	\alpha_i^{(n+1)} = \frac{2\pi(t_{n+1}-t_n)}{\sqrt{3}\,T_{\mathrm{c}}}\,.
\]
where $T_{\mathrm{c}}$ is the constant integral time scale of the flow \citep{CahuBou10}.
Changes occurring on time scales smaller than the smoothing scale $T_{\mathrm{c}}$ are suppressed in $[u_i]$. Compared to dynamic procedures, this algorithm is very easy to implement and computationally much cheaper.  
The shear-improved Smagorinsky model with exponential smoothing performs well in LES of plane-channel flow and reproduces data from direct numerical simulations \citep{LevTosch07}.
\citet{Semenov16} apply an exponential filter with $T_{\mathrm{c}}=10\;\mathrm{Myr}$ to filter out differential rotation in galaxy simulations.

\begin{figure}[htbp]
    \centerline{\includegraphics[width=\textwidth]{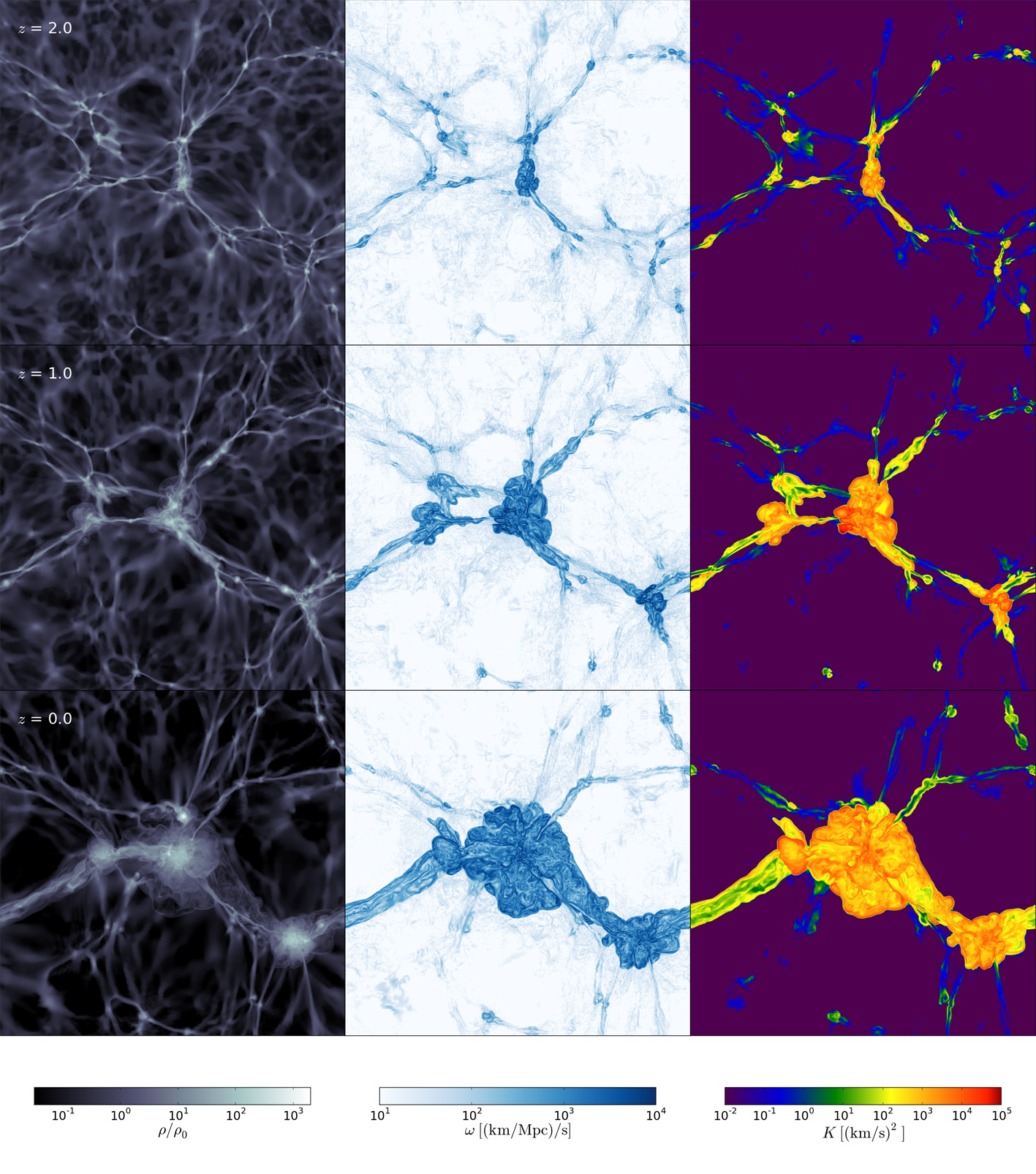}}
    \caption{Slices of the baryonic mass density (\emph{left}), vorticity
      modulus (\emph{middle}), and specific SGS turbulence energy (\emph{right}) at
      different redshifts for a simulation of the Santa Barbara
      cluster \citep{HeitRick05,SchmAlm13}. The box size is $64\mathrm{\ Mpc}\ h^{-1}$.
      Image reproduced with permission from \cite{SchmAlm13},
      copyright by the authors.}
    \label{fig:sb_slices}
\end{figure}

However, the simple exponential smoothing algorithm produces an estimate that lags behind the ensemble 
average if the mean flow evolves. To address this issue, \citet{CahuBou10} introduced the Kalman filtering technique.
A Kalman filter adapts itself to an unsteady mean flow by dynamically 
adjusting the weights of the recursive relation~(\ref{eq:filter_recursive}). A predictor-corrector scheme
is used to update the mean flow depending on the variances of the mean flow evolution, $[u_i]^{(n)}-[u_i]^{(n-1)}$,
and the detected deviation from the mean, $u_i^{(n)}-[u_i]^{(n)}$. The adaptive Kalman filter has an 
additional control parameter $u_{\mathrm{c}}$. In a statistically stationary state, 
it becomes effectively an exponential filter and the resulting velocity fluctuations 
should be of the order $u_{\mathrm{c}}$. The parameters $T_{\mathrm{c}}$ and $u_{\mathrm{c}}$ have to be chosen such that $u_{\mathrm{c}}$ is roughly the integral velocity of turbulence if the flow enters a steady state and
$T_{\mathrm{c}}$ is the characteristic time scale over which the flow evolves. 
\cite{CahuBou11} showed that LES of turbulence produced by the flow past a cylinder agree well with experimental data if Kalman filtering is applied with $T_{\mathrm{c}}$ and $u_{\mathrm{c}}$ being set to the inverse of the expected vortex-shedding frequency and upstream velocity, respectively. In general, an appropriate prior choice of the filter parameters might not be obvious. Nevertheless, they can be calibrated by performing low-resolution test runs. 
This was demonstrated for cosmological AMR simulations of the so-called Santa Barbara cluster by \citet{SchmAlm13} using the SGS turbulence energy model in combination with AMR (see Sect.~\ref{sec:consrv})
Figure~\ref{fig:sb_slices} illustrates the growth of the cluster in slices through its center at different redshifts.
The left column shows the density of the baryonic gas. As an indicator of turbulence, the vorticity of the velocity field is shown in the middle column. One can see that turbulence is produced inside the cluster and in thick filaments. The processes driving turbulence are the accretion of gas and mergers of smaller clusters to more massive ones. The SGS turbulence energy shown in the right column is clearly correlated with the turbulent flow. There is a sharp drop at the outer accretion shocks, where gas that falls at high speed into the gravitational well is heated and compressed. Owing to the shear-improved model, virtually no spurious SGS turbulence energy is found in the void (i.e.\ in low-density gas outside of the cluster).

\begin{figure}[tbp]
  	\centerline{\includegraphics[width=\textwidth]{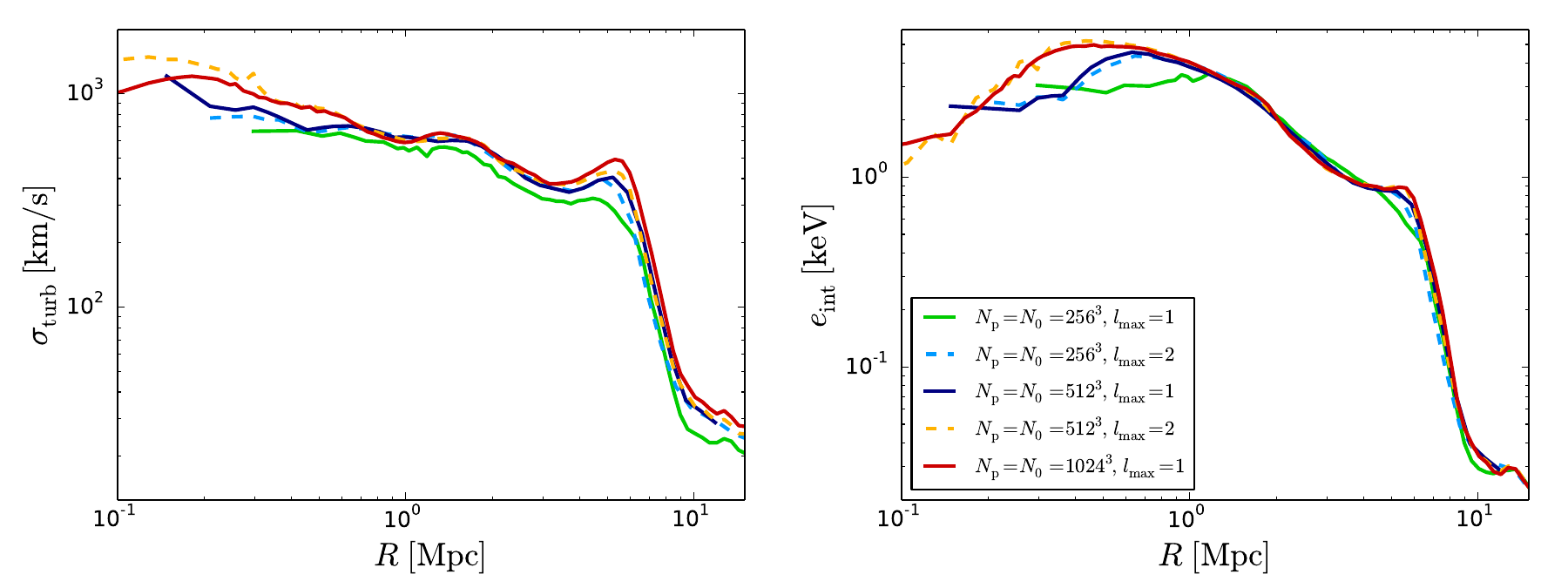}}
    \caption{Radial profiles of the turbulent velocity dispersion defined by
      Eq.~(\ref{eq:sigma_turb}) (\emph{left plot}) and the specific internal energy (\emph{right plot}) for 
      a selected dark-matter halo at different numerical resolutions. 
      Starting from a uniform root-grid with $N_0$ grid cells per spatial dimension, refinement by
      overdensity and vorticity modulus up to $l_{\max}$ levels is applied.
      Image reproduced with permission from \cite{schmidt_hot_2016},
      copyright by OUP.
	}
    \label{fig:sigma_profiles}
\end{figure}

\begin{figure}[tbp]
  	\centerline{
          \includegraphics[width=0.5\textwidth]{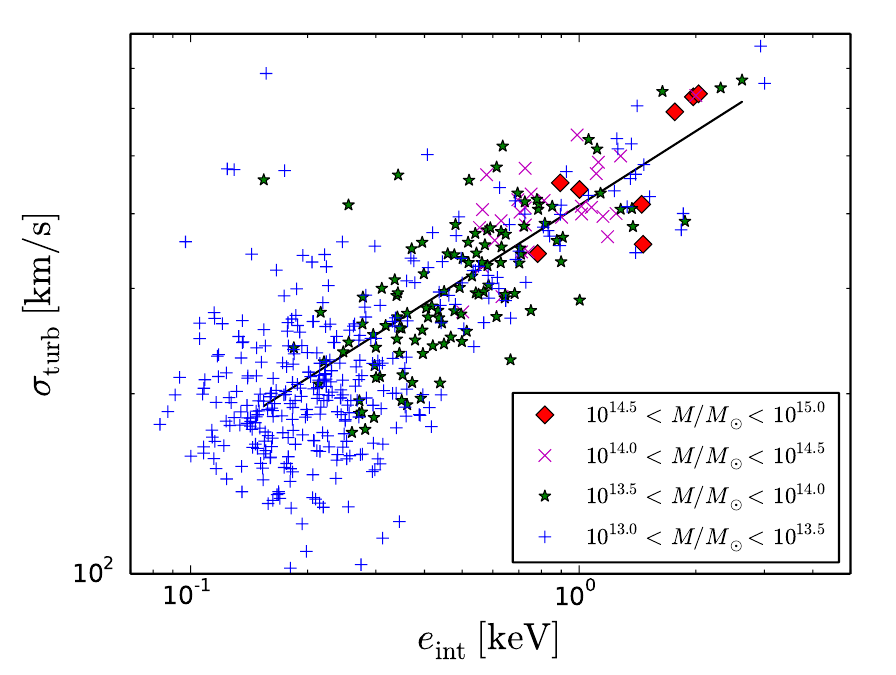}\quad
          \includegraphics[width=0.5\textwidth]{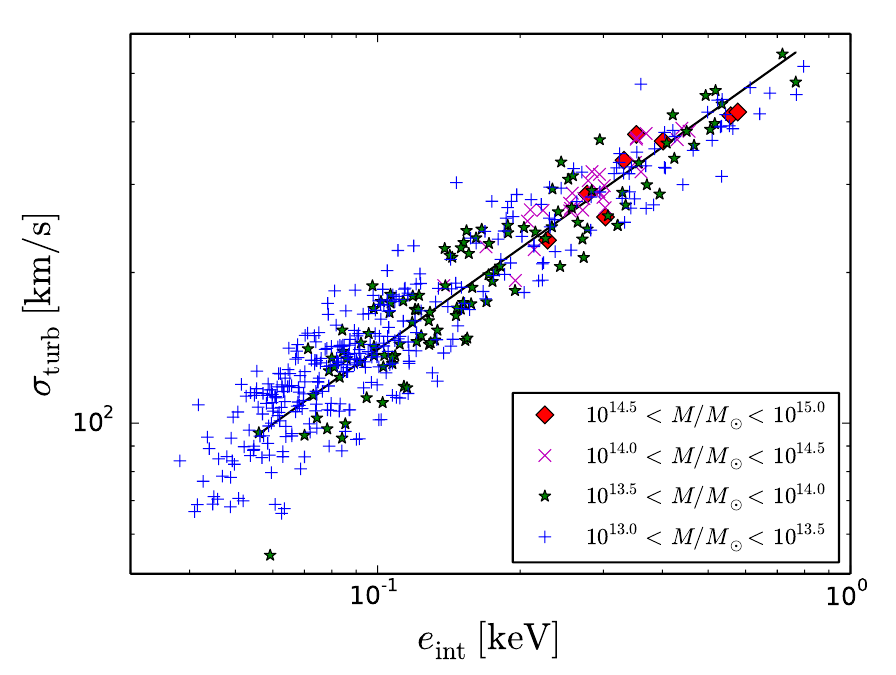}
        }
    \caption{\emph{Left plot:} Volume-weighted turbulent velocity dispersion vs.\
    		mean thermal energy in the ICM ($\rho/\rho_0 > 500$, $T>10^5\,$K). 
			\emph{Right plot:} Volume-weighted turbulent velocity dispersion vs.\
    		mean thermal energy in the WHIM ($\rho/\rho_0 < 500$, $T>10^5\,$K).
    		Both plots show haloes at redshift $z=0.0$. 
    		The different symbols correspond to the halo mass intervals specified in the legend. The
    		thick black lines show power-law fits to the data points with $M>10^{13.5}\,M_\odot$
      	Image reproduced with permission from \cite{schmidt_hot_2016},
      	copyright by OUP.
	}
    \label{fig:sigma_mean}
\end{figure}

In contrast to dynamic procedures, the shear-improved model has the merit of separating the turbulent flow in clusters from the gravity-driven bulk flow. This is utilized by \citet{SchmAlm13} to define the turbulent velocity dispersion as
\begin{equation}
  \label{eq:sigma_turb}
   \sigma_{\mathrm{turb}} = \sqrt{U^{\prime\,2} + 2K}\,,
\end{equation}
where $\vect{U}^\prime$ is the fluctuating component of the proper peculiar velocity~(\ref{eq:prop_peculiar}) computed with the Kalman filter. Thus, both numerically resolved and SGS turbulence contribute to $\sigma_{\mathrm{turb}}$.
The smoothed component of the proper peculiar velocity field, $[\vecU]$, is interpreted as non-turbulent bulk flow. 
Figure~\ref{fig:sigma_profiles} shows radial profiles of $\sigma_{\mathrm{turb}}$ for a dark-matter halo of mass $M=6.18\times 10^{14}\,M_\odot$ in AMR simulations based on a realistic cosmological model ($\Lambda$CDM cosmology with non-adiabatic physics; see \citealt{schmidt_hot_2016}). As indicated in the legend, the effective resolution, which corresponds to the maximum refinement level, varies between $512^3$ and $2048^3$. Except for the lowest-resolution case, the radial profiles of $\sigma_{\mathrm{turb}}$ show little sensitivity to numerical resolution. This is an important property of the turbulent velocity dispersion defined by Eq.~(\ref{eq:sigma_turb}). While the profiles are nearly flat for the ICM with a typical velocity dispersion close to $1000\,\mathrm{km/s}$, there is a distinctive drop in the cluster outskirts between $5$ and $10\;$Mpc. The whole box has a size of about $150\;$Mpc (at redshift zero). By averaging $\sigma_{\mathrm{turb}}$ and the internal energy of the gas for each halo in the box, correlations between the non-thermal and thermal energy reservoirs of clusters can be investigated. As shown in Fig.~\ref{fig:sigma_mean}, there are power-law relations: Although the scatter is relatively high for the high-density ICM, there is a correlation for haloes of mass $\sim 10^{14}\,M_\odot$ or higher. The correlation becomes even tighter for the WHIM, which is typically located in the outer regions of clusters. This suggests that the heating of gas in clusters and the production of turbulence are related processes \citep[see also][]{Vazza11,Miniati15,Wang24}.

However, galactic outflows driven by supernova explosion and active galactic nuclei (AGN) also contribute to the energy budget of clusters. Since feedback processes cannot be resolved in large-scale cosmological simulations, detailed models of these processes have been implemented in various codes \citep[][to mention just a few examples]{Hopkins14,weinberger16}. The properties of galactic outflows and their impact on the gaseous medium outside of galaxies were investigated, for instance, by \citet{nelson19}. 
For galactic haloes in filaments, \citet{schmidt_turbulence_2021-2} found that the relation between turbulent and thermal energy is roughly linear, but the additional energy injection by stellar feedback results in larger variations compared to clusters. 

\subsection{Turbulent mixing of metals}
\label{sec:mixing_metals}

\citet{Su17} investigated the impact of anisotropic conductivity and viscosity in isolated disk galaxy simulations and in zoom-ins from cosmological simulations. Since they use a mesh-free Lagrangian code without numerical diffusion, they add a diffusion term with eddy-diffusivity $\kappa_{\mathrm{sgs}}=C\Delta^2|S|$ for the turbulent mixing of metals (see Sect.~\ref{sec:mixing}). Although this closure is based on the Smagorinsky model, their simulations are not LES because gas dynamics is modelled by the Navier-Stokes equations with microscopic viscosity. If the flow is assumed to be numerically resolved, metals should be mixed entirely by advection on resolved scales. Consequently, their SGS mixing prescription has negligible effects.

A similar conclusion regarding the mixing of metals in cosmological AMR simulations was drawn by \citet{engels_modelling_2019}. In their simulations, the SGS turbulence energy model outlined in Sect.~\ref{sec:clusters} was complemented by a diffusivity $\kappa_{\mathrm{sgs}}=C\Delta\sqrt{K}$ for chemical species and thermal energy. They found that the circumgalactic medium (CGM) of galactic haloes is not significantly enriched by SGS mixing if the full model is applied. Interestingly, the strongest effect was seen if all diffusion terms in the SGS model were deactivated. In that case, stronger outflows transport a larger amount of metals into the CGM, which is in turn caused by enhanced stellar feedback. Additional diffusion, on the other hand, appears to reduce stellar feedback. This suggests that different effects in combination do not necessarily result in enhanced downgradient transport.

\begin{figure}[htbp]
  	\centerline{\includegraphics[width=0.9\textwidth]{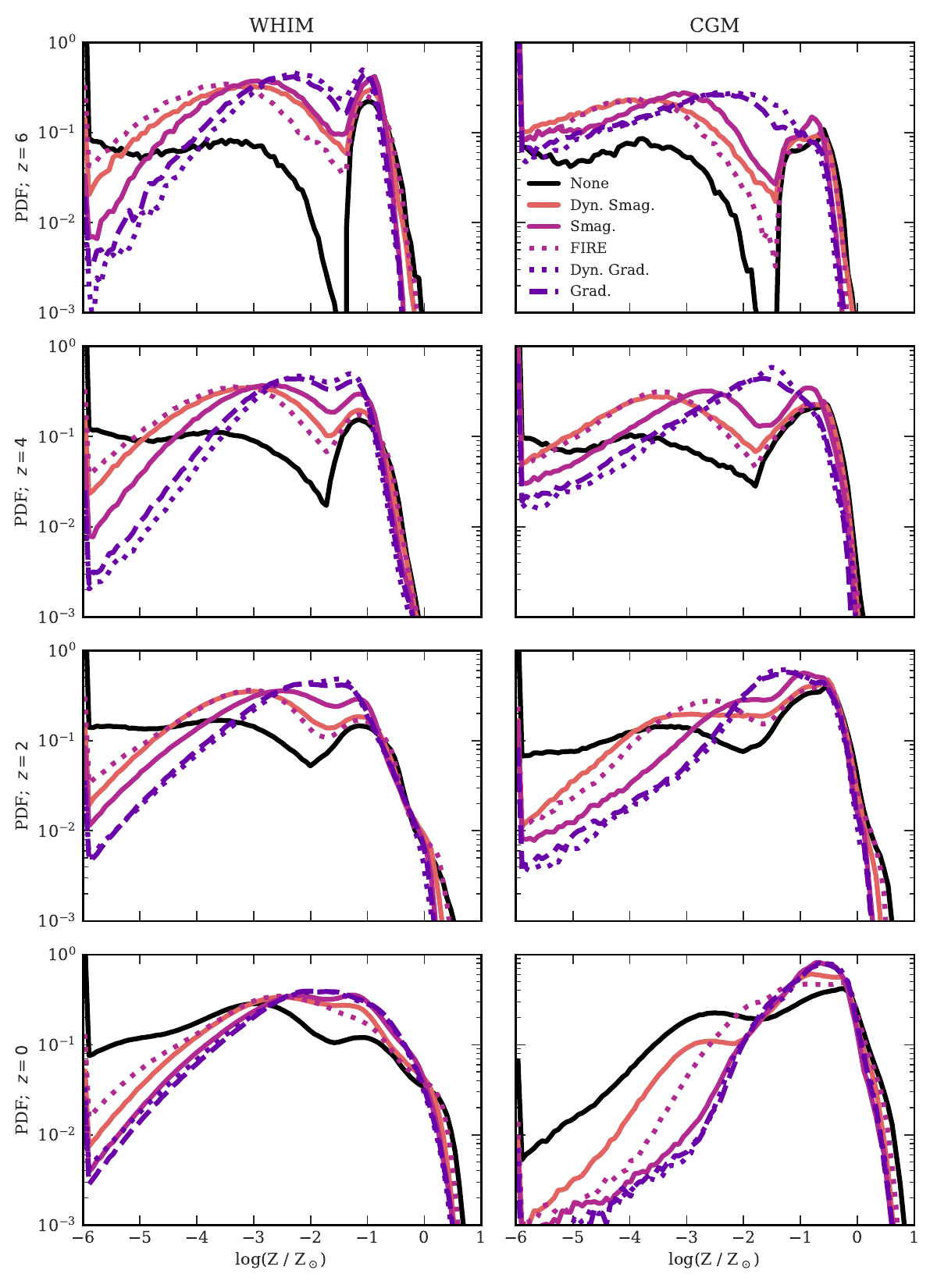}}
    \caption{Probability density functions of the metallicity $Z/Z_\odot$ in the WHIM (\emph{left}) and 
        CGM (\emph{right}) at different redshifts (from top to bottom).
      	Image reproduced with permission from \cite{rennehan_mixing_2021-1},
      	copyright by OUP.
	}
    \label{fig:mdf}
\end{figure}

\citet{rennehan_dynamic_2019-1} improved the Smagorinsky-like gradient model by introducing a dynamic procedure for the determination of the eddy-diffusivity coefficient $C$  (see Sect.~\ref{sec:dyn_proc}). This approach helps to avoid spurious mixing, for example, due to shear in a galactic disk. By analysing metal distribution functions (i.e.\ probability density functions of the metallicity $Z$) in mesh-free cosmological simulations, indications of enrichment due to turbulent mixing in metal-poor regions of the CGM and in hot halo gas were found. A model with a full treatment of SGS turbulence stresses and a variant of the anisotropic diffusivity defined by Eq.~(\ref{eq:diff_anisotropic}) is employed in \citet{rennehan_mixing_2021-1}. Figure~\ref{fig:mdf} shows metal distribution functions for distinct phases. At high redshift (top plots), there are pronounced peaks for $Z\sim 0.1 Z_\odot$ ($Z_\odot$ is the solar metallicity), which are produced by enriched gas that is expelled from star-forming galaxies into the CGM and beyond into the WHIM. The peak in the WHIM broadens and is shifted to lower $Z$ as high- and low-metallicity gas gradually mix over time. In the CGM, the peak grows due to the supply from ongoing star formation. However, in both phases the details are quite sensitive to the applied mixing model. The tightest distributions with well-defined maxima are obtained with the anisotropic diffusion model (labelled "Grad." for gradient model). For the interpretation of these results, it is important to keep in mind that Lagrangian codes have no intrinsic diffusion and, consequently, particles cannot exchange metals (labelled "None" in Fig.~\ref{fig:mdf}).

The influence of turbulent mixing on galactic outflows is further explored in \citet{steinwandel_pumping_2024}. For computational efficiency, they use a simplified Smagorinsky-like diffusivity. Their findings show that the model reduces the separation between hot, metal-rich gas and the warm ambient medium in phase space. The resulting phase space structure with a more gradual transition is in better agreement with grid-base codes with numerical diffusion. Without an SGS model, the hot phase is more enriched and tends to cool more rapidly. 

\subsection{Magnetic field amplification in collapsing or merging objects}
\label{sec:collapse_merge}

Capturing the small-scale dynamo that drives the amplification of the magnetic field in turbulent plasmas poses a significant challenge in numerical simulations. Turbulent eddies stretch, twist and
fold the magnetic field until an approximate equipartition between magnetic and kinetic energy
is reached and the dynamo saturates \citep[see, for example,][]{BrandSub05,Shukurov_Subramanian_2021}. 
As discussed by \citet{schober_magnetic_2013}, the largest growth rate of the magnetic field
is initially produced by the smallest scale on which dissipation effects can be neglected.
The production of magnetic energy at the expense of turbulent kinetic energy is then gradually
shifted toward larger scales until saturation is reached. The required numerical resolution to fully 
resolve this process is usually not affordable. Consequently, the numerically computed
turbulent dynamo is limited by the smallest resolved scale. This immediately raises the question
whether a SGS model could be applied to model the sub-resolution dynamo and, thus, obtain a
numerically converged magnetic field amplification.

A first step in this direction was the application of the non-linear structural MHD SGS model (see Sect.~\ref{sec:struc}) in
cosmological simulations of magnetized atomic cooling haloes by \citet{grete_intermittent_2019}. The cooling and 
collapse of gas in these haloes is a possible formation mechanism of supermassive black hole seeds 
\citep{latif_black_2013}. \citet{latif_small-scale_2013} and \citet{grete_intermittent_2019} 
found evidence of small-scale dynamo action, resulting in dynamically relevant magnetic field strengths. 
However, extremely high resolution is necessary to reach the regime where
turbulent amplification beyond simple compression of frozen-in magnetic fields in the collapsing
gas sets in. Even with the SGS model, MHD turbulence is not sufficiently resolved. By comparing
radial profiles of quantities such as the magnetic field strength, turbulent Mach numbers, and pressure 
between ILES (without explicit SGS model) and LES, it became clear that statistical variations among different halo realizations
with similar properties dominate any changes caused by the structural SGS model.

\begin{figure}[tbp]
        \centering
        \includegraphics[width=1.0\textwidth]{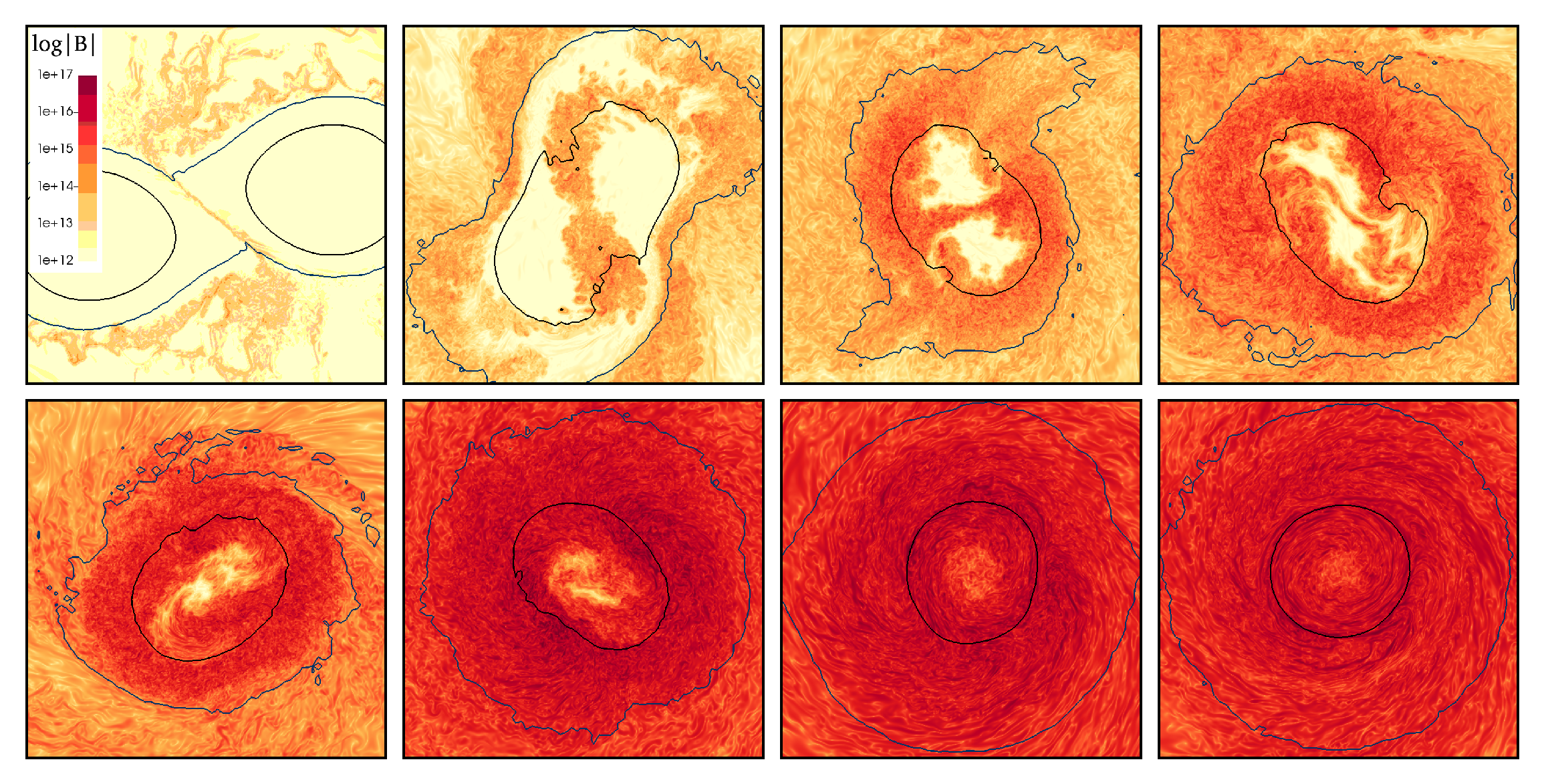}
        \caption{Magnetic field strength in Gauss 0.5, 1.5, 2.0, 2.5, 3.5, 5.0, 10, and 15 ms (from upper left to lower right) after the merger of binary neutron star. The white lines show contours of constant density, roughly corresponding to the bulk-envelope boundary (outer lines) and dense core regions (inner lines). Image reproduced with permission from \cite{palenzuela_turbulent_2022}, copyright by APS.
        }
        \label{fig:bns_merger_slices}
\end{figure}

\begin{figure}[htbp]
  	\centerline{\includegraphics[width=0.7\textwidth]{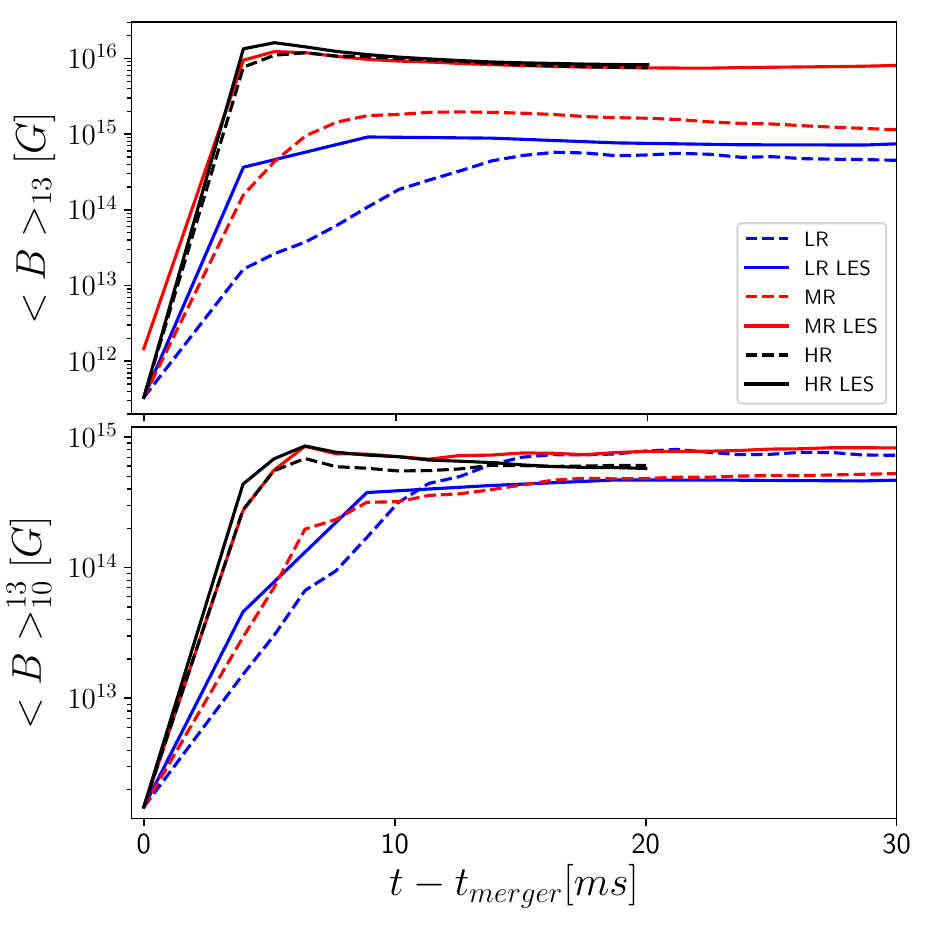}}
    \caption{Volume-averaged magnetic field strength in the bulk (\emph{top plot}) and in the envelope (\emph{bottom plot}) of the remnant produced by the merger of a binary neutron star for spatial resolutions of $120$, $60$, and $30\;$m. Solid and dashed lines show the results for LES with explicit SGS model and standard simulations, respectively.
      Image reproduced with permission from \cite{palenzuela_turbulent_2022},
      copyright by APS.
	}
    \label{fig:bns_merger_energy}
\end{figure}

Another example is binary neutron star mergers, where the small-scale dynamo is triggered
by Kelvin-Helmholtz instabilities. In this case, the equations of general relativistic MHD (GRMHD)
have to be solved. By combining a high-order numerical scheme, high resolution, and
a relativistic generalization of the structural SGS model introduced in Sect.~\ref{sec:struc},
\citet{palenzuela_turbulent_2022} successfully simulated the first 50 milliseconds after the merger. As illustrated in Fig.~\ref{fig:bns_merger_slices}, Kelvin-Helmholtz instabilities develop in the shear layer
between the neutron star cores. Subsequently, turbulent eddies spread throughout the merger remnant, leading to a significant amplification of the magnetic field by several orders of magnitude. Figure~\ref{fig:bns_merger_energy} demonstrates the typical challenge associated with the turbulent amplification of the field in numerical simulations: the growth rate increases with resolution due to the dominance of the smallest resolved scale, which is the grid scale. Even the saturated field at later time, which can be seen as plateau, is not fully converged.
Based on the numerical studies of \citet{aguilera-miret_turbulent_2020} and \citet{palenzuela_large_2022}, 
SGS terms were incorporated into the GRMHD equations to perform LES of binary neutron star mergers. 
However, unlike the model coefficients derived from \emph{a priori} tests \citep{GreteVlay16,vigano_general_2020},
coefficients associated with magnetic field amplification were boosted, while reducing or neglecting
diffusive SGS fluxes that compete with the significant numerical viscosity of the solver.

For the first time, they obtained a converged magnetic field amplification in merging neutron stars. 
The results in Fig.~\ref{fig:bns_merger_energy} show that LES reaches the saturated field strength
in a simulation with the maximal feasible resolution already at two times lower resolution, which 
implies a significant reduction of computational cost. 
The improved convergence of the magnetic field also helps to analyse kinetic and magnetic energy spectra.
While the kinetic energy follows the expected $k^{-5/3}$ power law on scales smaller than the 
energy-containing range, a $k^{3/2}$ Kazantsev spectrum is found on scales larger than 
the magnetic coherence scale for the magnetic energy spectrum \citep{palenzuela_turbulent_2022}. 
Below the coherence scale, magnetic energy almost reaches equipartition with kinetic energy at late times. 
Moreover, \citet{aguilera-miret_universality_2022} demonstrated that the average field strength 
and the spectral distribution of magnetic energy after the merger is insensitive to the 
initial configuration of the magnetic field in the two neutron stars. This is interpreted as 
universality of the turbulent magnetic field generated by the small-scale dynamo.
However, the authors emphasize that the initial field strength should not be arbitrarily high, 
which is often the \emph{ad hoc} solution to the amplification problem in numerical simulations.

\subsection{Concluding remarks}

A particular difficulty of validating LES in astrophysics is that neither DNS nor sufficiently accurate experimental data exist. For most instances of astrophysical turbulence, DNS are infeasible because of the very high Reynolds numbers (probably with the exception of turbulence in the ICM; see \citealt{parrish_viscosity_2012,roediger_helmholtz_2013,schmidt_viscosity_2017-1}). 
Observations are usually not sensitive enough to clearly discriminate between different numerical models. As a consequence, the standards for model validation in astrophysics are different compared to engineering or atmospheric sciences, where direct comparisons with measurements or DNS data are commonly made. 
Although the correct solution is unknown, we know for sure that there are couplings between numerically resolved scales and subgrid scales in simulations, which are caused by the turbulent stresses on the grid scale. The calculation of these stresses on the basis of scale separation (Sect.~\ref{sec:separation}) and testable closures (Sects.~\ref{sec:subgrid} and \ref{sec:closure}) is physically better motivated and provides a more accurate approximation than purely numerical truncation errors.

Statistical properties of homogeneous turbulence such as two-point statistics, turbulence energy spectra, or the distribution of mass density fluctuations are not very sensitive to the application of an explicit SGS model, provided that the numerical resolution is high enough \citep{grete_comparative_2017,Su17,rennehan_dynamic_2019-1}. This is in line with the scale locality of 
energy transfer, which implies that properties related to inertial-range dynamics on some scale must be largely independent of
effects on much smaller scales. From this point of view, numerical dissipation in ILES is sufficient to model turbulent
astrophysical flows. However, the additional dissipative effect of a SGS model can improve inertial-range scaling, higher-order statistics, and mixing of metals between different gas phases, particularly if higher-order schemes with particularly low dissipation or mesh-free methods are applied \citep{grete_comparative_2017,carrasco_gradient_2020,vigano_general_2020,rennehan_mixing_2021-1,steinwandel_pumping_2024} 
Moreover, SGS turbulence energy acts as a buffer between numerically resolved kinetic and internal energy, which helps to reduce artificial manipulations of the energy balance between grid levels in AMR simulations \citep{SchmAlm13}. This concept has recently 
been extended by \citet{semenov_capturing_2024}. By assuming that the rate of SGS turbulence energy production is implicitly given by the
dissipation of kinetic energy on numerically resolved scales, the local difference between the total and kinetic energies after each cell update can be put into the SGS turbulence energy reservoir, which in turn dissipates into heat as in the SGS turbulence energy equation model. This hybrid approach, which is called semi-implicit LES (SLES), effectively reproduces properties of decaying supersonic turbulence at lower resolution.

MHD turbulence remains a challenging problem. In contrast to hydrodynamical turbulence, magnetic field fluctuations are anisotropic and magnetic energy can build up from small scales if a small-scale dynamo is active. As a consequence, the basic assumption of scale locality can only be applied to a limited extent or even becomes invalid. For magnetic field evolution in galaxies, \citet{liu_subgrid_2022} proposed a SGS model based on the well known $\alpha$ dynamo effect \citep{brandenburg_galactic_2023}. They use an \emph{ad hoc} closure that is only justified by its positive impact on reaching expected magnetic field strengths for Milky Way-like star-forming galaxies. But \citet{GreteVlay16} showed that the correlation between the $\alpha$ term and the electromotive force is poor for driven MHD turbulence. In simulations of objects undergoing a gravitational collapse, such as atomic cooling haloes in the early universe, the magnetic field is amplified by a small-scale dynamo on top of compression. \cite{latif_small-scale_2013} demonstrated that the amplification depends strongly on numerical resolution. Since the magnetic field supports the gas against gravity, this in turn impacts fragmentation. Results from hydrodynamical LES suggest that the additional turbulent viscosity influences angular momentum transport in accretion disks, favouring the formation of higher-mass objects \citep{latif_black_2013}. However, studies of magnetized atomic cooling haloes with the non-linear structural MHD SGS model by \citet{grete_intermittent_2019} do not show any clear trend compared to ILES. The competition with numerical diffusivity is clearly a problem. Apart from that, the statistical scatter of similar haloes evolving from random initial condition outshines any effects caused by the SGS model. However, it is likely that an SGS model for MHD turbulence will pay off in combination with higher-order schemes. But strongly diffusive schemes are still favoured in astrophysics because of their robustness against numerical instabilities. An important step in this direction is the successful application to binary neutron star mergers by \citet{palenzuela_turbulent_2022} and \citet{aguilera-miret_role_2023}. With some tuning of the model parameters, they achieved converged magnetic field amplification just below the highest feasible resolution. 

From the current perspective, the main utility of SGS models in astrophysics is the treatment of complex sub-resolution physics.
The calculation of the turbulent burning speed in type Ia supernova simulations is an early example \citep{SchmNie06a,RoepHille07}. By now, it has become clear that turbulent deflagrations occur only in a subclass of this type of supernova, but the the lessons learned in the modelling of SGS turbulence turned out to be useful in other applications as well. 
For example, feedback from supernovae in galaxy simulations can be implemented as a source term in the SGS turbulence energy equation in addition to heating \citet{Braun14}. This offers an alternative to the often employed kinetic feedback, which produces resolved gas motions. In addition, turbulence-regulated star formation efficiencies can be computed on the basis of a SGS model \citep{Braun15,Semenov16}. This approach improves the modelling of star formation in highly dynamical evolutionary stages, when galaxies are formed or during interactions \citep{semenov_how_2024,semenov_uv-bright_2024}.



\phantomsection
\addcontentsline{toc}{section}{References}
\bibliography{refs}

\begin{thebibliography}{158}
\providecommand{\natexlab}[1]{#1}
\providecommand{\url}[1]{{#1}}
\providecommand{\urlprefix}{URL }
\providecommand{\doi}[1]{\url{https://doi.org/#1}}
\providecommand{\eprint}[2][]{\url{#2}}
 \bibcommenthead

\bibitem[{Aguilera-Miret et~al(2020)Aguilera-Miret, Vigan{\`o}, Carrasco,
  Mi{\~n}ano, and Palenzuela}]{aguilera-miret_turbulent_2020}
Aguilera-Miret R, Vigan{\`o} D, Carrasco F, et~al (2020) Turbulent
  magnetic-field amplification in the first 10 milliseconds after a binary
  neutron star merger: {Comparing} high-resolution and large-eddy simulations.
  Phys Rev D 102(10):103006. \doi{10.1103/PhysRevD.102.103006}

\bibitem[{{Aguilera-Miret} et~al(2022){Aguilera-Miret}, {Vigan{\`o}}, and
  {Palenzuela}}]{aguilera-miret_universality_2022}
{Aguilera-Miret} R, {Vigan{\`o}} D, {Palenzuela} C (2022) {Universality of the
  turbulent magnetic field in hypermassive neutron stars produced by binary
  mergers}. \apjl 926(2):L31. \doi{10.3847/2041-8213/ac50a7}

\bibitem[{Aguilera-Miret et~al(2023)Aguilera-Miret, Palenzuela, Carrasco, and
  Vigan{\`o}}]{aguilera-miret_role_2023}
Aguilera-Miret R, Palenzuela C, Carrasco F, et~al (2023) Role of turbulence and
  winding in the development of large-scale, strong magnetic fields in
  long-lived remnants of binary neutron star mergers. Phys Rev D
  108(10):103001. \doi{10.1103/PhysRevD.108.103001}

\bibitem[{Almgren et~al(2013)Almgren, Bell, Lijewski, Luki{\'c}, and
  Van~Andel}]{AlmBell13}
Almgren AS, Bell JB, Lijewski MJ, et~al (2013) Nyx: A massively parallel {AMR}
  code for computational cosmology. \apj 765:39.
  \doi{10.1088/0004-637X/765/1/39}

\bibitem[{Aluie(2011)}]{Aluie11}
Aluie H (2011) Compressible turbulence: The cascade and its locality. Phys Rev
  Lett 106:174502. \doi{10.1103/PhysRevLett.106.174502}

\bibitem[{Aluie(2013)}]{Aluie13}
Aluie H (2013) Scale decomposition in compressible turbulence. Physica D
  247:54--65. \doi{10.1016/j.physd.2012.12.009}

\bibitem[{Balarac et~al(2013)Balarac, Le~Sommer, Meunier, and
  Vollant}]{balarac_dynamic_2013}
Balarac G, Le~Sommer J, Meunier X, et~al (2013) A dynamic regularized gradient
  model of the subgrid-scale scalar flux for large eddy simulations. Phys
  Fluids p 075107. \doi{10.1063/1.4813812}

\bibitem[{Beattie et~al(2024)Beattie, Federrath, Klessen, Cielo, and
  Bhattacharjee}]{beattie_magnetized_2024}
Beattie JR, Federrath C, Klessen RS, et~al (2024) Magnetized compressible
  turbulence with a fluctuation dynamo and {Reynolds} numbers over a million.
  arXiv e-prints {\href{https://arxiv.org/abs/2405.16626}{{2405.16626}}}
  {[astro-ph]}

\bibitem[{Berger and Colella(1989)}]{BergCol89}
Berger MJ, Colella P (1989) Local adaptive mesh refinement for shock
  hydrodynamics. J Chem Phys 82:64--84. \doi{10.1016/0021-9991(89)90035-1}

\bibitem[{Berger and Oliger(1984)}]{BergOli84}
Berger MJ, Oliger J (1984) Adaptive mesh refinement for hyperbolic partial
  differential equations. J Chem Phys 53:484--512.
  \doi{10.1016/0021-9991(84)90073-1}

\bibitem[{Bertoldi and McKee(1992)}]{BertMcKee92}
Bertoldi F, McKee CF (1992) Pressure-confined clumps in magnetized molecular
  clouds. \apj 395:140--157. \doi{10.1086/171638}

\bibitem[{{Bland-Hawthorn} et~al(2024){Bland-Hawthorn}, {Tepper-Garcia},
  {Agertz}, and {Federrath}}]{bland-hawthorn_turbulent_2024}
{Bland-Hawthorn} J, {Tepper-Garcia} T, {Agertz} O, et~al (2024) Turbulent
  gas-rich disks at high redshift: {B}ars \& bulges in a radial shear flow.
  \apj 968(2):86. \doi{10.3847/1538-4357/ad4118}

\bibitem[{Borgani and Kravtsov(2011)}]{BorKravt11}
Borgani S, Kravtsov A (2011) Cosmological simulations of galaxy clusters. Adv
  Sci Lett 4:204--227. \doi{10.1166/asl.2011.1209}

\bibitem[{Boussinesq(1877)}]{Boussin77}
Boussinesq J (1877) Essai sur la th{\'e}orie des eaux courantes, M\'em.
  Pr\'esent\'es par divers savants Acad. Sci. Inst. Fr., vol~23. Imprimerie
  Nationale, Paris

\bibitem[{Brandenburg and Ntormousi(2023)}]{brandenburg_galactic_2023}
Brandenburg A, Ntormousi E (2023) Galactic dynamos. \araa 61(Volume 61,
  2023):561--606. \doi{10.1146/annurev-astro-071221-052807}

\bibitem[{Brandenburg and Subramanian(2005)}]{BrandSub05}
Brandenburg A, Subramanian K (2005) Astrophysical magnetic fields and nonlinear
  dynamo theory. Phys Rep 417:1--209. \doi{10.1016/j.physrep.2005.06.005},
  {\href{https://arxiv.org/abs/astro-ph/0405052}{{astro-ph/0405052}}}

\bibitem[{Braun and Schmidt(2012)}]{BraunSchm12}
Braun H, Schmidt W (2012) A semi-analytic model of the turbulent multi-phase
  interstellar medium. \mnras 421:1838--1860.
  \doi{10.1111/j.1365-2966.2011.19889.x}

\bibitem[{{Braun} and {Schmidt}(2015)}]{Braun15}
{Braun} H, {Schmidt} W (2015) {The small and the beautiful: how the star
  formation law affects galactic disc structure}. \mnras 454(2):1545--1555.
  \doi{10.1093/mnras/stv1856} {[astro-ph.GA]}

\bibitem[{{Braun} et~al(2014){Braun}, {Schmidt}, {Niemeyer}, and
  {Almgren}}]{Braun14}
{Braun} H, {Schmidt} W, {Niemeyer} JC, et~al (2014) {Large-eddy simulations of
  isolated disc galaxies with thermal and turbulent feedback}. \mnras
  442(4):3407--3426. \doi{10.1093/mnras/stu1119}

\bibitem[{Br{\"u}ggen(2013)}]{Brueggen13}
Br{\"u}ggen M (2013) Magnetic fields in galaxy clusters. Astron Nachr 334:543.
  \doi{10.1002/asna.201311895}

\bibitem[{Br{\"u}ggen and Vazza(2015)}]{Brueggen2015}
Br{\"u}ggen M, Vazza F (2015) Turbulence in the intracluster medium. In:
  Lazarian A, de~Gouveia Dal~Pino EM, Melioli C (eds) Magnetic Fields in
  Diffuse Media. Springer, Berlin, Heidelberg, p 599--614,
  \doi{10.1007/978-3-662-44625-6_21}

\bibitem[{B{\"u}chner(2007)}]{Buech07}
B{\"u}chner J (2007) Astrophysical reconnection and collisionless dissipation.
  Plasma Phys Control Fusion 49:325. \doi{10.1088/0741-3335/49/12B/S30}

\bibitem[{Cahuzac et~al(2010)Cahuzac, Boudet, Borgnat, and
  L{\'e}v{\^e}que}]{CahuBou10}
Cahuzac A, Boudet J, Borgnat P, et~al (2010) Smoothing algorithms for mean-flow
  extraction in large-eddy simulation of complex turbulent flows. Phys Fluids
  22:125104. \doi{10.1063/1.3490063}

\bibitem[{Cahuzac et~al(2011)Cahuzac, Boudet, Borgnat, and
  L{\'e}v{\^e}que}]{CahuBou11}
Cahuzac A, Boudet J, Borgnat P, et~al (2011) Dynamic {Kalman} filtering to
  separate low-frequency instabilities from turbulent fluctuations: Application
  to the large-eddy simulation of unsteady turbulent flows. J Phys: Conf Ser
  318:042047. \doi{10.1088/1742-6596/318/4/042047}

\bibitem[{Canuto(1994)}]{Canuto94}
Canuto VM (1994) Large eddy simulation of turbulence: A subgrid scale model
  including shear, vorticity, rotation, and buoyancy. \apj 428:729--752.
  \doi{10.1086/174281}

\bibitem[{Canuto(1997)}]{Canuto97}
Canuto VM (1997) Compressible turbulence. \apj 482:827. \doi{10.1086/304175}

\bibitem[{Carrasco et~al(2020)Carrasco, Vigan{\`o}, and
  Palenzuela}]{carrasco_gradient_2020}
Carrasco F, Vigan{\`o} D, Palenzuela C (2020) Gradient subgrid-scale model for
  relativistic {MHD} large-eddy simulations. Phys Rev D 101(6):063003.
  \doi{10.1103/PhysRevD.101.063003}

\bibitem[{Ciaraldi-Schoolmann et~al(2009)Ciaraldi-Schoolmann, Schmidt,
  Niemeyer, R{\"o}pke, and Hillebrandt}]{CirSchm08}
Ciaraldi-Schoolmann F, Schmidt W, Niemeyer JC, et~al (2009) Turbulence in a
  three-dimensional deflagration model for type {Ia} supernovae. {I}. {S}caling
  properties. \apj 696:1491--1497. \doi{10.1088/0004-637X/696/2/1491}

\bibitem[{Ciaraldi-Schoolmann et~al(2013)Ciaraldi-Schoolmann, Seitenzahl, and
  R{\"o}pke}]{CiaSeit13}
Ciaraldi-Schoolmann F, Seitenzahl IR, R{\"o}pke FK (2013) A subgrid-scale model
  for deflagration-to-detonation transitions in type {Ia} supernova explosion
  simulations. numerical implementation. \aap 559:A117.
  \doi{10.1051/0004-6361/201321480}

\bibitem[{Colella and Woodward(1984)}]{ColWood84}
Colella P, Woodward PR (1984) The piecewise parabolic method {(PPM)} for
  gas-dynamical simulations. J Chem Phys 54:174--201.
  \doi{10.1016/0021-9991(84)90143-8}

\bibitem[{Dobler et~al(2003)Dobler, Haugen, Yousef, and
  Brandenburg}]{DobHaug03}
Dobler W, Haugen NE, Yousef TA, et~al (2003) Bottleneck effect in
  three-dimensional turbulence simulations. Phys Rev E 68:26304.
  \doi{10.1103/PhysRevE.68.026304}

\bibitem[{Dolag et~al(2008)Dolag, Bykov, and Diaferio}]{DoBy08}
Dolag K, Bykov AM, Diaferio A (2008) Non-thermal processes in cosmological
  simulations. Space Sci Rev 134:311--335. \doi{10.1007/s11214-008-9319-2}

\bibitem[{Engels et~al(2019)Engels, Schmidt, and
  Niemeyer}]{engels_modelling_2019}
Engels JF, Schmidt W, Niemeyer J (2019) Modelling turbulent effects of stellar
  feedback in cosmological simulations. \mnras 482(4):4654--4672.
  \doi{10.1093/mnras/sty3037}

\bibitem[{Evoli and Ferrara(2011)}]{EvoFerr11}
Evoli C, Ferrara A (2011) Turbulence in the intergalactic medium. \mnras
  413:2721--2734. \doi{10.1111/j.1365-2966.2011.18343.x}

\bibitem[{Falkovich(1994)}]{Falko94}
Falkovich G (1994) Bottleneck phenomenon in developed turbulence. Phys Fluids
  6:1411--1414. \doi{10.1063/1.868255}

\bibitem[{Federrath(2013)}]{federrath_universality_2013}
Federrath C (2013) On the universality of supersonic turbulence. \mnras
  436(2):1245--1257. \doi{10.1093/mnras/stt1644}

\bibitem[{Federrath(2015)}]{Federrath15}
Federrath C (2015) {Inefficient star formation through turbulence, magnetic
  fields and feedback}. \mnras 450(4):4035--4042. \doi{10.1093/mnras/stv941}

\bibitem[{Federrath and Klessen(2012)}]{FederKless12}
Federrath C, Klessen RS (2012) The star formation rate of turbulent magnetized
  clouds: Comparing theory, simulations, and observations. \apj 761:156.
  \doi{10.1088/0004-637X/761/2/156}

\bibitem[{Federrath et~al(2010)Federrath, Roman-Duval, Klessen, Schmidt, and
  Mac~Low}]{FederRom10}
Federrath C, Roman-Duval J, Klessen RS, et~al (2010) Comparing the statistics
  of interstellar turbulence in simulations and observations. {S}olenoidal
  versus compressive turbulence forcing. \aap 512:A81.
  \doi{10.1051/0004-6361/200912437}

\bibitem[{{Federrath} et~al(2021){Federrath}, {Klessen}, {Iapichino}, and
  {Beattie}}]{federrath_sonic_2021}
{Federrath} C, {Klessen} RS, {Iapichino} L, et~al (2021) {The sonic scale of
  interstellar turbulence}. Nature Astronomy 5:365--371.
  \doi{10.1038/s41550-020-01282-z}

\bibitem[{Ferrari et~al(2008)Ferrari, Govoni, Schindler, Bykov, and
  Rephaeli}]{FerrGov08}
Ferrari C, Govoni F, Schindler S, et~al (2008) Observations of extended radio
  emission in clusters. Space Sci Rev 134:93--118.
  \doi{10.1007/s11214-008-9311-x}

\bibitem[{Frisch(1995)}]{Frisch}
Frisch U (1995) Turbulence: The Legacy of A. N. Kolmogorov. Cambridge
  University Press, Cambridge; New York

\bibitem[{Galtier and Banerjee(2011)}]{GalBan11}
Galtier S, Banerjee S (2011) Exact relation for correlation functions in
  compressible isothermal turbulence. Phys Rev Lett 107:134501.
  \doi{10.1103/PhysRevLett.107.134501}

\bibitem[{Garnier et~al(2009)Garnier, Adams, and Sagaut}]{Garnier}
Garnier E, Adams N, Sagaut P (2009) Large Eddy Simulation for Compressible
  Flows. Scientific Computation, Springer, Berlin; New York,
  \doi{10.1007/978-90-481-2819-8}

\bibitem[{Germano(1992)}]{Germano92}
Germano M (1992) Turbulence: the filtering approach. J Fluid Mech 238:325--336.
  \doi{10.1017/S0022112092001733}

\bibitem[{Germano et~al(1991)Germano, Piomelli, Moin, and Cabot}]{GerPio91}
Germano M, Piomelli U, Moin P, et~al (1991) A dynamic subgrid-scale eddy
  viscosity model. Phys Fluids 3:1760--1765. \doi{10.1063/1.857955}

\bibitem[{Ghosal et~al(1995)Ghosal, Lund, Moin, and Akselvoll}]{GhoLund95}
Ghosal S, Lund TS, Moin P, et~al (1995) A dynamic localization model for
  large-eddy simulation of turbulent flows. J Fluid Mech 286:229--255.
  \doi{10.1017/S0022112095000711}

\bibitem[{Gnedin et~al(2009)Gnedin, Tassis, and Kravtsov}]{GnedTass09}
Gnedin NY, Tassis K, Kravtsov AV (2009) Modeling molecular hydrogen and star
  formation in cosmological simulations. \apj 697:55--67.
  \doi{10.1088/0004-637X/697/1/55}

\bibitem[{Grete et~al(2016)Grete, Vlaykov, Schmidt, and
  Schleicher}]{GreteVlay16}
Grete P, Vlaykov DG, Schmidt W, et~al (2016) A nonlinear structural
  subgrid-scale closure for compressible {MHD}. {II}. \textit{{A} priori}
  comparison on turbulence simulation data. Phys Plasmas 23(6):062317.
  \doi{10.1063/1.4954304}

\bibitem[{{Grete} et~al(2017){Grete}, {O'Shea}, {Beckwith}, {Schmidt}, and
  {Christlieb}}]{grete_transfer_2017}
{Grete} P, {O'Shea} BW, {Beckwith} K, et~al (2017) {Energy transfer in
  compressible magnetohydrodynamic turbulence}. Phys Plasmas 24(9):092311.
  \doi{10.1063/1.4990613}

\bibitem[{Grete et~al(2017)Grete, Vlaykov, Schmidt, and
  Schleicher}]{grete_comparative_2017}
Grete P, Vlaykov DG, Schmidt W, et~al (2017) Comparative statistics of selected
  subgrid-scale models in large-eddy simulations of decaying, supersonic
  magnetohydrodynamic turbulence. Phys Rev E 95(3):033206.
  \doi{10.1103/PhysRevE.95.033206}

\bibitem[{Grete et~al(2019)Grete, Latif, Schleicher, and
  Schmidt}]{grete_intermittent_2019}
Grete P, Latif MA, Schleicher DRG, et~al (2019) Intermittent fragmentation and
  statistical variations during gas collapse in magnetized atomic cooling
  haloes. \mnras 487(4):4525--4535. \doi{10.1093/mnras/stz1568}

\bibitem[{{Grete} et~al(2023){Grete}, {O'Shea}, and
  {Beckwith}}]{grete_dynamical_range_2023}
{Grete} P, {O'Shea} BW, {Beckwith} K (2023) {As a matter of dynamical range -
  scale dependent energy dynamics in MHD turbulence}. \apjl 942(2):L34.
  \doi{10.3847/2041-8213/acaea7}

\bibitem[{Haugen and Brandenburg(2006)}]{HauBrand06}
Haugen NEL, Brandenburg A (2006) Hydrodynamic and hydromagnetic energy spectra
  from large eddy simulations. Phys Fluids 18:075106. \doi{10.1063/1.2222399}

\bibitem[{Heitmann et~al(2005)Heitmann, Ricker, Warren, and Habib}]{HeitRick05}
Heitmann K, Ricker PM, Warren MS, et~al (2005) Robustness of cosmological
  simulations. {I}. {L}arge-scale structure. \apjs 160:28--58.
  \doi{10.1086/432646}

\bibitem[{Hennebelle and Chabrier(2011)}]{HenneChab11}
Hennebelle P, Chabrier G (2011) Analytical star formation rate from
  gravoturbulent fragmentation. \apjl 743:L29.
  \doi{10.1088/2041-8205/743/2/L29}

\bibitem[{Hennebelle and Falgarone(2012)}]{HenneFalg12}
Hennebelle P, Falgarone E (2012) Turbulent molecular clouds. \aapr 20:55.
  \doi{10.1007/s00159-012-0055-y}

\bibitem[{{Hillebrandt} and {Kupka}(2009)}]{hille_lnp_2009}
{Hillebrandt} W, {Kupka} F (2009) {Interdisciplinary Aspects of Turbulence},
  Lecture Notes in Physics, vol 756. Springer, \doi{10.1007/978-3-540-78961-1}

\bibitem[{Hillebrandt et~al(2013)Hillebrandt, Kromer, R{\"o}pke, and
  Ruiter}]{HilleKrom13}
Hillebrandt W, Kromer M, R{\"o}pke FK, et~al (2013) Towards an understanding of
  type {Ia} supernovae from a synthesis of theory and observations. Front Phys
  8:116--143. \doi{10.1007/s11467-013-0303-2}

\bibitem[{{Hopkins} et~al(2014){Hopkins}, {Kere{\v{s}}}, {O{\~n}orbe},
  {Faucher-Gigu{\`e}re}, {Quataert}, {Murray}, and {Bullock}}]{Hopkins14}
{Hopkins} PF, {Kere{\v{s}}} D, {O{\~n}orbe} J, et~al (2014) {Galaxies on FIRE
  (Feedback In Realistic Environments): stellar feedback explains
  cosmologically inefficient star formation}. \mnras 445(1):581--603.
  \doi{10.1093/mnras/stu1738}

\bibitem[{Iapichino et~al(2011)Iapichino, Schmidt, Niemeyer, and
  Merklein}]{IapSchm11}
Iapichino L, Schmidt W, Niemeyer JC, et~al (2011) Turbulence production and
  turbulent pressure support in the intergalactic medium. \mnras
  414:2297--2308. \doi{10.1111/j.1365-2966.2011.18550.x}

\bibitem[{Iapichino et~al(2013)Iapichino, Viel, and Borgani}]{IapViel13}
Iapichino L, Viel M, Borgani S (2013) Turbulence driven by structure formation
  in the circumgalactic medium. \mnras 432:2529--2540.
  \doi{10.1093/mnras/stt611}

\bibitem[{Jha(2017)}]{Jha2017}
Jha SW (2017) Type {Iax} supernovae. In: Alsabti AW, Murdin P (eds) Handbook of
  Supernovae. Springer, Cham, p 375--401, \doi{10.1007/978-3-319-21846-5_42}

\bibitem[{Joung et~al(2009)Joung, Mac~Low, and Bryan}]{JoungLow09}
Joung MR, Mac~Low MM, Bryan GL (2009) Dependence of interstellar turbulent
  pressure on supernova rate. \apj 704:137--149.
  \doi{10.1088/0004-637X/704/1/137}

\bibitem[{Khokhlov et~al(1997)Khokhlov, Oran, and Wheeler}]{KhokhOran97}
Khokhlov AM, Oran ES, Wheeler JC (1997) Deflagration-to-detonation transition
  in thermonuclear supernovae. \apj 478:678. \doi{10.1086/303815}

\bibitem[{Kim et~al(2016)Kim, Agertz, Teyssier, Butler, Ceverino, Choi,
  Feldmann, Keller, Lupi, Quinn, Revaz, Wallace, Gnedin, Leitner, Shen, Smith,
  Thompson, Turk, Abel, Arraki, Benincasa, Chakrabarti, DeGraf, Dekel,
  Goldbaum, Hopkins, Hummels, Klypin, Li, Madau, Mandelker, Mayer, Nagamine,
  Nickerson, O'Shea, Primack, Roca-F{\`a}brega, Semenov, Shimizu, Simpson,
  Todoroki, Wadsley, Wise, and {AGORA Collaboration}}]{kim_agora_2016}
Kim Jh, Agertz O, Teyssier R, et~al (2016) The {AGORA} high-resolution galaxy
  simulations comparison project. {II}. {Isolated} disk test. \apj 833:202.
  \doi{10.3847/1538-4357/833/2/202}

\bibitem[{Kim and Menon(1999)}]{KimMen99}
Kim WW, Menon S (1999) An unsteady incompressible navier-stokes solver for
  large eddy simulation of turbulent flows. Int J Numer Meth Fluids
  31:983--1017.
  \doi{10.1002/(SICI)1097-0363(19991130)31:6<983::AID-FLD908>3.0.CO;2-Q}

\bibitem[{Kraichnan(1976)}]{Kraich76}
Kraichnan RH (1976) Eddy viscosity in two and three dimensions. J Atmos Sci
  33:1521--1536. \doi{10.1175/1520-0469(1976)033<1521:EVITAT>2.0.CO;2}

\bibitem[{Kravtsov and Borgani(2012)}]{KravtBor12}
Kravtsov AV, Borgani S (2012) Formation of galaxy clusters. \araa 50:353--409.
  \doi{10.1146/annurev-astro-081811-125502}

\bibitem[{Kritsuk et~al(2007)Kritsuk, Norman, Padoan, and Wagner}]{KritNor07}
Kritsuk AG, Norman ML, Padoan P, et~al (2007) The statistics of supersonic
  isothermal turbulence. \apj 665:416--431. \doi{10.1086/519443}

\bibitem[{Kritsuk et~al(2013)Kritsuk, Wagner, and Norman}]{KritWag13}
Kritsuk AG, Wagner R, Norman ML (2013) Energy cascade and scaling in supersonic
  isothermal turbulence. J Fluid Mech 729:R1. \doi{10.1017/jfm.2013.342}

\bibitem[{Krumholz and McKee(2005)}]{KrumMcKee05}
Krumholz MR, McKee CF (2005) A general theory of turbulence-regulated star
  formation, from spirals to ultraluminous infrared galaxies. \apj
  630:250--268. \doi{10.1086/431734}

\bibitem[{Krumholz et~al(2009)Krumholz, McKee, and Tumlinson}]{KrumMcKee09}
Krumholz MR, McKee CF, Tumlinson J (2009) The star formation law in atomic and
  molecular gas. \apj 699:850--856. \doi{10.1088/0004-637X/699/1/850},
  {\href{https://arxiv.org/abs/0904.0009}{{arXiv:0904.0009}}}

\bibitem[{{Latif} et~al(2013{\natexlab{a}}){Latif}, {Schleicher}, {Schmidt},
  and {Niemeyer}}]{latif_black_2013}
{Latif} MA, {Schleicher} DRG, {Schmidt} W, et~al (2013{\natexlab{a}}) {Black
  hole formation in the early Universe}. \mnras 433(2):1607--1618.
  \doi{10.1093/mnras/stt834}

\bibitem[{{Latif} et~al(2013{\natexlab{b}}){Latif}, {Schleicher}, {Schmidt},
  and {Niemeyer}}]{latif_small-scale_2013}
{Latif} MA, {Schleicher} DRG, {Schmidt} W, et~al (2013{\natexlab{b}}) {The
  small-scale dynamo and the amplification of magnetic fields in massive
  primordial haloes}. \mnras 432(1):668--678. \doi{10.1093/mnras/stt503}

\bibitem[{L{\'e}v{\^e}que et~al(2007)L{\'e}v{\^e}que, Toschi, Shao, and
  Bertoglio}]{LevTosch07}
L{\'e}v{\^e}que E, Toschi F, Shao L, et~al (2007) Shear-improved {Smagorinsky}
  model for large-eddy simulation of wall-bounded turbulent flows. J Fluid Mech
  570:491--502. \doi{10.1017/S0022112006003429}

\bibitem[{Liu et~al(1994)Liu, Meneveau, and Katz}]{LiuMen94}
Liu S, Meneveau C, Katz J (1994) On the properties of similarity subgrid-scale
  models as deduced from measurements in a turbulent jet. J Fluid Mech
  275:83--119. \doi{10.1017/S0022112094002296}

\bibitem[{Liu et~al(2022)Liu, Kretschmer, and Teyssier}]{liu_subgrid_2022}
Liu Y, Kretschmer M, Teyssier R (2022) A subgrid turbulent mean-field dynamo
  model for cosmological galaxy formation simulations. \mnras
  513(4):6028--6041. \doi{10.1093/mnras/stac1266}

\bibitem[{Maier et~al(2009)Maier, Iapichino, Schmidt, and
  Niemeyer}]{MaierIap09}
Maier A, Iapichino L, Schmidt W, et~al (2009) Adaptively refined large eddy
  simulations of a galaxy cluster: Turbulence modeling and the physics of the
  intracluster medium. \apj 707:40--54. \doi{10.1088/0004-637X/707/1/40}

\bibitem[{Malone et~al(2014)Malone, Nonaka, Woosley, Almgren, Bell, Dong, and
  Zingale}]{MalNona13}
Malone CM, Nonaka A, Woosley SE, et~al (2014) The deflagration stage of
  chandrasekhar mass models for type ia supernovae. {I}. {E}arly evolution.
  \apj 782:11. \doi{10.1088/0004-637X/782/1/11},
  {\href{https://arxiv.org/abs/1309.4042}{{arXiv:1309.4042}}}

\bibitem[{{Miniati} and {Beresnyak}(2015)}]{Miniati15}
{Miniati} F, {Beresnyak} A (2015) {Self-similar energetics in large clusters of
  galaxies}. \nat 523(7558):59--62. \doi{10.1038/nature14552}

\bibitem[{Moin et~al(1991)Moin, Squires, Cabot, and Lee}]{MoinSqui91}
Moin P, Squires K, Cabot W, et~al (1991) A dynamic subgrid-scale model for
  compressible turbulence and scalar transport. Phys Fluids 3:2746--2757.
  \doi{10.1063/1.858164}

\bibitem[{Murante et~al(2014)Murante, Monaco, Borgani, Tornatore, Dolag, and
  Goz}]{Murante14}
Murante G, Monaco P, Borgani S, et~al (2014) {Simulating realistic disc
  galaxies with a novel sub-resolution ISM model}. \mnras 447(1):178--201.
  \doi{10.1093/mnras/stu2400}

\bibitem[{Nelson et~al(2019)Nelson, Pillepich, Springel, Pakmor, Weinberger,
  Genel, Torrey, Vogelsberger, Marinacci, and Hernquist}]{nelson19}
Nelson D, Pillepich A, Springel V, et~al (2019) First results from the {TNG50}
  simulation: galactic outflows driven by supernovae and black hole feedback.
  \mnras 490(3):3234--3261. \doi{10.1093/mnras/stz2306}

\bibitem[{Niemeyer and Hillebrandt(1995)}]{NieHille95}
Niemeyer JC, Hillebrandt W (1995) Turbulent nuclear flames in type {Ia}
  supernovae. \apj 452:769. \doi{10.1086/176345}

\bibitem[{Niemeyer and Kerstein(1997)}]{NieKer97}
Niemeyer JC, Kerstein AR (1997) Burning regimes of nuclear flames in {SN} {Ia}
  explosions. New Astron 2:239--244. \doi{10.1016/S1384-1076(97)00017-1}

\bibitem[{Nonaka et~al(2012)Nonaka, Aspden, Zingale, Almgren, Bell, and
  Woosley}]{NonaAsp12}
Nonaka A, Aspden AJ, Zingale M, et~al (2012) High-resolution simulations of
  convection preceding ignition in type {Ia} supernovae using adaptive mesh
  refinement. \apj 745:73. \doi{10.1088/0004-637X/745/1/73}

\bibitem[{Osher and Sethian(1988)}]{OshSeth88}
Osher S, Sethian JA (1988) Fronts propagating with curvature-dependent speed:
  Algorithms based on {Hamilton-Jacobi} formulations. J Chem Phys 79:12--49.
  \doi{10.1016/0021-9991(88)90002-2}

\bibitem[{Padoan and Nordlund(2002)}]{Padoan_2002}
Padoan P, Nordlund {\AA} (2002) The stellar initial mass function from
  turbulent fragmentation. \apj 576(2):870. \doi{10.1086/341790}

\bibitem[{Padoan and Nordlund(2011)}]{Padoan11}
Padoan P, Nordlund {\AA} (2011) The star formation rate of supersonic
  magnetohydrodynamic turbulence. \apj 730(1):40.
  \doi{10.1088/0004-637X/730/1/40}

\bibitem[{Padoan et~al(2012)Padoan, Haugb{\o}lle, and Nordlund}]{PadHaug12}
Padoan P, Haugb{\o}lle T, Nordlund {\AA} (2012) A simple law of star formation.
  \apjl 759:L27. \doi{10.1088/2041-8205/759/2/L27}

\bibitem[{Palenzuela et~al(2022{\natexlab{a}})Palenzuela, Aguilera-Miret,
  Carrasco, Ciolfi, Kalinani, Kastaun, Mi{\~n}ano, and
  Vigan{\`o}}]{palenzuela_turbulent_2022}
Palenzuela C, Aguilera-Miret R, Carrasco F, et~al (2022{\natexlab{a}})
  Turbulent magnetic field amplification in binary neutron star mergers. Phys
  Rev D 106(2):023013. \doi{10.1103/PhysRevD.106.023013}

\bibitem[{Palenzuela et~al(2022{\natexlab{b}})Palenzuela, Liebling, and
  Mi{\~n}ano}]{palenzuela_large_2022}
Palenzuela C, Liebling S, Mi{\~n}ano B (2022{\natexlab{b}}) Large eddy
  simulations of magnetized mergers of neutron stars with neutrinos. Phys Rev D
  105(10):103020. \doi{10.1103/PhysRevD.105.103020}

\bibitem[{Pan et~al(2009)Pan, Padoan, and Kritsuk}]{PanPad09}
Pan L, Padoan P, Kritsuk AG (2009) Dissipative structures in supersonic
  turbulence. Phys Rev Lett 102:034501. \doi{10.1103/PhysRevLett.102.034501}

\bibitem[{{Parrish} et~al(2012){Parrish}, {McCourt}, {Quataert}, and
  {Sharma}}]{parrish_viscosity_2012}
{Parrish} IJ, {McCourt} M, {Quataert} E, et~al (2012) {The effects of
  anisotropic viscosity on turbulence and heat transport in the intracluster
  medium}. \mnras 422(1):704--718. \doi{10.1111/j.1365-2966.2012.20650.x}

\bibitem[{Peacock(1999)}]{Peacock}
Peacock JA (1999) Cosmological Physics. Cambridge University Press, Cambridge;
  New York

\bibitem[{Peters(1999)}]{Peters99}
Peters N (1999) The turbulent burning velocity for large-scale and small-scale
  turbulence. J Fluid Mech 384:107--132. \doi{10.1017/S0022112098004212}

\bibitem[{Piomelli(1993)}]{Pio93}
Piomelli U (1993) High {Reynolds} number calculations using the dynamic
  subgrid-scale stress model. Phys Fluids 5:1484--1490. \doi{10.1063/1.858586}

\bibitem[{Piomelli and Liu(1995)}]{PioLiu95}
Piomelli U, Liu J (1995) Large-eddy simulation of rotating channel flows using
  a localized dynamic model. Phys Fluids 7:839--848. \doi{10.1063/1.868607}

\bibitem[{Pocheau(1994)}]{Poch94}
Pocheau A (1994) Scale invariance in turbulent front propagation. Phys Rev E
  49:1109--1122. \doi{10.1103/PhysRevE.49.1109}

\bibitem[{Poludnenko et~al(2019)Poludnenko, Chambers, Ahmed, Gamezo, and
  Taylor}]{poludnenko_detonation_19}
Poludnenko AY, Chambers J, Ahmed K, et~al (2019) A unified mechanism for
  unconfined deflagration-to-detonation transition in terrestrial chemical
  systems and type {Ia} supernovae. Science 366(6465):eaau7365.
  \doi{10.1126/science.aau7365}

\bibitem[{{Polzin} et~al(2024){Polzin}, {Kravtsov}, {Semenov}, and
  {Gnedin}}]{polzin_universality_2024}
{Polzin} A, {Kravtsov} AV, {Semenov} VA, et~al (2024) {On the universality of
  star formation efficiency in galaxies}. Open J Astrophys 7:114.
  \doi{10.33232/001c.127042}

\bibitem[{Pope(2000)}]{Pope}
Pope SB (2000) Turbulent flows. Cambridge University Press, Cambridge; New York

\bibitem[{Reinecke et~al(1999)Reinecke, Hillebrandt, Niemeyer, Klein, and
  Gr{\"o}bl}]{ReinHille99}
Reinecke M, Hillebrandt W, Niemeyer JC, et~al (1999) A new model for
  deflagration fronts in reactive fluids. \aap 347:724--733

\bibitem[{Reinecke et~al(2002)Reinecke, Hillebrandt, and
  Niemeyer}]{ReinHille02}
Reinecke M, Hillebrandt W, Niemeyer JC (2002) Three-dimensional simulations of
  type {Ia} supernovae. \aap 391:1167--1172. \doi{10.1051/0004-6361:20020885}

\bibitem[{Renaud et~al(2013)Renaud, Bournaud, Emsellem, Elmegreen, Teyssier,
  Alves, Chapon, Combes, Dekel, Gabor, Hennebelle, and Kraljic}]{Renaud13}
Renaud F, Bournaud F, Emsellem E, et~al (2013) {A sub-parsec resolution
  simulation of the Milky Way: global structure of the interstellar medium and
  properties of molecular clouds}. \mnras 436(2):1836--1851.
  \doi{10.1093/mnras/stt1698}

\bibitem[{Rennehan(2021)}]{rennehan_mixing_2021-1}
Rennehan D (2021) Mixing matters. \mnras 506(2):2836--2852.
  \doi{10.1093/mnras/stab1813}

\bibitem[{Rennehan et~al(2019)Rennehan, Babul, Hopkins, Dav{\'e}, and
  Moa}]{rennehan_dynamic_2019-1}
Rennehan D, Babul A, Hopkins PF, et~al (2019) Dynamic localized turbulent
  diffusion and its impact on the galactic ecosystem. \mnras 483(3):3810--3831.
  \doi{10.1093/mnras/sty3376}

\bibitem[{{Roediger} et~al(2013){Roediger}, {Kraft}, {Forman}, {Nulsen}, and
  {Churazov}}]{roediger_helmholtz_2013}
{Roediger} E, {Kraft} RP, {Forman} WR, et~al (2013) {Kelvin-Helmholtz}
  instabilities at the sloshing cold fronts in the {Virgo} cluster as a measure
  for the effective intracluster medium viscosity. \apj 764(1):60.
  \doi{10.1088/0004-637X/764/1/60}

\bibitem[{R{\"o}pke(2007)}]{Roepke07}
R{\"o}pke FK (2007) Flame-driven deflagration-to-detonation transitions in type
  {Ia} supernovae? \apj 668:1103--1108. \doi{10.1086/520830}

\bibitem[{R{\"o}pke and Hillebrandt(2005)}]{RoepHille05}
R{\"o}pke FK, Hillebrandt W (2005) The distributed burning regime in type {Ia}
  supernova models. \aap 429:L29--L32. \doi{10.1051/0004-6361:200400100}

\bibitem[{R\"opke and Schmidt(2009)}]{RoepSchm09}
R\"opke FK, Schmidt W (2009) Turbulent combustion in thermonuclear supernoave.
  In: Hillebrandt W, Kupka F (eds) Interdisciplinary Aspects of Turbulence,
  Lecture Notes in Physics, vol 756. Springer, Berlin; New York, pp 255--289,
  \doi{10.1007/978-3-540-78961-1_7}

\bibitem[{{R{\"o}pke} and {Sim}(2018)}]{Roepke18}
{R{\"o}pke} FK, {Sim} SA (2018) Models for type {Ia} supernovae and related
  astrophysical transients. \ssr 214(4):72. \doi{10.1007/s11214-018-0503-8}

\bibitem[{R{\"o}pke et~al(2007)R{\"o}pke, Hillebrandt, Schmidt, Niemeyer,
  Blinnikov, and Mazzali}]{RoepHille07}
R{\"o}pke FK, Hillebrandt W, Schmidt W, et~al (2007) A three-dimensional
  deflagration model for type {Ia} supernovae compared with observations. \apj
  668:1132--1139. \doi{10.1086/521347}

\bibitem[{Sagaut(2006)}]{Sagaut}
Sagaut P (2006) Large Eddy Simulation for Incompressible Flows: An
  Introduction, 3rd edn. Springer, Berlin; New York, \doi{10.1007/b137536}

\bibitem[{Scannapieco and Br{\"u}ggen(2010)}]{ScanBruegg10}
Scannapieco E, Br{\"u}ggen M (2010) Simulating supersonic turbulence in galaxy
  outflows. \mnras 405:1634--1653. \doi{10.1111/j.1365-2966.2010.16599.x}

\bibitem[{Schmidt(2004)}]{SchmidtPhD}
Schmidt W (2004) Turbulent thermonuclear combustion in degenerate stars. PhD
  thesis, Max Planck Institute for Astrophysics and Department of Physics /
  Technical University Munich, Garching; Munich

\bibitem[{Schmidt(2007)}]{Schm07}
Schmidt W (2007) On the applicability of the level set method beyond the
  flamelet regime in thermonuclear supernova simulations. \aap 465:263--269.
  \doi{10.1051/0004-6361:20066510}

\bibitem[{Schmidt and Federrath(2011)}]{SchmFeder11}
Schmidt W, Federrath C (2011) A fluid-dynamical subgrid scale model for highly
  compressible astrophysical turbulence. \aap 528:A106.
  \doi{10.1051/0004-6361/201015630}

\bibitem[{{Schmidt} and {Grete}(2019)}]{schmidt_kinetic_2019}
{Schmidt} W, {Grete} P (2019) {Kinetic and internal energy transfer in implicit
  large-eddy simulations of forced compressible turbulence}. \pre
  100(4):043116. \doi{10.1103/PhysRevE.100.043116}

\bibitem[{Schmidt and Niemeyer(2006)}]{SchmNie06a}
Schmidt W, Niemeyer JC (2006) Thermonuclear supernova simulations with
  stochastic ignition. \aap 446:627--633. \doi{10.1051/0004-6361:20054145}

\bibitem[{Schmidt et~al(2006{\natexlab{a}})Schmidt, Hillebrandt, and
  Niemeyer}]{SchmHille06}
Schmidt W, Hillebrandt W, Niemeyer JC (2006{\natexlab{a}}) Numerical
  dissipation and the bottleneck effect in simulations of compressible
  isotropic turbulence. Comput Fluids 35:353--371.
  \doi{10.1016/j.compfluid.2005.03.002}

\bibitem[{Schmidt et~al(2006{\natexlab{b}})Schmidt, Niemeyer, and
  Hillebrandt}]{SchmNie06b}
Schmidt W, Niemeyer JC, Hillebrandt W (2006{\natexlab{b}}) A localised subgrid
  scale model for fluid dynamical simulations in astrophysics. {I}. {T}heory
  and numerical tests. \aap 450:265--281. \doi{10.1051/0004-6361:20053617}

\bibitem[{Schmidt et~al(2006{\natexlab{c}})Schmidt, Niemeyer, Hillebrandt, and
  R{\"o}pke}]{SchmNie06c}
Schmidt W, Niemeyer JC, Hillebrandt W, et~al (2006{\natexlab{c}}) A localised
  subgrid scale model for fluid dynamical simulations in astrophysics. {II}.
  {A}pplication to type {Ia} supernovae. \aap 450:283--294.
  \doi{10.1051/0004-6361:20053618}

\bibitem[{Schmidt et~al(2008)Schmidt, Federrath, and Klessen}]{SchmFeder08}
Schmidt W, Federrath C, Klessen R (2008) Is the scaling of supersonic
  turbulence universal? Phys Rev Lett 101:194505.
  \doi{10.1103/PhysRevLett.101.194505}

\bibitem[{Schmidt et~al(2009)Schmidt, Federrath, Hupp, Kern, and
  Niemeyer}]{SchmFeder09}
Schmidt W, Federrath C, Hupp M, et~al (2009) Numerical simulations of
  compressively driven interstellar turbulence. {I}. {I}sothermal gas. \aap
  494:127--145. \doi{10.1051/0004-6361:200809967}

\bibitem[{Schmidt et~al(2014)Schmidt, Almgren, Braun, Engels, Niemeyer, Schulz,
  Mekuria, Aspden, and Bell}]{SchmAlm13}
Schmidt W, Almgren AS, Braun H, et~al (2014) Cosmological fluid mechanics with
  adaptively refined large eddy simulations. \mnras 440:3051--3077.
  \doi{10.1093/mnras/stu501}

\bibitem[{Schmidt et~al(2016)Schmidt, Engels, Niemeyer, and
  Almgren}]{schmidt_hot_2016}
Schmidt W, Engels JF, Niemeyer JC, et~al (2016) Hot and turbulent gas in
  clusters. \mnras 459(1):701--719. \doi{10.1093/mnras/stw632}

\bibitem[{Schmidt et~al(2017)Schmidt, Byrohl, Engels, Behrens, and
  Niemeyer}]{schmidt_viscosity_2017-1}
Schmidt W, Byrohl C, Engels JF, et~al (2017) Viscosity, pressure and support of
  the gas in simulations of merging cool-core clusters. \mnras 470(1):142--156.
  \doi{10.1093/mnras/stx1274}

\bibitem[{Schmidt et~al(2021)Schmidt, Schmidt, and
  Grete}]{schmidt_turbulence_2021-2}
Schmidt W, Schmidt JP, Grete P (2021) Turbulence in the intragroup and
  circumgalactic medium. \aap 654:A115. \doi{10.1051/0004-6361/202140920}

\bibitem[{Schober et~al(2013)Schober, Schleicher, and
  Klessen}]{schober_magnetic_2013}
Schober J, Schleicher DRG, Klessen RS (2013) Magnetic field amplification in
  young galaxies. \aap 560:A87. \doi{10.1051/0004-6361/201322185}

\bibitem[{Schumann(1975)}]{Schu75}
Schumann U (1975) Subgrid scale model for finite difference simulations of
  turbulent flows in plane channels and annuli. J Comput Phys 18:376--404.
  \doi{10.1016/0021-9991(75)90093-5}

\bibitem[{Seitenzahl et~al(2012)Seitenzahl, Ciaraldi-Schoolmann, R{\"o}pke,
  Fink, Hillebrandt, Kromer, Pakmor, Ruiter, Sim, and
  Taubenberger}]{Seitenzahl12}
Seitenzahl IR, Ciaraldi-Schoolmann F, R{\"o}pke FK, et~al (2012)
  Three-dimensional delayed-detonation models with nucleosynthesis for type
  {Ia} supernovae. \mnras 429(2):1156--1172. \doi{10.1093/mnras/sts402}

\bibitem[{Semenov(2024)}]{semenov_capturing_2024}
Semenov VA (2024) Capturing turbulence with numerical dissipation: a simple
  dynamical model for unresolved turbulence in hydrodynamic simulations. arXiv
  e-prints \doi{10.48550/arXiv.2410.23339},
  {\href{https://arxiv.org/abs/2410.23339}{{2410.23339}}}

\bibitem[{Semenov et~al(2016)Semenov, Kravtsov, and Gnedin}]{Semenov16}
Semenov VA, Kravtsov AV, Gnedin NY (2016) Nonuniversal star formation
  efficiency in turbulent {ISM}. \apj 826(2):200.
  \doi{10.3847/0004-637X/826/2/200}

\bibitem[{{Semenov} et~al(2025{\natexlab{a}}){Semenov}, {Conroy}, and
  {Hernquist}}]{semenov_uv-bright_2024}
{Semenov} VA, {Conroy} C, {Hernquist} L (2025{\natexlab{a}}) {From UV-bright
  Galaxies to Early Disks: The Importance of Turbulent Star Formation in the
  Early Universe}. \apj 989(2):219. \doi{10.3847/1538-4357/ade22d}

\bibitem[{{Semenov} et~al(2025{\natexlab{b}}){Semenov}, {Conroy}, {Smith},
  {Puchwein}, and {Hernquist}}]{semenov_how_2024}
{Semenov} VA, {Conroy} C, {Smith} A, et~al (2025{\natexlab{b}}) {How Early
  Could the Milky Way's Disk Form?} \apj 990(1):7.
  \doi{10.3847/1538-4357/addf48}

\bibitem[{Shukurov and Subramanian(2021)}]{Shukurov_Subramanian_2021}
Shukurov A, Subramanian K (2021) Elements of Magnetohydrodynamics, Cambridge
  University Press, pp 5--67. Cambridge Astrophysics,
  \doi{10.1017/9781139046657.003}

\bibitem[{Smagorinsky(1963)}]{Smago63}
Smagorinsky J (1963) General circulation experiments with the primitive
  equations. Mon Wea Rev 91:99--164.
  \doi{10.1175/1520-0493(1963)091<0099:GCEWTP>2.3.CO;2}

\bibitem[{{Speziale} et~al(1988){Speziale}, {Hussaini}, {Erlebacher}, and
  {Zang}}]{speziale_subgrid_1988}
{Speziale} CG, {Hussaini} MY, {Erlebacher} G, et~al (1988) {The subgrid-scale
  modeling of compressible turbulence}. Phys Fluids 31(4):940--942.
  \doi{10.1063/1.866778}

\bibitem[{Springel and Hernquist(2003)}]{SpringHern03}
Springel V, Hernquist L (2003) Cosmological smoothed particle hydrodynamics
  simulations: a hybrid multiphase model for star formation. \mnras
  339:289--311. \doi{10.1046/j.1365-8711.2003.06206.x}

\bibitem[{{Spruit}(2016)}]{Spruit16}
{Spruit} HC (2016) Essential magnetohydrodynamics for astrophysics. arXiv
  e-prints arXiv:1301.5572. \doi{10.48550/arXiv.1301.5572},
  {\href{https://arxiv.org/abs/1301.5572}{{arXiv:1301.5572}}} {[astro-ph.IM]}

\bibitem[{Steinwandel et~al(2024)Steinwandel, Rennehan, Orr, Fielding, and
  Kim}]{steinwandel_pumping_2024}
Steinwandel UP, Rennehan D, Orr ME, et~al (2024) Pumping iron: {How} turbulent
  metal diffusion impacts multiphase galactic outflows. arXiv e-prints
  \doi{10.48550/arXiv.2407.14599}

\bibitem[{Su et~al(2017)Su, Hopkins, Hayward, Faucher-Gigu{\`e}re, Kere{\v s},
  Ma, and Robles}]{Su17}
Su KY, Hopkins PF, Hayward CC, et~al (2017) {Feedback first: the surprisingly
  weak effects of magnetic fields, viscosity, conduction and metal diffusion on
  sub-L* galaxy formation}. \mnras 471(1):144--166. \doi{10.1093/mnras/stx1463}

\bibitem[{Sytine et~al(2000)Sytine, Porter, Woodward, Hodson, and
  Winkler}]{SyPort00}
Sytine IV, Porter DH, Woodward PR, et~al (2000) Convergence tests for the
  piecewise parabolic method and {Navier-Stokes} solutions for homogeneous
  compressible turbulence. J Comput Phys 158:225--238.
  \doi{10.1006/jcph.1999.6416}

\bibitem[{Taubenberger(2017)}]{Taubenberger2017}
Taubenberger S (2017) The extremes of thermonuclear supernovae. In: Alsabti AW,
  Murdin P (eds) Handbook of Supernovae. Springer, Cham, p 317--373,
  \doi{10.1007/978-3-319-21846-5_37}

\bibitem[{Vasilyev et~al(1998)Vasilyev, Lund, and Moin}]{VasilLund98}
Vasilyev OV, Lund TS, Moin P (1998) A general class of commutative filters for
  les in complex geometries. J Comput Phys 146:82--104.
  \doi{10.1006/jcph.1998.6060}

\bibitem[{{Vazza} et~al(2011){Vazza}, {Brunetti}, {Gheller}, {Brunino}, and
  {Br{\"u}ggen}}]{Vazza11}
{Vazza} F, {Brunetti} G, {Gheller} C, et~al (2011) {Massive and refined. II.
  The statistical properties of turbulent motions in massive galaxy clusters
  with high spatial resolution}. \aap 529:A17.
  \doi{10.1051/0004-6361/201016015}

\bibitem[{Vigan{\`o} et~al(2020)Vigan{\`o}, Aguilera-Miret, Carrasco,
  Mi{\~n}ano, and Palenzuela}]{vigano_general_2020}
Vigan{\`o} D, Aguilera-Miret R, Carrasco F, et~al (2020) General relativistic
  {MHD} large eddy simulations with gradient subgrid-scale model. Phys Rev D
  101(12):123019. \doi{10.1103/PhysRevD.101.123019}

\bibitem[{Vlaykov et~al(2016)Vlaykov, Grete, Schmidt, and
  Schleicher}]{VlayGrete16}
Vlaykov DG, Grete P, Schmidt W, et~al (2016) A nonlinear structural
  subgrid-scale closure for compressible {MHD}. {I}. {Derivation} and energy
  dissipation properties. Phys Plasmas 23(6):062316. \doi{10.1063/1.4954303}

\bibitem[{Vollant et~al(2016)Vollant, Balarac, and
  Corre}]{vollant_dynamic_2016}
Vollant A, Balarac G, Corre C (2016) A dynamic regularized gradient model of
  the subgrid-scale stress tensor for large-eddy simulation. Phys Fluids
  28(2):025114. \doi{10.1063/1.4941781}

\bibitem[{Wagner et~al(2012)Wagner, Falkovich, Kritsuk, and Norman}]{WagFalk12}
Wagner R, Falkovich G, Kritsuk AG, et~al (2012) Flux correlations in supersonic
  isothermal turbulence. J Fluid Mech 713:482--490. \doi{10.1017/jfm.2012.470}

\bibitem[{Wang et~al(2010)Wang, Klessen, Dullemond, van~den Bosch, and
  Fuchs}]{wang_equilibrium_2010}
Wang HH, Klessen RS, Dullemond CP, et~al (2010) Equilibrium initialization and
  stability of three-dimensional gas discs. \mnras 407(2):705--720.
  \doi{10.1111/j.1365-2966.2010.16942.x}

\bibitem[{Wang and He(2024)}]{Wang24}
Wang Y, He P (2024) Turbulence, thermal pressure, and their dynamical effects
  on cosmic baryonic fluid. \mnras 534(1):L14--L20.
  \doi{10.1093/mnrasl/slae073}

\bibitem[{Weinberger et~al(2016)Weinberger, Springel, Hernquist, Pillepich,
  Marinacci, Pakmor, Nelson, Genel, Vogelsberger, Naiman, and
  Torrey}]{weinberger16}
Weinberger R, Springel V, Hernquist L, et~al (2016) Simulating galaxy formation
  with black hole driven thermal and kinetic feedback. \mnras
  465(3):3291--3308. \doi{10.1093/mnras/stw2944}

\bibitem[{Woodward et~al(2006)Woodward, Porter, Anderson, Fuchs, and
  Herwig}]{WoodPort06}
Woodward PR, Porter DH, Anderson S, et~al (2006) Large-scale simulations of
  turbulent stellar convection flows and the outlook for petascale computation.
  J Phys: Conf Ser 46:370--384. \doi{10.1088/1742-6596/46/1/052}

\bibitem[{Yoshizawa(1991)}]{Yoshi91}
Yoshizawa A (1991) Subgrid-scale modeling of compressible turbulent flows. Phys
  Fluids A 3:714. \doi{10.1063/1.858078}

\bibitem[{Zweibel and Yamada(2009)}]{ZweiYa09}
Zweibel EG, Yamada M (2009) Magnetic reconnection in astrophysical and
  laboratory plasmas. \araa 47:291--332.
  \doi{10.1146/annurev-astro-082708-101726}

\end{thebibliography}

\end{document}